\pdfoutput=1
\documentclass[aps,prX,amssymb,superscriptaddress,nofootinbib,twocolumn,nopreprintnumbers]{revtex4-1}  

\usepackage{graphicx}%,bm}
\usepackage{picture}
\usepackage{placeins}
\usepackage{float}
\DeclareGraphicsExtensions{.pdf} %.png
\usepackage{amsmath,scalefnt}
\usepackage{amssymb}
\usepackage{verbatim}
\usepackage{amsmath,amsfonts,bbm}
\usepackage{amsbsy}
\usepackage{color}
\usepackage{cancel}
\usepackage{soul}
\usepackage{dsfont}
\usepackage[svgnames]{xcolor} 

\usepackage[breaklinks=true,colorlinks=true,linkcolor=DarkBlue,urlcolor=DarkBlue,citecolor=DarkBlue]{hyperref}

\usepackage{subfigure}

\def\bs#1{\boldsymbol{#1}}

\begin{document}

\title{Quantum decoherence of phonons in Bose-Einstein condensates}

\author{Richard Howl}
\email[Author to whom correspondence should be addressed: ]{richard.howl@univie.ac.at}
	\affiliation{University of Vienna, Faculty of Physics, Boltzmanngasse 5, 1090 Wien, Austria}

\author{Carlos Sab\'{i}n}
\affiliation{Instituto de F\'{i}sica Fundamental, CSIC, Serrano 113-bis 28006 Madrid, Spain}

\author{Lucia Hackerm\"{u}ller}
\affiliation{School of Physics and Astronomy,
University of Nottingham,
University Park,
Nottingham NG7 2RD,
United Kingdom}

\author{Ivette Fuentes}
\affiliation{University of Vienna, Faculty of Physics, Boltzmanngasse 5, 1090 Wien, Austria}
  \affiliation{School of Mathematical Sciences,
University of Nottingham,
University Park,
Nottingham NG7 2RD,
United Kingdom}

\begin{abstract} 
We apply modern techniques from quantum optics and quantum information science to Bose-Einstein Condensates (BECs) in order to  study, for the first time, the quantum decoherence of phonons of isolated BECs. In the last few years, major advances in the manipulation and control of phonons have highlighted their potential as carriers of quantum information in quantum technologies, particularly in quantum processing and quantum communication.  Although most of these studies have focused on trapped ion and crystalline systems, another promising system that has remained relatively unexplored is BECs.  The potential benefits in using this system have been emphasized recently with proposals of relativistic quantum devices that exploit quantum states of phonons in BECs to achieve, in principle, superior performance over standard non-relativistic devices. Quantum decoherence is often the limiting factor in the practical realization of quantum technologies, but here we show that quantum decoherence of phonons is not expected to heavily constrain the performance of these proposed relativistic quantum devices.
\end{abstract}
 
\maketitle

\section*{Introduction} \label{sec:Intro}

Quantum decoherence, the environment-induced dynamical destruction of quantum coherence, has a wide range of applications.  For example, it plays an important role in fundamental questions of quantum mechanics where it continues to provide insights into a potential resolution of the non-observation of quantum superpositions at macroscopic scales \cite{Decoherence1970,RevModPhys.76.1267}. Furthermore, its control is imperative to the operation and physical realization of quantum technologies that could soon revolutionize our technological world.  This is because the time in which a quantum system decoheres is often found to be much shorter than the time that characterizes its energy relaxation \cite{Joos85_2,Zurek1986,RevModPhys.75.715,2003quant.ph..6072Z}. 

BECs are considered to be promising candidates for the implementation of various quantum technologies since they can usually be well-isolated from their surroundings and so offer relatively long coherence times.  In particular, two-component Bose-Einstein condensates, either as bosons condensed in two different spatial sites  \cite{Cataliotti03082001,refId0} or as condensed bosons in different hyperfine levels 
\cite{PhysRevLett.81.1539,PhysRevLett.81.1543}, have been  widely considered for the implementation of certain quantum technologies.  A primary application for such two-component BECs is quantum metrology \cite{2010Natur.464.1165G,PhysRevLett.94.090405,2001Natur.413..498H,2013NatCo...4E2077B,PhysRevLett.100.080405,Schumm06,2000PhRvL..84.4749F}, such as atomic clocks and accelerometers, but other applications, such as quantum computation and communication, have also been investigated \cite{1367-2630-15-9-093019,PhysRevA.85.040306,2001Natur.413..498HComp,2000PhRvL..84.4749F,2001Natur.409...63S,Buluta02102009,PhysRevA.85.040306}.  In particular, a major advancement in this area has been the development of BECs on atom chips, which facilitates the control of many BECs \cite{RevModPhys.79.235,PhysRevLett.80.645,PhysRevLett.84.1124}. 

Recently, BECs have been considered in quantum technology from a fundamentally different viewpoint.  Instead of the atomic states of the BECs being used as the carriers of quantum information, the collective quantum excitations of the atoms, phonons, were used  \cite{Accelerometer,GWDetectorFirst,GWDetectorThermal} (for a recent review see e.g., \cite{ReviewPaper}).  In the last few years, the emerging field of phonon-based quantum technologies has attracted considerable attention \cite{PhysRevLett.107.235502,PhysRevB.88.064308,PhononQuantumNetworks,PhysRevX.5.031031,Gustafsson207,PhysRevLett.109.013603,PhononicComputingReview}.  For example, in \cite{PhysRevB.88.064308} a phononic equivalent of circuit-QED was considered for quantum processing, and in \cite{PhononQuantumNetworks} phononic quantum networks were proposed.   Phonons behave in similar ways to photons but their advantages include the fact that they can be localized and made to interact with each other while still maintaining long coherence times  \cite{PhononicComputingReview}.  They also have, in general, much shorter wavelengths than the photons that are created in laboratory settings, allowing for regimes of atomic physics to be explored which cannot be reached in photonic systems.  

Phonon-based quantum technologies have principally concentrated on crystalline and ion trap systems rather than BEC systems.  However, the devices proposed in \cite{Accelerometer,GWDetectorFirst,GWDetectorThermal} illustrate the potential benefits of BEC systems. 
These proposed devices exploit the fact that phonons behave similarly to photons but propagate with much slower speeds.  This has already been utilized in analogue gravity experiments and has culminated in the recent observation of an acoustic analogue of the elusive Hawking radiation of a black hole where, in this case, it is sound waves rather than light waves that cannot escape \cite{2010PhRvL.105x0401L,Steinhauer2014,Steinhauer2016}.  However, in \cite{Accelerometer,GWDetectorFirst} it was demonstrated  that the slow propagation speeds can also be used to enhance real rather than just analogue spacetime effects, allowing for the development of quantum metrology devices.  For example, a gravitational wave (GW) detector was proposed in \cite{GWDetectorFirst} where the GW modifies the phononic field in a measurable way since the slow propagation speeds allow for a resonance process in a micrometre sized system for promising GW signals.  Interestingly, initial calculations suggest that these phononic quantum devices should have a precision that is, in principle, orders of magnitude superior to the state of the art \cite{GWDetectorFirst,GWDetectorThermal}.

However, understanding the quantum decoherence of phonons in BECs will be crucial in  assessing the practical realization of these proposed phononic quantum technologies that utilise BECs.  Although quantum decoherence has been considered for the condensed atoms of a BEC, it has not, as far as we are aware, been considered for the phonons of BECs.  Instead, previous studies have concentrated on the energy relaxation time of the phonons in isolated \cite{Landau:1946jc,Beliaev1958aSet2,Beliaev1958bSet2,Hohenberg1965291,PhysicalKinetics,	KONDOR1974393,LiuAndShieve,PitaevskiiDamping,Liu,Giorgini1998,DampingTrapped,PhysRevA.58.3146,PhysRevA.58.R3391,Langevin,PhysRevA.59.3851,LBReview,CastinAppendix,PhysRevLett.81.5262,PhysRevA.62.023609,2012arXiv1203.4458C,PhysRevA.70.033615,2008PhDT........70E,PhysRevLett.102.110401}  and open BEC systems \cite{PhysRevA.93.033634}, which can be used to set only an upper bound for the time scale of \emph{quantum} decoherence.    

Here we consider the quantum decoherence processes for a single-mode phononic state of an isolated BEC. We start with a quantum field theory description of the BEC from which the phonons and their interactions can be derived.  After determining how the reduced density matrix of  a single-mode phononic state evolves in the Born-Markov approximation, we then employ techniques from quantum optics and quantum information science to determine how certain global entropic measures and nonclassical indicators evolve for this state when it is of Gaussian form, which can be used to quantify quantum decoherence. For illustrative purposes, we apply these general techniques to the specific case of a three-dimensional and uniform version of the phononic-based GW detector proposed in \cite{GWDetectorFirst}, and find that the quantum decoherence of phonons does not significantly constrain the performance of the device.

\section{Time evolution of the phonon density operator}
\subsection{Phonons in the Bogoliubov approximation} \label{sec:BogoApprox}

The quantum field Hamiltonian for a rarefied, interacting, non-relativistic Bose gas   is (see, e.g., \cite{PitaevskiiBook})
\begin{align} \nonumber
\hat{H} =  &\int  d\boldsymbol{r} \hat{\psi}^{\dagger}({\boldsymbol{r}}) \Big[-\frac{\hbar^2}{2m} \nabla^2 + \mathcal{V}(\boldsymbol{r}) \Big]\hat{\psi}({\boldsymbol{r}}) \\ \label{eq:fullH}
&+ \frac{1}{2} \int  d\boldsymbol{r} d\boldsymbol{r}^{\prime}\hat{\psi}^{\dagger}({\boldsymbol{r}}) \hat{\psi}^{\dagger}({\boldsymbol{r}^{\prime}})  \mathcal{U}(\boldsymbol{r}^{\prime} - \boldsymbol{r}) \hat{\psi}({\boldsymbol{r}}) \hat{\psi}({\boldsymbol{r}^{\prime}}),
\end{align}
where $\hat{\psi}^{\dagger}({\boldsymbol{r}})$ and $\hat{\psi}({\boldsymbol{r}})$ are the field operators creating and annihilating a bosonic atom at position $\boldsymbol{r}$; $\mathcal{U}(\boldsymbol{r})$ is the two-body potential; and $\mathcal{V}(\boldsymbol{r})$ is the trapping potential. 

 For clarity, we first consider a uniform gas occupying a volume $V$ and later discuss how the analysis is straightforwardly extended to trapped systems.  Although BECs can be created in a uniform trap \cite{PhysRevLett.110.200406}, they are usually constrained in harmonic potentials.  However, since studying uniform BECs is generally much simpler, these systems are often used as a first step in the theoretical analysis of trapped systems to describe certain properties through methods such as the local density approximation (see e.g., \cite{RevModPhys.71.463}) or by only probing small central sections of the trapped ultracold gas \cite{PhysRevLett.107.190403,PhysRevA.86.031601,PhysRevLett.109.220402}. 
 
 Considering a uniform system, we can substitute the plane-wave solutions $\hat{\psi} (\boldsymbol{r}) = \frac{1}{\sqrt{V}} \sum_{\boldsymbol{k}} \hat{a}_{\boldsymbol{k}} e^{i \boldsymbol{k}.\boldsymbol{r}}$ with $\boldsymbol{k} = 2 \pi  \boldsymbol{n} / L$; $\boldsymbol{n} = (n_x, n_y, n_z)$; and $n_x,n_y,n_z \in \mathbb{N}$ into \eqref{eq:fullH} to obtain a complete description of the gas in terms of annihilation and creation operators in momentum space:
\begin{equation} \label{eq:fullBEC}
\hat{H} = \sum_{\boldsymbol{k}} \frac{p^2}{2m} \hat{a}^{\dagger}_{\boldsymbol{k}} \hat{a}_{\boldsymbol{k}} + \frac{1}{2V} \sum_{\boldsymbol{k}, \boldsymbol{k^{\prime}}, \boldsymbol{q} } \mathcal{U}_{\boldsymbol{q}} \hat{a}^{\dagger}_{\boldsymbol{k}+\boldsymbol{q}} \hat{a}^{\dagger}_{\boldsymbol{k^{\prime}}-\boldsymbol{q}} \hat{a}_{\boldsymbol{k^{\prime}}}  \hat{a}_{\boldsymbol{k}},
\end{equation}
where $\mathcal{U}_{\boldsymbol{q}} = \int \mathcal{U}(\boldsymbol{r})e^{-i\boldsymbol{q}.\boldsymbol{r} } d\boldsymbol{r}$.

Assuming the condensate to be macroscopically occupied, we next apply the Bogoliubov approximation \cite{Bogo1947}.  This involves replacing $\hat{a}_0$ and $\hat{a}^{\dagger}_0$ with the c-number $\sqrt{N}$ and only retaining terms quadratic in $a_{\boldsymbol{k}\neq 0}$ and $a^{\dagger}_{\boldsymbol{k}\neq 0}$ (the higher-order terms are suppressed since they have fewer factors of $\sqrt{N} \gg 1$), where $N$ is the number of atoms.  In this approximation one also replaces the microscopic potential $\mathcal{U}$ with an effective soft potential and expands the $\boldsymbol{q}=0$ component up to quadratic terms in the coupling constant $g = 4 \pi \hbar^2 a / m$  where $a$ is the s-wave scattering length and $m$ is the atomic mass \cite{PitaevskiiBook,LandauLifshitzQM}.  

The resulting Hamiltonian in $a_{\boldsymbol{k}\neq 0}$ and $a^{\dagger}_{\boldsymbol{k}\neq 0}$ can then be diagonalized by applying the following Bogoliubov transformation (see, e.g., \cite{PitaevskiiBook})
\begin{align} \label{eq:Bogotrans}
&\hat{a}_{\boldsymbol{k}} = u_k \hat{b}_{\boldsymbol{k}} + v_{k} \hat{b}^{\dagger}_{-\boldsymbol{k}},\\ 
&\hat{a}^{\dagger}_{\boldsymbol{k}} = u_{k} \hat{b}^{\dagger}_{\boldsymbol{k}} + v_{k} \hat{b}_{-\boldsymbol{k}},
\end{align}
to obtain
\begin{align} \label{eq:bogoH}
\hat{H} = \epsilon_0 + \sum_{\boldsymbol{k}\neq0} \hbar \omega_k \hat{b}^{\dagger}_{\boldsymbol{k}} \hat{b}_{\boldsymbol{k}},
\end{align}
where $\hat{b}^{\dagger}_{\boldsymbol{k}}$ and $\hat{b}_{\boldsymbol{k}}$ are the creation and annihilation operators for quasi-particles; $\epsilon_0$ is their ground state energy; $c_s$ and $\omega_k$ are respectively the speed of sound and quasi-particle frequency
\begin{align}
\label{eq:ep}
(\hbar \omega_k)^2 &:=  (c_s \hbar k)^2 + (\hbar^2 k^2 /2m)^2,\\
\label{eq:c} c^2_s &:= gn / m,
\end{align}
and $u_k$ and $v_k$ must satisfy
\begin{equation} \label{eq:uv}
u_k, v_k = \pm \Big(\frac{\hbar^2 k^2/2m + m c^2_s \pm \hbar \omega_k }{2 \hbar \omega_k}  \Big)^{\frac{1}{2}},
\end{equation}
where $n$ is the number density of the gas.

As illustrated by \eqref{eq:bogoH}, the Bose gas in this Bogoliubov approximation can be described by a non-interacting  gas of quasi-particles, where the low momentum modes ($\hbar k \ll m c_s$) are phonons travelling at speed $c_s$ since $\omega_k \approx c_s k$ in this regime.  Given that we have an ideal gas of quasi-particles, in thermal equilibrium the average occupation number $N_{\boldsymbol{k}}$ of quasi-particles carrying momentum $\hbar \boldsymbol{k}$ must satisfy
\begin{equation} \label{eq:Nth}
N_{\boldsymbol{k}}:= \langle \hat{b}^{\dagger}_{\boldsymbol{k}} \hat{b}_{\boldsymbol{k}} \rangle =   \frac{1}{e^{\beta_k} - 1},
\end{equation}
where $\beta_k := \hbar \omega_k / k_B T$.    

\subsection{Phonon interactions}

In the Bogoliubov approximation the phonons are non-interacting and so have infinite lifetimes.  However, this approximation only keeps terms that are quadratic in $\hat{a}_{\boldsymbol{k}}$ and $\hat{a}^{\dagger}_{\boldsymbol{k}}$.  If we instead also include the (more suppressed) cubic and quartic terms then, after the above Bogoliubov transformation \eqref{eq:Bogotrans}, these terms will provide interactions between the quasi-particles, resulting in finite lifetimes.

Concentrating on just a single momentum mode $\boldsymbol{q}$ of the phonons, the cubic interaction terms for this mode are \cite{PitaevskiiDamping,Giorgini1998,CastinAppendix}
\begin{align} \label{eq:HI}
\hat{H}_I = \hat{b}_{\boldsymbol{q}} \hat{E}^{\dagger}_{\boldsymbol{q}} + \hat{b}^{\dagger}_{\boldsymbol{q}} \hat{E}_{\boldsymbol{q}},
\end{align}
where
\begin{align} \label{eq:E}
\hat{E}_{\boldsymbol{q}} &:= \hat{A} + \hat{B} + \hat{L},\\
\hat{A}^{\dagger}_{\boldsymbol{q}} &:= g \sqrt{\frac{n}{V}} \sum_{\boldsymbol{k},\boldsymbol{k}^{\prime} \neq \{\boldsymbol{0},\boldsymbol{k}\}} \mathcal{A}_{k,k^{\prime}}  \hat{b}_{\boldsymbol{k}} \hat{b}_{\boldsymbol{k}^{\prime}} \delta_{-\boldsymbol{q},\boldsymbol{k} + \boldsymbol{k}^{\prime}},\\
\hat{B}^{\dagger}_{\boldsymbol{q}} &:= g \sqrt{\frac{n}{V}}
 \sum_{\boldsymbol{k},\boldsymbol{k}^{\prime} \neq \{\boldsymbol{0},\boldsymbol{q}\}} \mathcal{B}_{k,k^{\prime}} 
\hat{b}^{\dagger}_{\boldsymbol{k}} \hat{b}^{\dagger}_{\boldsymbol{k}^{\prime}} \delta_{\boldsymbol{q},\boldsymbol{k} + \boldsymbol{k}^{\prime}},\\
\hat{L}^{\dagger}_{\boldsymbol{q}} &:= g \sqrt{\frac{n}{V}}
 \sum_{\boldsymbol{k},\boldsymbol{k}^{\prime} \neq \{\boldsymbol{0},\boldsymbol{q}\}} \mathcal{L}_{k,k^{\prime}}  \hat{b}_{\boldsymbol{k}} \hat{b}^{\dagger}_{\boldsymbol{k}^{\prime}} \delta_{\boldsymbol{q},\boldsymbol{k}^{\prime} -\boldsymbol{k}},\\  \nonumber
\mathcal{A}_{k,k^{\prime}} &:=~ u_q ( v_k v_{k^{\prime}} + u_k v_{k^{\prime}} + v_k u_{k^{\prime}})\\ \label{eq:Acoeff}
 &~~~~~+ v_q ( u_k v_{k^{\prime}} + v_k u_{k^{\prime}} + u_k u_{k^{\prime}}),\\ \nonumber
\mathcal{B}_{k,k^{\prime}} &:=~ u_q ( u_k u_{k^{\prime}} + v_k u_{k^{\prime}} + u_k v_{k^{\prime}})\\ \label{eq:Bcoeff}
 &~~~~~+ v_q ( v_k v_{k^{\prime}} + v_k u_{k^{\prime}} + u_k v_{k^{\prime}}),\\ \nonumber
 \mathcal{L}_{k,k^{\prime}} /2 &:=~ u_q ( v_k u_{k^{\prime}} + u_k u_{k^{\prime}} + v_k v_{k^{\prime}})\\ \label{eq:Ccoeff}
 &~~~~~+ v_q ( u_k v_{k^{\prime}} + u_k u_{k^{\prime}} + v_k v_{k^{\prime}}).
\end{align}
The resonant interactions $ \hat{b}_{\boldsymbol{q}} \hat{L}^{\dagger}_{\boldsymbol{q}}$ and $\hat{b}_{\boldsymbol{q}} \hat{B}^{\dagger}_{\boldsymbol{q}}$ are the Landau and Beliaev interactions \cite{Landau:1946jc,Beliaev1958aSet2,Beliaev1958bSet2,Hohenberg1965291,PhysicalKinetics,KONDOR1974393,LiuAndShieve,PitaevskiiDamping,Liu,Giorgini1998,DampingTrapped,PhysRevA.58.3146,PhysRevA.58.R3391,Langevin,PhysRevA.59.3851,LBReview}.   In the Landau process $ \hat{b}_{\boldsymbol{q}} \hat{L}^{\dagger}$, a quasi-particle from mode $\boldsymbol{q}$ collides with a quasi-particle from another mode to create a higher-energy quasi-particle.  Since this requires the thermal occupation of a quasi-particle mode, the process vanishes at zero temperature.  On the other hand, in the Beliaev process $\hat{b}_{\boldsymbol{q}} \hat{B}^{\dagger}$, a quasi-particle of mode $\boldsymbol{q}$ spontaneously annihilates into two new quasi-particles with lower energies, which is analogous to parametric down-conversion in quantum optics \cite{SPDC} and  can occur at absolute zero. Since the processes originate from \eqref{eq:fullBEC}, they can also be considered from the point-of-view of four-body interactions between the condensate and non-condensate atoms.

\subsection{Markov quantum master equation for a single-mode phonon state} \label{sec:Markov}

Treating the above single phonon mode $\bs{q}$ as an open quantum system, and the rest of the quasi-particle modes as its environment, we can decompose the Hamiltonian for the full system $\hat{H}$ in the following way:
\begin{equation} \label{eq:FullSystem}
\hat{H} = \hat{H}_{S} + \hat{H}_{E} + \hat{H}_I,
\end{equation}
where $\hat{H}_S$ and $\hat{H}_E$ derive from \eqref{eq:bogoH}, and respectively describe the free Hamiltonian of the considered single-mode phonon system and all other quasi-particle modes
\begin{align} \label{eq:HS}
\hat{H}_S &:=  \hbar \omega_q \hat{b}^{\dagger}_{\boldsymbol{q}} \hat{b}_{\boldsymbol{q}} ,\\
\hat{H}_E &:=  \sum_{\boldsymbol{k} \neq \{\bs{0},\bs{k}\} } \hbar \omega_k \hat{b}^{\dagger}_{\boldsymbol{k}} \hat{b}_{\boldsymbol{k}},
\end{align}
and $\hat{H}_I$ is defined in \eqref{eq:HI}.  We explicitly ignore the  interaction terms between the states of the large system $E$, which describes all quasi-particle modes with momentum $\hbar \boldsymbol{k} \neq \hbar \boldsymbol{q}  \neq 0$ and is assumed to be in thermal equilibrium. 

To determine the evolution of the single-mode phonon system,  we assume that the initial state of the full system is a product state $\hat{\rho}(0) = \hat{\rho}_S(0) \otimes \hat{\rho}_E (0)$ and perform the Born-Markov approximation.  That is, we assume that the coupling between $E$ and $S$ is weak, which is well justified for a rarefied Bose gas at low temperatures, and that the   future evolution of $\hat{\rho}_S (t)$ does not depend on its past history. The latter is satisfied when the environment correlation time $\tau_E$ is much shorter than the time scale for significant change in $S$, which occurs when the environment is a large system maintained in thermal equilibrium, as expected here.  The evolution of the single-mode system is then defined by the Lindblad master equation which, in diagonal form, is given by
\begin{equation} \label{eq:lindblad}
\frac{d \hat{\rho}_S }{dt} = -\frac{i}{\hbar} [\hat{H}^{\prime}_S, \hat{\rho}_S] + \sum_{i=1}^{i=4} \bigg(  \hat{c}_i \hat{\rho}_S \hat{c}^{\dagger}_i - \frac{1}{2} \{ \hat{c}^{\dagger}_i \hat{c}_i, \hat{\rho}_S \} \bigg),
\end{equation}
where $\hat{H}_S^{\prime}$ is the renormalized free Hamiltonian of the single-mode phonon system.  Assuming the environment to be in thermal equilibrium, we have $\hat{c}_1 = \sqrt{\gamma_{1}} \hat{b}_{\boldsymbol{q}}$, $\hat{c}_2 = \sqrt{\gamma_{2}} \hat{b}^{\dagger}_{\boldsymbol{q}}$ where the rates $\gamma_1$ and $\gamma_2$ are related to environment correlation functions
\begin{align} \label{eq:gamma1}
\gamma_{1} 
&= \frac{1}{\hbar^2} \int^{\infty}_{-\infty} dt^{\prime} e^{i\omega_{q} t^{\prime}} \langle \tilde{E}_{\boldsymbol{q}} (t) \tilde{E}^{\dagger}_{\boldsymbol{q}} (t - t^{\prime}) \rangle_E, \\ \label{eq:gamma2}
\gamma_{2}
&= \frac{1}{\hbar^2} \int^{\infty}_{-\infty} dt^{\prime} e^{-i\omega_{q} t^{\prime}} \langle \tilde{E}^{\dagger}_{\boldsymbol{q}} (t) \tilde{E}_{\boldsymbol{q}} (t - t^{\prime}) \rangle_E,
\end{align}  
and $\tilde{E}_{\bs{q}}$ is the operator defined in \eqref{eq:E} but now in the interaction picture.  The master equation \eqref{eq:lindblad} therefore describes the time evolution of a single-mode quasi-particle of a BEC in thermal equilibrium when taking into account Landau and Beliaev interactions (see also \cite{CastinAppendix}).

\section{Time evolution of the phonon covariance matrix}
\subsection{The covariance matrix formalism}

By restricting our analysis to Gaussian states we can use the covariance matrix formalism to conveniently describe the dynamics of the single-mode phonon system.  Such states are assumed in the relativistic quantum devices proposed in \cite{Accelerometer,GWDetectorFirst,GWDetectorThermal} that exploit the phonons of BECs, and can be readily created in BECs.  For example, such states can be generated using Bragg spectroscopy \cite{PhysRevLett.83.2876,PhysRevLett.88.060402,PhysRevLett.88.120407,FirstSoundWaves1,FirstSoundWaves2} and in processes that are acoustic analogues to the dynamical Casimir effect (DCE) \cite{DCETheory,DCEWestbrook}.  

The covariance matrix formalism is often used in quantum optics and continuous-variable quantum information science, and is a convenient mathematical framework in quantum phase space for describing Gaussian states and their dynamics.  It utilizes the fact that Gaussian states are completely defined by just their first $\boldsymbol{d}$ and second $\boldsymbol{\sigma}$ statistical moments
\begin{align} \label{eq:d}
d_i &:= \langle\hat{x}_i\rangle ~=~ Tr(\hat{x}_i \hat{\rho}_S),\\ \nonumber
\sigma_{ij} &:= \frac{1}{2} \langle \{\hat{x}_i, \hat{x}_j\} \rangle - \langle \hat{x}_i \rangle \langle \hat{x}_j \rangle \\ \label{eq:cov}
&~\equiv  \frac{1}{2} Tr(\{\hat{x}_i - d_i, \hat{x}_j - d_j\} \hat{\rho}_S),
\end{align}
where $\hat{x}_i$ are the quadrature phase space operators which, for the single-mode system, are defined as
\[\hat{\boldsymbol{x}} := \left( \begin{array}{c}
 \hat{x}_1 \\ 
\hat{x}_2 \end{array} \right)
:= \frac{1}{2 \kappa} \left( \begin{array}{cc}
1 & 1\\
-i & i \end{array} \right)
\left( \begin{array}{c}
\hat{b}_{\boldsymbol{q}} \\
\hat{b}^{\dagger}_{\boldsymbol{q}} \end{array} \right).
 \]
These phase space operators then obey the commutation relations $[\hat{x}_i, \hat{x}_j ] = \frac{i}{2 \kappa^2} \Omega_{ij}$ where $\boldsymbol{\Omega}$ is of symplectic form, and $\kappa$ is a constant that is often taken to be $1 / \sqrt{2}$ in quantum optics.  By assuming Gaussian states we have, therefore, reduced the complete description of the state down from the infinite dimensional space of the density matrix to terms of the two-dimensional matrices $\boldsymbol{\sigma}$ and $\boldsymbol{d}$, which are referred to as the covariance matrix and displacement matrix of the state respectively.

\subsection{Time evolution of statistical moments} \label{sec:EvStatMoments}

To determine how the covariance and displacement matrices evolve for the single-mode phonon system, we first transform the Lindblad equation for the density matrix \eqref{eq:lindblad} into the phase space basis $\hat{x}_i$ and then use this equation in differential versions of  \eqref{eq:d} and \eqref{eq:cov} \cite{Isar1994,Sandulescu1987,OptimalFeedbackControl}.  In fact it is possible to derive the time evolution equations for the covariance and displacement matrices of a general $M$-dimensional Gaussian state whose density matrix obeys \eqref{eq:lindblad}: taking a general quadratic Hamiltonian\footnote{$\hat{H}_S$ must be at most quadratic to preserve Gaussianity \cite{Schumaker1986317}.} $\hat{H}_S = H_0 + \kappa \hat{\boldsymbol{x}}^T \boldsymbol{H}_1 + \kappa^2 \hat{\boldsymbol{x}}^T \boldsymbol{H}_2 \hat{\boldsymbol{x}}$ where $H_0$ is a constant; $\boldsymbol{H}_1$ is a $2M$-dimensional column-vector; and $\boldsymbol{H}_2$ is a $2M\times 2M$ real-symmetric matrix, it is straightforward to show that  the covariance and displacement matrices  evolve in the following way: 
\begin{align} \label{eq:dd}
\frac{d \boldsymbol{d}}{dt} &= \boldsymbol{\mathcal{H}}_1 + \boldsymbol{A} \boldsymbol{d},\\
\label{eq:dcov}
\frac{d \boldsymbol{\sigma}}{dt} &= \boldsymbol{A}\boldsymbol{\sigma} + \boldsymbol{\sigma} \boldsymbol{A^T} + \boldsymbol{D},
\end{align}
where $\boldsymbol{\mathcal{H}}_1 := \boldsymbol{\boldsymbol{\Omega}} \boldsymbol{H}_1 / \hbar$.  These equations have been derived previously in quantum optics for the case when $\boldsymbol{H}_1 = 0$ \cite{Isar1994,Sandulescu1987,OptimalFeedbackControl,OptimalThermalization}.  In these quantum optics studies, the matrix $\boldsymbol{A}$ and symmetric matrix $\boldsymbol{D}$ are  referred to as the drift and diffusion matrices respectively \cite{OptimalFeedbackControl}.  They are defined as
\begin{align} \label{eq:drift}
\boldsymbol{D} &:= \frac{1}{4 \kappa^4} \boldsymbol{\Omega} Re(\boldsymbol{C^{\dagger} C}) \boldsymbol{\Omega^T},\\ \label{eq:diffusion}
\boldsymbol{A} &:= \boldsymbol{\mathcal{H}}_2 + \boldsymbol{\mathcal{K}},
\end{align}
where
\begin{align}
\boldsymbol{\mathcal{K}} &:= \frac{1}{2\kappa^2} \boldsymbol{\Omega} Im (\boldsymbol{C^{\dagger}C}),\\
\boldsymbol{\mathcal{H}}_2 &:= \frac{1}{\hbar} \boldsymbol{\boldsymbol{\Omega}} \boldsymbol{H}_2,
\end{align}
and the matrix $\boldsymbol{C}$ is defined by $\hat{c}_i = C_{ij} \hat{x}_j$. 

The general solution of \eqref{eq:dd} is
\begin{align} \label{eq:generalSoldXY}
\boldsymbol{d}(t) = \boldsymbol{X}(t) \boldsymbol{d}_0 + \boldsymbol{Y}(t),
\end{align}
where, when $\boldsymbol{A}$ is independent of time,
\begin{align} \label{eq:timeIndd1}
\boldsymbol{X}(t) &= e^{\boldsymbol{A}t},\\ \nonumber
\boldsymbol{Y}(t) &= \int^t_0 e^{\boldsymbol{A} (t-s)} ds \boldsymbol{H}_1 \\ \label{eq:timeIndd2}
&=\boldsymbol{A}^{-1} (e^{\boldsymbol{A} t} - 1) \mathcal{\boldsymbol{H}}_1,
\end{align}
so that the evolution of the displacement matrix is given by %\cite{My thesis?}:
\begin{align} \label{eq:generalSold}
\boldsymbol{d}(t) = e^{\boldsymbol{A}t} \boldsymbol{d}_0 + \int^t_0 e^{\boldsymbol{A} (t-s)} ds~ \boldsymbol{H}.
\end{align}

The equation of motion for the covariance matrix \eqref{eq:dcov} is, in general, a time varying differential Lyapunov matrix equation, of which the general solution is
\begin{align} \label{eq:generalSolCovXY}
\boldsymbol{\sigma}(t) = \boldsymbol{X}(t) \boldsymbol{\sigma}_0 \boldsymbol{X}^T(t) + \boldsymbol{Z}(t),
\end{align}
where
\begin{align}
\boldsymbol{X}(t) &= \boldsymbol{\Phi}(t,0),\\
\boldsymbol{Z}(t) &= \int^t_0 \boldsymbol{\Phi}(t,s) \boldsymbol{D} \boldsymbol{\Phi}^T(t,s) ds.
\end{align}
When $\boldsymbol{A}$ is independent of time,\footnote{There is no general analytic expression for the transition matrix $\boldsymbol{\Phi}(t,s)$ when $\boldsymbol{A}$ is time dependent  and in this case a numerical method is then the only way to obtain a solution \cite{LyapunovBook}.} $\boldsymbol{\Phi}(t,s) = e^{\boldsymbol{A}(t - s)}$ and so the solution to  \eqref{eq:dcov} can be written  as\footnote{Another option is to convert \eqref{eq:dcov} to a vector-valued ODE, which can then be readily solved, using vectorization of the matrix $\boldsymbol{\sigma}(t)$ \cite{CtsTimeLyapunov}.}
\begin{equation} \label{eq:generalSolCov}
\boldsymbol{\sigma}(t) = e^{ \boldsymbol{A} t} \boldsymbol{\sigma}_0 e^{ \boldsymbol{A}^T t} + \int^t_0 e^{ \boldsymbol{A} (t-s)} \boldsymbol{D} e^{ \boldsymbol{A}^T (t-s)} ds,
\end{equation}
for which an analytical expression can be obtained when $\boldsymbol{A}$ is diagonalizable \cite{DiagonalSolution}.   

A connection can be made with the evolution of the displacement vector and covariance matrix generated by a Gaussian unitary  by neglecting all dissipative effects ($\bs{D}=0$, $\bs{A}=\bs{\mathcal{H}}_2$).  In this case, assuming for convenience that $\bs{H}_2$ and $\bs{H}_1$ are independent of time, the solutions  \eqref{eq:generalSoldXY} and \eqref{eq:generalSolCovXY} reduce to
\begin{align}
\boldsymbol{d} (t) &= \boldsymbol{S} \boldsymbol{d}_0 + \boldsymbol{e}, \\
\boldsymbol{\sigma} (t) &= \boldsymbol{S}\boldsymbol{\sigma}_0 \boldsymbol{S}^T,
\end{align}
where $\boldsymbol{S} := e^{\boldsymbol{\mathcal{H}}_2 t}$ is the symplectic transformation corresponding to the free unitary evolution of the system, and $\bs{e} := \bs{B} \bs{\mathcal{H}}_1$ is a real vector with $\bs{B}:= \boldsymbol{\mathcal{H}}^{-1}_2 (e^{\boldsymbol{\mathcal{H}}_2 t} - 1)$. The matrix $\boldsymbol{\mathcal{H}}_2 = \boldsymbol{\Omega} \boldsymbol{H}_2 / \hbar$ thus forms a symplectic algebra so that  $\bs{\mathcal{H}}_2$ is a (real) Hamiltonian matrix and $\bs{H}_2$ is a (real) symmetric matrix, which is also required by the Hermitian property of $\hat{H}$.

For our single-mode phonon system, $\hat{H}_S$ is given by \eqref{eq:HS} and thus $H_0  = 0$, $\boldsymbol{H}_1 = \bs{0}$ and $\boldsymbol{H}_2 = \hbar \omega^{\prime}_q \boldsymbol{I}_2$ where $\boldsymbol{I}_2$ is the two-dimensional identity matrix.  The environment is also assumed to be in thermal equilibrium, and so the diffusion and drift matrices are given by
\begin{align} \label{eq:D}
\boldsymbol{D}&= \frac{1}{4\kappa^2} \gamma_T \boldsymbol{I}_2, \\ \label{eq:A}
 \boldsymbol{A} &= -\frac{1}{2} \gamma \boldsymbol{I}_2 + \omega^{\prime}_{\boldsymbol{q}} \boldsymbol{\Omega},
\end{align}
where $\gamma_T := \gamma_{1} + \gamma_{2}$; $\gamma := \gamma_{2} - \gamma_{1}$; and $\omega^{\prime}_q$ is the renormalized frequency of the single-mode phonon system.  

Since the environment is in thermal equilibrium, the rates $\gamma_{1}$ and $\gamma_{2}$ are not independent but instead satisfy $\gamma_{1} = e^{\beta_q} \gamma_{2}$  \cite{CastinAppendix,OpenBook}.
We can, therefore, write $\gamma_{T}$ as
\begin{equation} \label{eq:gammaT}
\gamma_T := \gamma \coth ( \frac{1}{2} \beta_q) = \gamma (1 + 2N^{th}_q),
\end{equation}
where $N^{th}_q$ is the average thermal occupation defined in \eqref{eq:Nth}.  The matrix $\boldsymbol{D}$ is then given by
\begin{align}
\boldsymbol{D} =  \frac{(1 + 2N_q)}{4 \kappa^2} \gamma  \boldsymbol{I}_2 := \gamma \boldsymbol{\sigma}_{\infty},
\end{align}
where $\boldsymbol{\sigma}_{\infty}$ is the covariance matrix of a single-mode thermal state. 

Substituting the above drift and diffusion matrices for the single-mode phonon system into the general time-independent solutions \eqref{eq:generalSoldXY} (using \eqref{eq:timeIndd1} and \eqref{eq:timeIndd2}) and \eqref{eq:generalSolCov}, the displacement vector and covariance matrix at time $t$ are given by
\begin{align} \label{eq:phononSold}
\boldsymbol{d} (t) &= e^{-\frac{1}{2} \gamma t} \boldsymbol{R}(t) \boldsymbol{d}_0, \\ \label{eq:phononSolCov}
\bs{\sigma}(t) &= e^{- \gamma t} \Big(\boldsymbol{R}(t) \boldsymbol{\sigma}_0 \boldsymbol{R}^T(t) \Big)  + (1 - e^{- \gamma t}) \boldsymbol{\sigma}_{\infty},
\end{align}
where $\boldsymbol{R}(t) := e^{\frac{1}{\hbar} \boldsymbol{\Omega} \boldsymbol{H}_2 t}$ is the symplectic transformation corresponding to the free unitary evolution of the single-mode phonon system $U = e^{-\frac{i}{\hbar} H_S t}$, and  thus $\boldsymbol{\Omega} \boldsymbol{H}_2$ forms the symplectic algebra (it is a Hamiltonian matrix).  

  Since $\boldsymbol{H}_2 = \hbar \omega^{\prime}_{q}\boldsymbol{I}_2$ for our phonon system, $\boldsymbol{R}(t) = \cos (\omega_q t)\boldsymbol{I}_2 + \sin (\omega_q t) \boldsymbol{\Omega}$, which is just the usual symplectic (and in this case rotational) transformation for the phase shift operator.  From \eqref{eq:phononSolCov}, due to the dissipative effects, this free evolution is now damped by $e^{-\gamma t}$ and the state asymptotically approaches $\boldsymbol{\sigma}_{\infty}$:
\begin{align} \label{eq:phononSoldNoR}
\bs{\sigma}(t) &= e^{- \gamma t} \boldsymbol{\sigma}_0   + (1 - e^{- \gamma t}) \boldsymbol{\sigma}_{\infty}.
\end{align}
This form of equation has been considered in quantum optics studies \cite{QOInPhaseSpace,GaussianStatesInContinuousQI,QuantifyingDecoherenceRef} but the rate $\gamma$ in those studies is not the same as that for the phononic system studied here.  This rate comes from the expressions for $\gamma_1$ and $\gamma_2$ \eqref{eq:gamma1}-\eqref{eq:gamma2} given in Section \ref{sec:Markov}.  From these expressions, and assuming a continuum of modes, it can be easily shown that $\gamma = \gamma_B + \gamma_L$ where
\small
\begin{align} \label{eq:gammaL} 
\gamma_L &:= \frac{g^2 n}{V\hbar^2} \int^{\infty}_0  \pi d\omega_k p_k \mathcal{L}_{kl}^2  \delta(\omega_q + \omega_k - \omega_l) (N^{th}_l - N^{th}_k), \\ \label{eq:gammaB}
\gamma_B &:= \frac{2g^2 n}{V\hbar^2} \int^{\infty}_0  \pi d\omega_k p_k \mathcal{B}_{kl}^2  \delta(\omega_q - \omega_k - \omega_l) (1 + N^{th}_l + N^{th}_k),
\end{align}
\normalsize
\iffalse
\begin{align} \label{eq:gammaL}
\gamma_L &:= \frac{g^2 n}{V\hbar^2} \int^{\infty}_0  \pi d\omega_k p_k \mathcal{L}_{kl}^2  \delta(\omega_q + \omega_k - \omega_l) (N^{th}_l - N^{th}_k), \\ \nonumber
\gamma_B &:= \frac{2g^2 n}{V\hbar^2} \int^{\infty}_0  \pi d\omega_k p_k \mathcal{B}_{kl}^2  \delta(\omega_q - \omega_k - \omega_l) \\ \label{eq:gammaB}
 &\hspace{2.5cm}\times (1 + N^{th}_l + N^{th}_k),
\end{align}
\fi
which are just the Landau and Beliaev damping rates \cite{Landau:1946jc,Beliaev1958aSet2,Beliaev1958bSet2,Hohenberg1965291,PhysicalKinetics,KONDOR1974393,LiuAndShieve,PitaevskiiDamping,Liu,Giorgini1998,DampingTrapped,PhysRevA.58.3146,PhysRevA.58.R3391,Langevin,PhysRevA.59.3851,LBReview,CastinAppendix}, where $p_k$ is an assumed density of states.  These rates have been explicitly calculated under various approximations.  For example, in \cite{Giorgini1998} the rates were calculated for a uniform three-dimensional BEC where it was found that, when $k_B T \ll\hbar \omega_q$ such that Beliaev damping dominates over Landau damping $\gamma_B \gg \gamma_L$, the total damping rate $\gamma$ can be approximated as \cite{Giorgini1998}
\begin{align} \label{eq:QuantumRegime}
\gamma \approx \gamma_B \approx \frac{3}{ 640 \pi} \frac{\hbar \omega^5_q}{m n c^5_s}   \Big[ 1 + \Big(\frac{k_B T}{\hbar \omega_q}\Big)^3 \Big],
\end{align}
which doesn't vanish at $T=0$.  On the other hand, in the opposite regime $k_B T \gg \hbar \omega_q$,   Landau damping dominates over Beliaev damping, and for very high
temperatures $k_B T \gg \mu \gg \hbar \omega_q$ (where $\mu = gn = m c^2_s$ is the chemical potential), the total damping rate was found to be given by \cite{Giorgini1998}
\begin{equation} \label{eq:ThermalRegimeHigh}
\gamma \approx \gamma_L \approx \frac{3 \pi}{8} \frac{ k_B T a}{\hbar c_s} \omega_q,\end{equation}
whereas, for temperatures such that $\mu \gg k_B T \gg \hbar \omega_q$, the damping rate is found to be \cite{Giorgini1998}
\begin{align} \label{eq:ThermalRegimeLow}
\gamma \approx \gamma_L \approx \frac{3 \pi^3}{40} \frac{ (k_B T)^4}{m n \hbar^3 c^5_s} \omega_q.
\end{align}

\subsection{Summary}

Starting from the full Hamiltonian of a Bose gas \eqref{eq:fullH}, we have derived how the covariance matrix of a single-mode Gaussian phonon mode evolves in an isolated BEC \eqref{eq:phononSoldNoR}. In the next section we investigate how this can then be used to find how quickly a prepared quantum state of phonons will decohere.  

We now mention a few important steps in the above derivation of \eqref{eq:phononSoldNoR} from \eqref{eq:fullH}. The above discussion concentrated on a uniform BEC, such as that created in experiments performed in \cite{PhysRevLett.110.200406}. However, this can also be straightforwardly extended to more common trapped BECs, such as an harmonic trap.  In this case we would start again from \eqref{eq:fullH} but with $\mathcal{V}(\bs{r})$ replaced with the particular trapping potential, such as $ \mathcal{V}(\bs{r}) = \frac{1}{2} m (\omega^2_x x^2 + \omega^2_y y^2 + \omega^2_z z^2)$ for an anisotropic harmonic trap.  The Bogoliubov transformations \eqref{eq:Bogotrans} can then be applied again to diagonalize the Hamiltonian in the Bogoliubov approximation \eqref{eq:bogoH}, but now the coefficients $u_k$ and $v_k$ will be different to the uniform case (see e.g., \cite{PhysRevA.58.3146,DampingTrapped}). Going to next order in $\sqrt{N}$ we would then also find Beliaev and Landau interaction terms between the phonons as in the uniform case \eqref{eq:HI} but again with different coefficients \cite{PhysRevA.58.3146,DampingTrapped}. Therefore, for a non-uniform trapped system  we will also end up with the  evolution equation \eqref{eq:phononSoldNoR} for the covariance matrix of the phonons but with the damping rate of that particular trapped system. For example, the damping rate for non-uniform trapped systems has  been analysed for low-energy excitations in \cite{PhysRevA.58.3146,DampingTrapped}.

We also note that, even though the interaction Hamiltonian for our phonon system \eqref{eq:HI} is cubic in field modes, Gaussianity of the state will be preserved under the Born-Markov approximation.  This is evident from the fact that the general solution to \eqref{eq:dcov} has the form of the transformation brought about by a general Gaussian channel, which is a trace-preserving completely positive map that maps Gaussian trace-class operators onto Gaussian trace-class operators \cite{Demoen97,Caves94,Lindblad2000,Preskill2001,Eisert02,Serafini04,Holevo99,Holevo01,Giovannetti03,Giovannetti03b,Multiplicativity}.  It is also illustrated by studying the Fokker-Planck equation for the Wigner function that derives from a general Markov master equation for the density operator, where it is found that the Wigner function continues to be Gaussian \cite{QOInPhaseSpace,GaussianStatesInContinuousQI}. The evolution of a Gaussian state in the Markov approximation is generally analysed for an interaction Hamiltonian that is bilinear in the field operators (see, e.g., \cite{GQI,QOInPhaseSpace,GaussianStatesInContinuousQI}). However, the above analysis for  a single-mode phonon state of a BEC clearly emphasizes that this isn't a necessary condition for the preservation of Gaussianity.  This is important since it means that, under realistic approximations, the simple description of the displacement and covariance matrices fully characterizing the state will persist throughout the state's evolution.

\section{Quantifying the quantum decoherence and relaxation of phonons of BECs} \label{sec:QuantifyingDecoherence}

The quantum decoherence of quantum optics systems has been extensively studied.  In particular, for single-mode Gaussian states,  the evolution of such states of electromagnetic radiation in thermal reservoirs was investigated in \cite{PhysRevA.47.4474,PhysRevA.47.4487} where certain properties, such as purity and squeezing of a displaced squeezed thermal mode, were analysed. Furthermore, in \cite{Purity,QuantifyingDecoherenceRef},  the quantum decoherence of a Gaussian quantum optics system was characterized by analysing the evolution of certain global entropic measures and nonclassical indicators using the covariance matrix formalism. We can apply the same quantum optical techniques to our phononic system since the phonon system is in a Gaussian state and the evolution of the covariance matrix \eqref{eq:phononSoldNoR} takes the same form as in \cite{Purity,QuantifyingDecoherenceRef}, with only the damping rates being different.  This similarity in the evolution of the state, despite the interactions being very different, is due to applying the Born-Markov approximation in both cases such that the quantum master equation for the phonons \eqref{eq:lindblad} is of a similar form to that of a quantum optics master equation as described in the previous section.  
 
\subsection{Evolution of purity} \label{sec:Purity}

For a single-mode Gaussian state, the purity $\mu := Tr(\rho^2_S)$ is given by
\begin{align} \nonumber
\mu &= \frac{1}{4 \kappa^2 \sqrt{\det\boldsymbol{\sigma}}}\\ \nonumber
&=  \frac{1}{4 \kappa^2 s}\\
&= \frac{1}{1 + 2\mathcal{N}},
\end{align}
where $\mathcal{N}$ is the so-called ``thermal'' occupation of the state and $s$ is its symplectic eigenvalue (the eigenvalue of the matrix $|i \boldsymbol{\Omega} \boldsymbol{\sigma}|$).  Using Williamson's theorem \cite{Williamson}, any single-mode covariance matrix can be written in the general form \cite{Adam95,QOInPhaseSpace,GaussianStatesInContinuousQI}
\begin{align} \label{eq:generalCov}
\scalefont{0.85}{\boldsymbol{\sigma} = \frac{1}{4 \kappa^2 \mu} \left( \begin{array}{cc} \cosh 2r + \sinh 2r \cos \psi &\mkern-15mu \sinh 2r \sin \psi \\
\sinh 2r \sin \psi &\mkern-15mu \cosh 2r - \sinh 2r \cos \psi \end{array} \right)},
\end{align}
where $r$ and $\psi$ are defined by $\xi = r e^{i\psi}$, which is the squeezing parameter for a single-mode squeezing transformation.  Using this general form of the covariance matrix in the solution \eqref{eq:phononSolCov} of its equation of motion, the purity of the single-mode state is found to evolve as \cite{PhysRevA.47.4487,QuantifyingDecoherenceRef}
\begin{align} \nonumber
\mu(t) = \mu_0 \Big( &e^{-2\gamma t} + \frac{\mu^2_0}{\mu^2_{\infty}} (1 - e^{-\gamma t})^2  \\
&+ \frac{2 \mu_0}{\mu_{\infty}} e^{-\gamma t} (1 - e^{-\gamma t}) \cosh 2 r_0 \Big)^{-\frac{1}{2}},
\end{align}
where $\mu_{\infty} = (1 + 2N^{th}_q)^{-1}= \tanh (\frac{1}{2} \beta_q)$ is the purity of the thermal state $\boldsymbol{\sigma}_{\infty}$ that the covariance matrix asymptotically approaches.  The purity will undergo a local minimum  for squeezed states for which $r_0 > \mathrm{max}[\mu_0 / \mu_{\infty}, \mu_{\infty} / \mu_0]$ at \cite{PhysRevA.47.4487,QuantifyingDecoherenceRef,Purity}
\begin{align} \label{eq:minPurity}
t_{min} = \frac{1}{\gamma} \ln \bigg( \frac{\frac{\mu_0}{\mu_{\infty}} + \frac{\mu_{\infty}}{\mu_{0}} - 2 \cosh(2r_0)}{\frac{\mu_0}{\mu_{\infty}} - \cosh(2r_0)}\bigg),
\end{align}
which can provide a good characterization of the decoherence time of such  states since this represents the point at which the state becomes most mixed, with any subsequent increase in the purity just reflecting the state being driven towards the state of the environment \cite{QuantifyingDecoherenceRef,Purity}.  To determine this decoherence time one just needs to know the initial purity and squeezing of the state; the temperature of the BEC (which defines $\mu_{\infty}$); and the damping rate $\gamma$.   As an example, the quantum decoherence time for phonons in a particular BEC is calculated and presented in Section \ref{sec:Example}. This BEC is that assumed in \cite{GWDetectorFirst,GWDetectorThermal} and the results can therefore be used to predict a quantum decoherence time for the phononic GW detector discussed in the \hyperref[sec:Intro]{Introduction}.

\subsection{Evolution of nonclassical depth}

An alternative characterization of the decoherence time can be provided by the nonclassical depth \cite{NonclassicalDepth}, which is a popular measure for quantifying  the nonclassicality of a quantum state.  This measure has the physical meaning of the number of thermal photons necessary to destroy the nonclassical nature of the quantum state \cite{Lee1992}.  For a general Gaussian state, the nonclassical depth $\tau$ detects the state as nonclassical if a canonical quadrature exists whose variance is below $1/2$ \cite{QuantifyingDecoherenceRef} and, for a single-mode Gaussian state, it is given by
\begin{align}
\tau = \mathrm{max} \Big[ \frac{1}{2} \Big( 1 - \frac{e^{-2r}}{\mu} \Big), 0 \Big].
\end{align}
From the time evolution of the covariance matrix \eqref{eq:phononSolCov}, the nonclassical depth $\tau$ of the single-mode can be shown to evolve as \cite{QuantifyingDecoherenceRef}
\begin{align} \label{eq:tau}
\tau(t) = \frac{1}{2 \mu_{\infty}} \Big[  e^{-\gamma t} \Big( 1 - \frac{\mu_{\infty}}{\mu_0} e^{-2 r_0} \Big) + \mu_{\infty} - 1  \Big].
\end{align}
Note that the time at which the state becomes ``classical'' ($\tau = 0$)  is given by
\begin{align} 
t_{\tau=0} &= \frac{1}{\gamma} \ln \bigg( \frac{1 - \frac{\mu_{\infty}}{\mu_0} e^{-2 r_0}}{1 - \mu_{\infty}} \bigg),
\end{align}
providing an alternative description for quantifying the decoherence time of quantum states.

\subsection{Evolution of squeezing}

As illustrated by the general covariance matrix of a single-mode Gaussian state \eqref{eq:gammaL}, such states are fully defined by their first-statistical moment, purity, and squeezing $\xi = r e^{i \psi}$. For a single-mode state whose covariance matrix evolves as in \eqref{eq:phononSolCov}, $\psi$ is a constant of motion and $r$ evolves as \cite{QuantifyingDecoherenceRef}   
\begin{align}
\cosh(2r(t)) &= \mu (t) \Big( e^{-\gamma t} \frac{\cosh 2 r_0}{\mu_0} + \frac{1 - e^{-\gamma t}}{\mu_{\infty}} \Big).
\end{align}

\subsection{Evolution of average occupation}
The average occupation of a single mode state $\boldsymbol{q}$ is defined as $N_{\boldsymbol{q}} := \langle \hat{b}^{\dagger}_{\boldsymbol{q}} \hat{b}_{\boldsymbol{q}} \rangle$.  From the Lindblad master equation for the density operator of the single-mode  state interacting with an environment in thermal equilibrium \eqref{eq:lindblad}, it is easy to show that the average occupation of the single-mode phonon state evolves as
\begin{align}
N_{\boldsymbol{q}}(t) = e^{-\gamma t} N_{\boldsymbol{q}}(0) + (1 - e^{-\gamma t}) N^{th}_{q},
\end{align}
where $N^{th}_{q} := ({e^{\beta \omega_q} - 1})^{-1}$ is the thermal occupation of the mode at temperature $T$. Since the average occupation is simply related to the average energy of the state, one can characterize the relaxation of the state from its evolution.  Furthermore, since the average occupation is also defined  by $N_q = \kappa^2  Tr(\boldsymbol{\sigma}) + \kappa^2 \boldsymbol{d}^T \boldsymbol{d} - 1/2$, a single-mode Gaussian state at a time $t$ is fully defined by  its average occupation $N_{\boldsymbol{q}}$, purity $\mu$ and squeezing $r$. 

\section{Example: Quantum decoherence in a phononic GW detector} \label{sec:Example}

\begin{figure*}[ht!]
     \begin{center}
        \subfigure{%
        \put(-6,110){(a)}
            \label{fig:first}
            \includegraphics[width=0.29\textwidth]{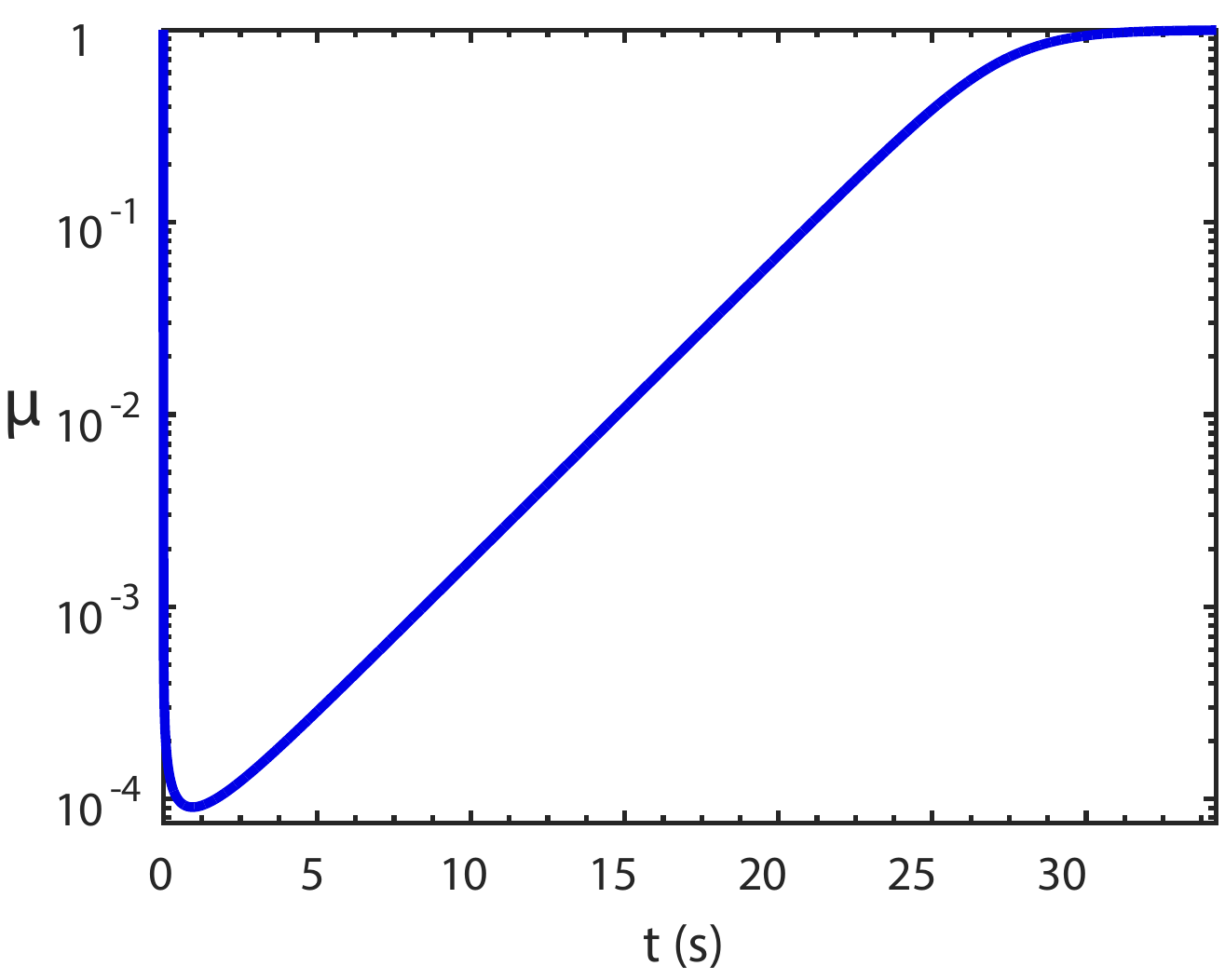}            
        }%
        ~~~
        \subfigure{%
        \put(-5,110){(b)}
           \label{fig:second}
           \includegraphics[width=0.29\textwidth]{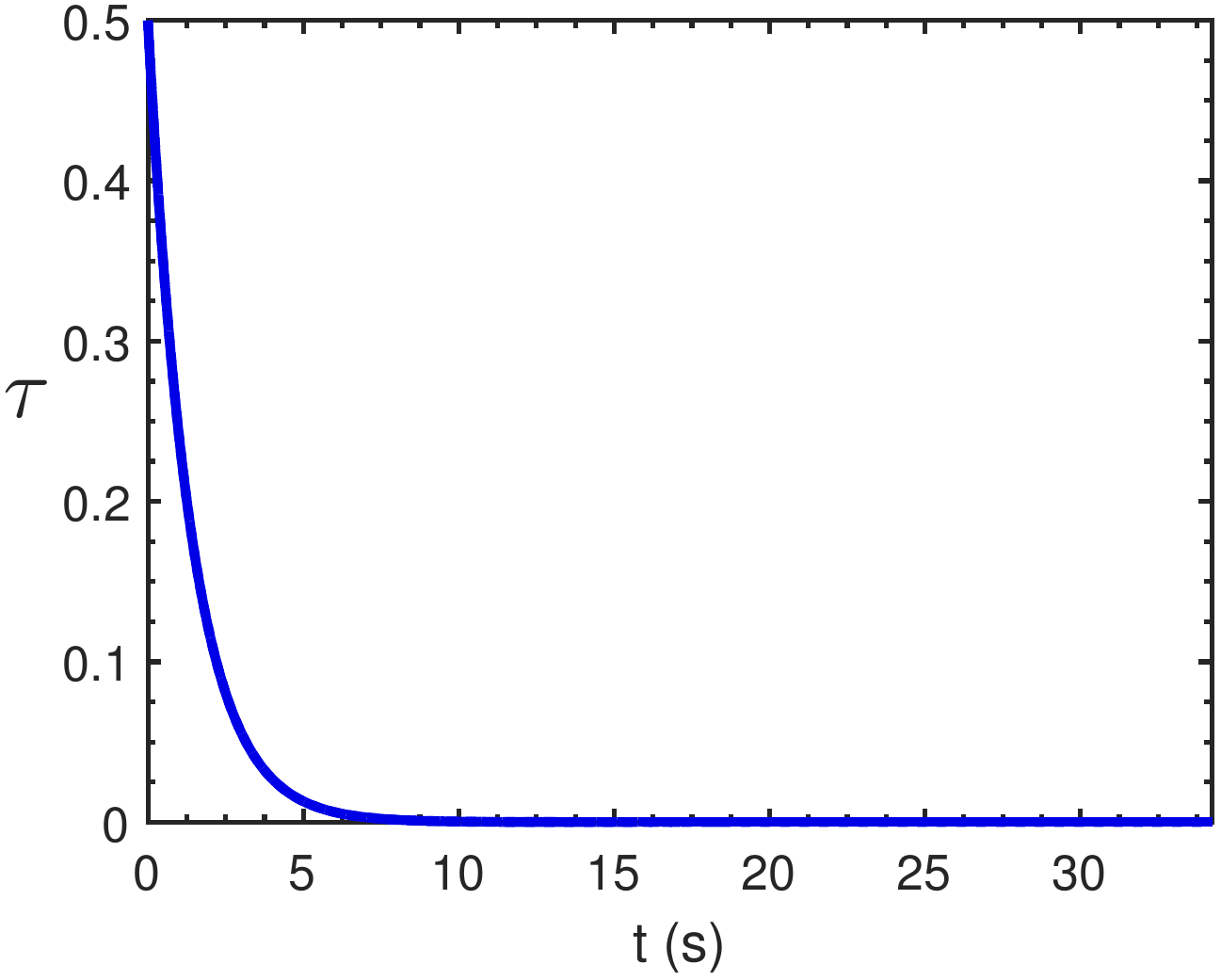}
        }\\ %  ------- End of the first row ----------------------%
        \subfigure{%
        \put(-5,110){(c)}
            \label{fig:third}
            \hspace{1.5mm}\includegraphics[width=0.28\textwidth]{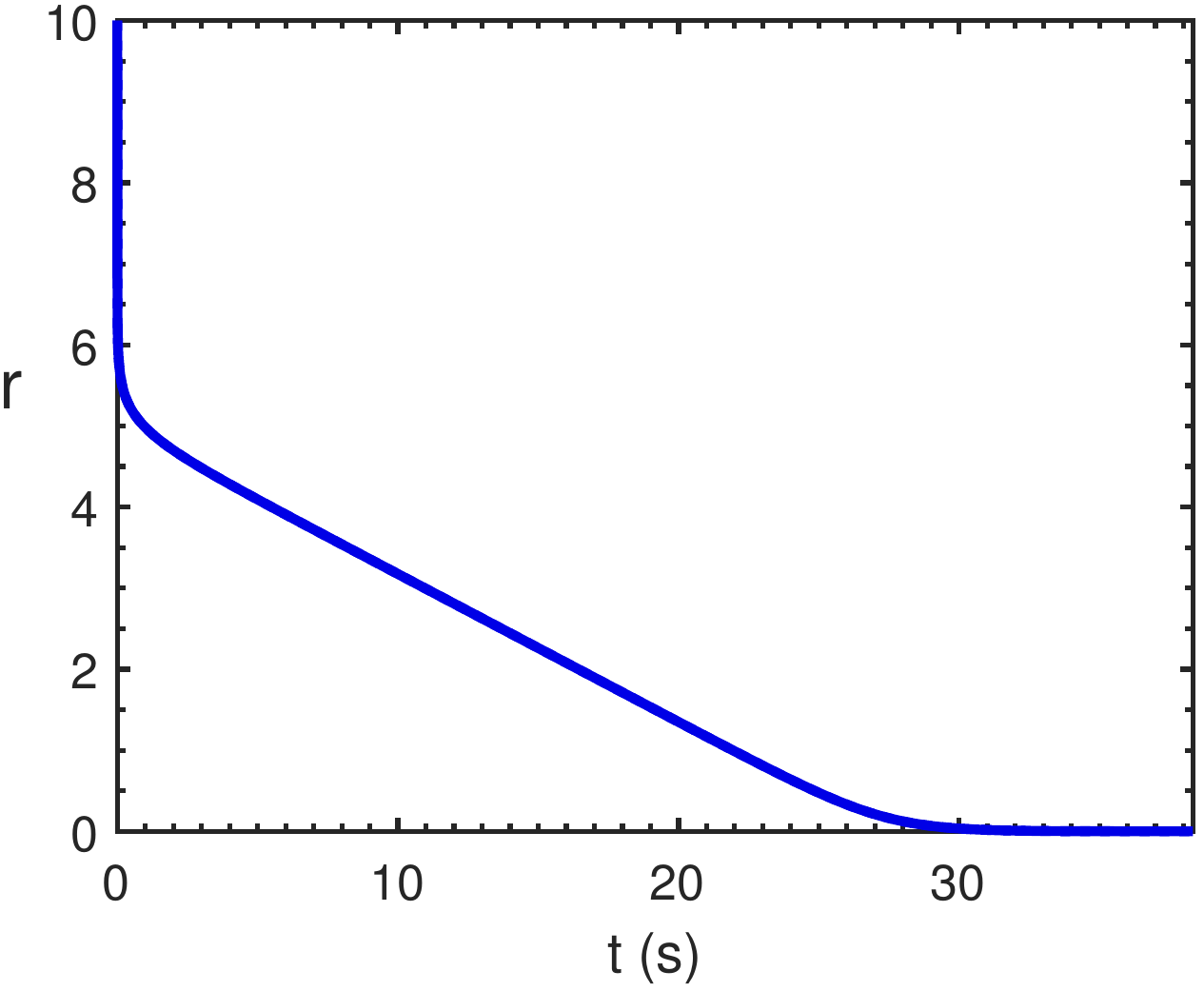}
        }%
        ~~~
        \subfigure{%
        \put(-5,110){(d)}
            \label{fig:fourth}
            \hspace{0mm}\includegraphics[width=0.29\textwidth]{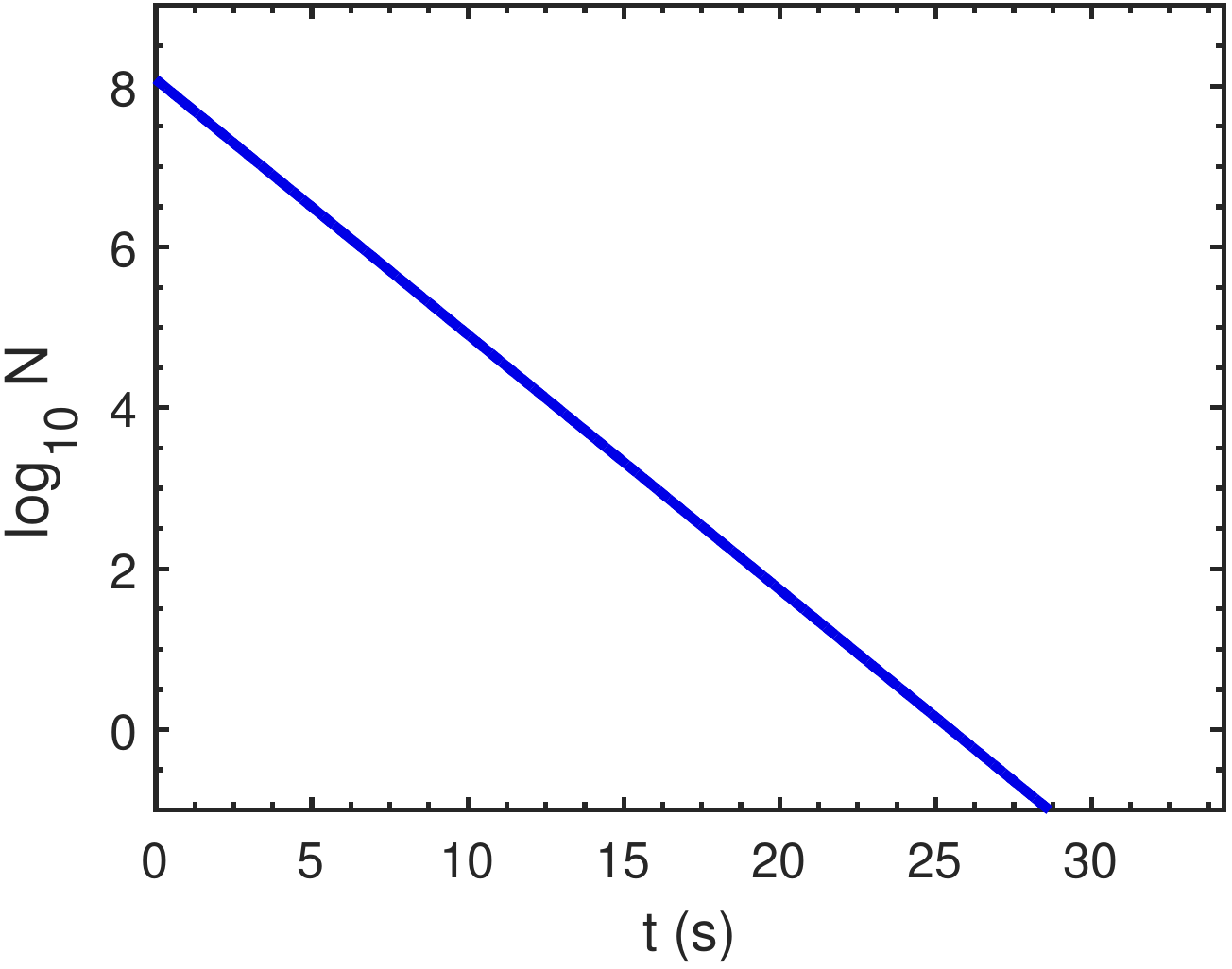}
        }%
    \end{center}
    \caption{Time evolution of (a) purity $\mu$; (b) nonclassical depth $\tau$; (c) squeezing parameter $r$; and (d) average occupation $N$.   A $^{87}\mathrm{Rb}$ BEC in a uniform trap with $c_s = 3.4\,\mathrm{mm}\,\mathrm{s}^{-1}$ and temperature $0.5$\,$\mathrm{nK}$ was assumed, with the single phonon mode having  frequency $\omega_q = 10$\,$ \mathrm{kHz}$ and initial squeezing $r=10$. For these parameters the Beliaev damping rate from \eqref{eq:QuantumRegime} is $0.73\,\mathrm{s^{-1}}$.
	}
   \label{fig:QuantumkHz}
\end{figure*}

In this section we apply the general techniques derived above for quantifying the quantum decoherence of phonons of BECs to the specific example of a three-dimensional version of the phononic GW detector proposed in  \cite{GWDetectorFirst,GWDetectorThermal}. The detector consists of a BEC constrained to a rigid trap, with a prepared quantum state of phonons, such as a two-mode squeezed state.\footnote{Methods for squeezing phonon states in experiments include  introducing measurement back action under weak continuous probing \cite{PhysRevLett.115.060401,PhysRevA.93.023610}, utilizing Beliaev damping \cite{BeliaevEntanglement}, or implementing acoustic versions of Hawking radiation \cite{Steinhauer2016} and the dynamical Casimir effect \cite{DCETheory,DCEWestbrook}.} This state is  modified by the passing of a GW and, if the frequency of the wave matches the sum of the frequencies of the two phonon modes, the transformation of the state is resonantly enhanced in a phenomenon resembling the DCE. This frequency matching is made possible, despite the much shorter length of  BEC in comparison to the GW wavelength, because of the low speeds of sound of a BEC, which are of order $\mathrm{mm}\,\mathrm{s}^{-1}$. Such a quantum resonance process is absent in laser interferometers since the frequencies of the GWs are far from the optical regime. However, using resonance to detect GWs was  the concept behind the first  GW detectors, Weber bars, which are generally made of large metal rods.  These have much larger speeds of sound compared to BECs and so are macroscopic devices that cannot be cooled to such low temperatures.  They are, therefore, essentially classical devices, whereas the BEC detector would be a quantum device. This  allows for the utilisation of  quantum metrology \cite{PhysRevLett.72.3439}, which can enable sensitivities that are not possible with classical devices, making the detection of GWs a possibility.

The phononic field of a BEC can be considered to obey a massless Klein-Gordon equation with an effective curved space-time metric \cite{AnalogueGravityReview,RelBECs}. This depends on the real space-time metric and an analogue metric coming from the condensate. The latter metric is used in analogue gravity experiments to mimic effects predicted by quantum field theory in curved spacetime, such as Hawking radiation \cite{Steinhauer2016}.  In \cite{GWDetectorFirst}, it was shown that a GW perturbs the effective metric of phonons, which results in a Bogoliubov transformation of the field modes. This  causes a change to the prepared quantum state of the phonons, which is resonantly enhanced with frequency matching as described above. Through determining the resonant change to the quantum state of the phonons, certain properties of the GW can then be extracted and, since this is a quantum process, it is possible to achieve beyond-classical scaling in the estimation process by, for example, using squeezed states of phonons. 

In \cite{GWDetectorThermal}, the effect of finite temperature on the performance of the device was analysed and found to be negligible. However, the quantum decoherence of the phononic states at various phononic frequencies and temperatures was not considered. Here we attempt to obtain an estimate for how the device would be affected by quantum decoherence by analysing the evolution of the purity and nonclassical depth as discussed in the previous section.  Additionally, we also investigate the relaxation of the system from the evolution of the average occupation, and characterize the evolution of the full state by further determining the evolution in squeezing. 

\begin{figure*}[ht!]
     \begin{center}
     {
            \includegraphics[width=0.375\textwidth]{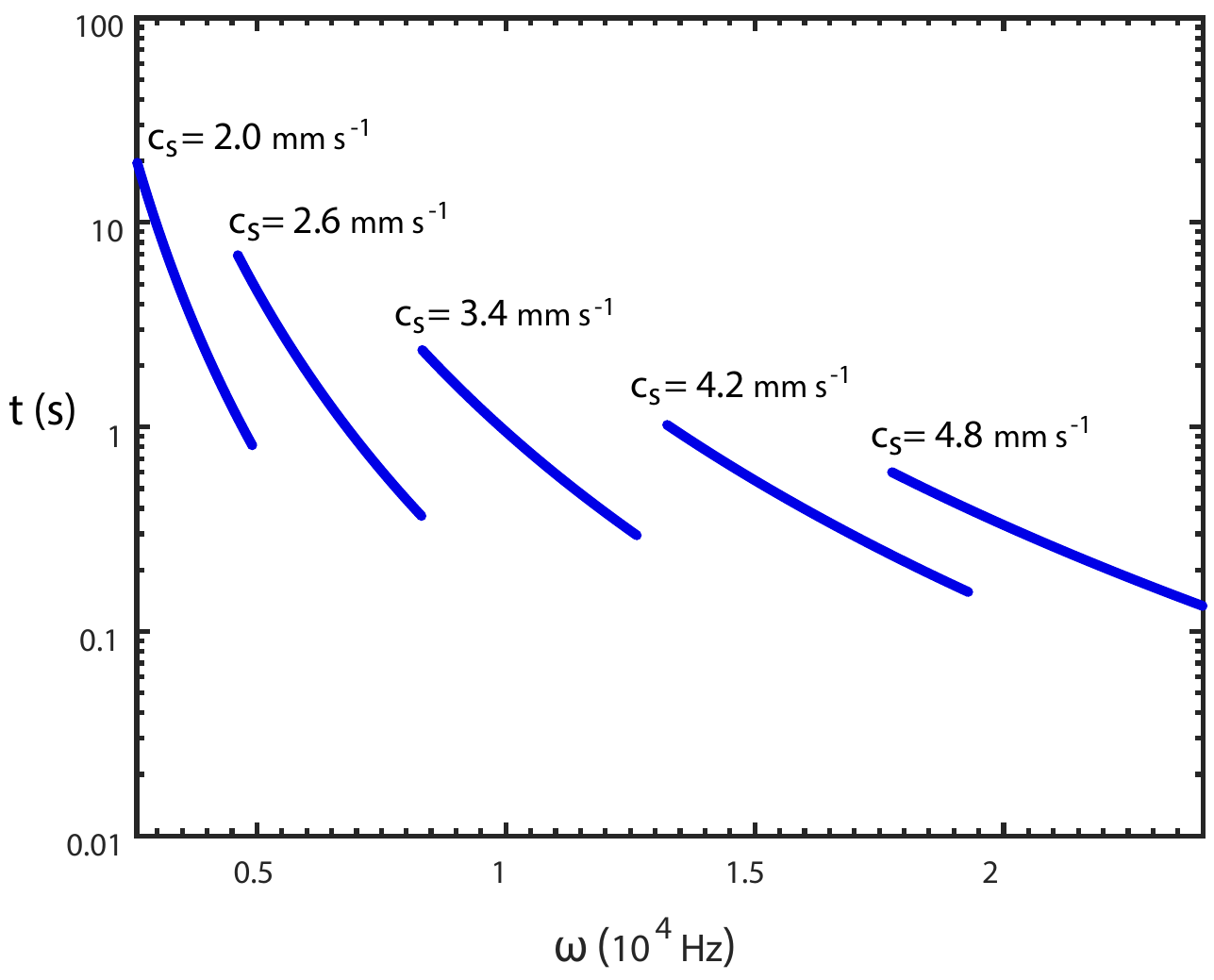}            
     }
    \end{center}
    \caption{Quantum decoherence time of phonons $t$ with frequency $\omega$ for several values of speed of sound $c_s$. Decoherence times are supplied up to the point where they match the half-life of the BEC, after which three-body recombination would be expected to provide the dominant source of decoherence. 
	}
   \label{fig:DecVsFreq}
\end{figure*} 

The most interesting phonon frequencies for the GW detector are likely to be around $10\, \mathrm{kHz}$ since this allows for the detection of GWs with frequencies that are either just beyond what LIGO can currently search for or for when the sensitivity in the device starts to rapidly tail off \cite{GWDetectorFirst,GWDetectorThermal,PhysRevD.93.112004Truncated}. Predicted signals with such frequencies include spinning neutron stars or neutron star mergers, which could inform us about the equation of state of such stars.  The optimal strain sensitivity of the detector for this frequency range was obtained for an initial two-mode squeezed state with squeezing $r = 10$.  A uniform trap for a $^{87}{\mathrm{Rb}}$ BEC was used for which temperatures as low as $0.5\,\mathrm{nK}$ have been reached \cite{Leanhardt1513,Steinhauer2014},  and  high phonon frequencies  can be generated  \cite{GWDetectorFirst}.  At $10\, \mathrm{kHz}$ phonon frequencies, $k_B T \ll\hbar \omega_q$ is likely to be satisfied for temperatures theoretically up to the critical temperature of condensation of the Bose gas. From Section \ref{sec:EvStatMoments}, this implies that Beliaev damping will dominate over Landau damping for this phonon mode in a uniform three-dimensional BEC, and the total damping rate $\gamma$ for the phonons can be approximated as \eqref{eq:QuantumRegime}.  

Figure \ref{fig:QuantumkHz} illustrates  the  time evolution of the purity, nonclassical depth, squeezing and mean occupation of a single-mode  squeezed vacuum phonon state with $r =10$ and frequency $\omega_q =  10^4\,\mathrm{Hz}$ in a $^{87}{\mathrm{Rb}}$ BEC at temperature $0.5\,\mathrm{nK}$ with speed of sound $c_s = 3.4\,\mathrm{mm}\,\mathrm{s}^{-1}$.\footnote{The speed of sound assumed here is less than that used in \cite{GWDetectorFirst} since  we are only interested in the most promising frequency of phonons for GW detection, and we want to minimise three-body decay as much as possible. Note that the value of the speed of sound $c_s$ does not directly enter the sensitivity of the detector \cite{GWDetectorFirst}, it is only used to determine what frequencies phonons can have as discussed at the end of Section \ref{sec:BogoApprox}.} From Fig. \ref{fig:QuantumkHz} the squeezed states undergo a minimum of purity before asymptomatically relaxing to the state of the environment, which is approximately in its vacuum state.  The time $t_{min}$ at which the minimum is attained is given by \eqref{eq:minPurity} and, as discussed in Section \ref{sec:Purity}, provides a good characterization of the decoherence time of such squeezed states \cite{QuantifyingDecoherenceRef}. 

The quantum decoherence time, as characterized by $t_{min}$, is around $1$\,$\mathrm{s}$ for this  high frequency phonon mode,  and would occur  even at absolute zero. A quantum decoherence time of $1$\,$\mathrm{s}$ would result in  a loss of just one  order of magnitude in strain sensitivity in comparison to that calculated using a phonon lifetime of $1000 \, \mathrm{s}$, which was assumed in \cite{GWDetectorFirst}.  This still allows for a sensitivity that improves upon that of advanced LIGO at the upper end of its frequency range (around $7 \, \mathrm{kHz}$)  \cite{GWDetectorFirst,PhysRevD.93.112004Truncated}. Note that, if we had instead chosen a thermal two-mode squeezed state, thermalized to $0.5\,\mathrm{nK}$, we would also obtain a decoherence time of $1\,\mathrm{s}$.

In Fig. \ref{fig:DecVsFreq} we also plot how the quantum decoherence time, as characterized by the minimum in purity, depends on the frequency of the phonons for different values of speed of sound.  We find that the decoherence time improves at lower frequencies and so the performance of the detector is less affected by phonon decoherence in its lower operating range of GW frequencies.

\subsection{Assumptions}

It should be emphasized that the above estimate for the quantum decoherence the GW detector was calculated under various assumptions. In particular, a three-dimensional uniform BEC is assumed, whereas the GW detector was originally investigated in a one-dimensional setting.  For one-dimensional Bose gases there is no condensation mechanism. Instead quasi one-dimensional BECs are investigated in experiments where the trapping frequencies in two of the dimensions are much greater than the third, making it possible to effectively integrate out the two dimensions to leave an approximate one-dimensional Gross-Pitaevskii equation for the condensate. Phonon damping has been investigated experimentally when moving from a three-dimensional setting to a quasi one-dimensional BEC \cite{PhysRevLett.109.195301,1367-2630-17-9-093041}, where it was found that the damping rate was reduced, with the reduction depending on the temperature  \cite{1367-2630-17-9-093041}.  Attempts to theoretically model damping in quasi one-dimensional BECs are provided in \cite{2011ChPhB..20g0307M,2015arXiv150308884T,PhysRevA.77.033610,PhysRevA.83.043625,YuenThesis}. 

A continuum of modes has also been assumed, which is an assumption that is often made in deriving phonon damping \cite{Landau:1946jc,Beliaev1958aSet2,Beliaev1958bSet2,Hohenberg1965291,PhysicalKinetics,KONDOR1974393,LiuAndShieve,PitaevskiiDamping,Liu,Giorgini1998,DampingTrapped,PhysRevA.58.3146,PhysRevA.58.R3391,Langevin,PhysRevA.59.3851,LBReview,CastinAppendix}. In particular, we've assumed that the phonon frequency of interest is ten times that of the fundamental frequency $\omega_1$ of the BEC, similar to that investigated in \cite{GWDetectorFirst}. A discrete spectrum is likely to lead to an increase in decoherence times due to the reduction in phase space for the output states. For example, the discretization of the phonon spectrum in the radial directions of an elongated trap was studied in \cite{YuenThesis} to explain the experimental results of \cite{1367-2630-17-9-093041}.\footnote{However, the axial spectrum was still assumed  to be continuous.}  A large difference can be observed compared to continuous calculations for high temperatures, but the difference reduces at low temperatures.  

The effect of the GW itself was also neglected in these calculations, which would be expected to extend the coherence life of the phonons of interest. How phonon interactions affect the squeezing created by the  dynamical Casimir effect, which is related to how the GWs create coherence of phonons \cite{GWDetectorFirst}, has been recently studied in \cite{2016arXiv160609400Z}.  

Another important assumption has been the neglect of how three-body recombination of the BEC affects the phonon modes. For a three-dimensional BEC, three-body interactions cause the density of the BEC to evolve as  \cite{PhysRevA.53.916}
\begin{align} \label{eq:ThreeBody}
\frac{d \rho (t)}{dt} = -\gamma(t) \rho(t)
\end{align}
where $\gamma(t) := L \rho^2(t)$ and $L$ has been found to be $5.8(1.9) \times 10^{-42} \,\mathrm{m^6\,s^{-1}}$ for a  $^{87}\mathrm{Rb}$ BEC in the $F=1$, $m_f = -1$ hyperfine state \cite{Rb3bodyRate}. From \eqref{eq:ThreeBody}, the half-life of the BEC is approximately $3 / (2 \gamma(0))$. Using $c_s = 3.4 \mathrm{mm}\,\mathrm{s}^{-1}$ we find that decay rate $\gamma(0)$ is less than the Beliaev damping rate $\gamma_B$, and that the half-life of the BEC is over twice as long as the predicted time for full quantum decoherence of the phonons. Unlike for three-dimensional BECs, for which three-body recombination is constant at ultracold energies \cite{PhysRevA.65.010705}, in two and one-dimensional Bose gases three-body recombination can be vanishing at ultracold energies \cite{PhysRevA.76.022711,PhysRevA.83.052703,PhysRevA.91.062710}.  Therefore, moving to a quasi one- or two-dimensional BEC operating at low temperatures would be expected to reduce the three-body recombination rate \cite{PhysRevA.91.062710}.

As well as removing atoms from the trap, three-body recombination will also cause the BEC to heat up. In \cite{PhysRevA.68.043607}, the way in which three-body recombination, and in general n-body inelastic interactions, affects the phonons of a BEC was studied using an open system approach. A similar master equation is found to that of \eqref{eq:lindblad} but  is derived using three-body inelastic interactions (and more generally n-body inelastic interactions), rather than the intrinsic two-body atomic interactions considered here, and the environment is the modes outside the trap. The relaxation rate was found to be comparable to the normal decay rate of three-body interactions, $\gamma(t)$ in \eqref{eq:ThreeBody}, which is less than the Beliaev decay rate considered here for the three-dimensional detector at $t=0$, as discussed above, and will continue to decrease as the particle number falls.     

Of course, now that an estimate for the quantum decoherence time has been calculated for this frequency range we can also begin to investigate how the detector could be modified in order to increase this time. For example, one option might be to increase $\omega_1$ such that the initial state involves a lower energy mode of the cavity for which Beliaev damping will be suppressed.  However, this would require modifications to $c_s$ and the trap geometry, such as increasing the speed of sound and reducing the effective size of the trap, which would lead to more rapid three body recombination. Other options to increase the quantum decoherence time could be to squeeze the environment \cite{QuantifyingDecoherenceRef}, use larger mass BECs such as $^{174}\mathrm{Yb}$, and to utilise lower dimensional traps as described above.

\section{Conclusions}

We have investigated the quantum decoherence time of phononic excitations of an isolated BEC in the Born-Markov approximation and assuming that the phonon states are Gaussian.  The results can be used to assess the resourcefulness of phonons of BECs as carriers of quantum information. In particular, we have estimated the quantum decoherence time of phonons for a three-dimensional version of the GW detector that was proposed in  \cite{GWDetectorFirst}, and found that this still allows for a very high sensitivity at the most promising GW frequencies for detection.

Although this work has been theoretical, it should also be possible to estimate the quantum decoherence of phonons experimentally.  In particular, the general results of Section \ref{sec:QuantifyingDecoherence} should be applicable to  generic BEC setups, not just the GW detector presented in Section \ref{sec:Example}, with the results just depending on a few experimental parameters such as temperature and frequency of the phonon mode.  One possible way to measure the quantum decoherence time of a squeezed single-mode phonon state would be to determine the time at which the state's purity reaches a minimum. Purity is a non-linear function of the state's density operator and so is not related to the expectation value of a single-system Hermitian operator or a single-system probability distribution that would be obtained from a positive operator-valued measure \cite{Purity}.  However, if the full quantum state of the system is known then the purity can be determined. For Gaussian states  this would  mean  having to determine their first two statistical moments, which can be measured by the joint detection of two conjugate quadratures via heterodyne and multi-port homodyne detection schemes from quantum optics \cite{Purity}. Similar methods have also been discussed for phonons of BECs in measurements of entanglement or squeezing  \cite{EntangledPhonons,PhysRevD.92.024043,BeliaevEntanglement,Steinhauer2016,PhysRevA.89.063606,PhysRevD.89.105024,PhysRevD.87.124018,2016arXiv161103904R}, which could also likely be tailored to study purity.

We investigated the quantum decoherence of phonons of BECs by analysing the evolution of the purity and nonclassical depth of a single-mode system. It would, however, also be instructive to determine the evolution of proper coherence measures \cite{2006quant.ph.12146A,QuantifyingCoherence,PhysRevA.92.022124,PhysRevLett.115.020403,2016arXiv160603181Y}, which have recently been applied to infinite dimensional systems and Gaussian states \cite{PhysRevA.93.012334,PhysRevA.93.032111}, for an analysis of quantum decoherence.  For multi-mode states, another useful measure for loss of quantumness of a state is the evolution of entanglement \cite{QuantifyingDecoherenceRef}.  Entanglement has recently been observed for phonon states in the emission of the acoustic analogue of Hawking radiation \cite{Steinhauer2016}, potentially allowing for future studies into how the entanglement degrades with time.  Understanding this de-entanglement processes could  dictate what is possible to measure in analogue experiments, and perhaps  provide potential clues to the information paradox in black hole physics \cite{Kiefer2001,Kiefer2004}.%\\ %We leave such proposals, however,  to future work.

\section*{Acknowledgements}

We thank Sabrina Maniscalco, Gerardo Adesso and Denis Boiron for useful discussions and comments. R.H. and I.F. would like to acknowledge that this project was made possible through the support of the grant `Leaps in cosmology: gravitational wave detection with quantum systems' (No. 58745) from the John Templeton Foundation. The opinions expressed in this publication are those of the authors and do not necessarily reflect the views of the John Templeton Foundation.  Financial support by Fundaci\'on General CSIC (Programa ComFuturo) is acknowledged by C.S.


\begin{thebibliography}{148}%
\makeatletter
\providecommand \@ifxundefined [1]{%
 \@ifx{#1\undefined}
}%
\providecommand \@ifnum [1]{%
 \ifnum #1\expandafter \@firstoftwo
 \else \expandafter \@secondoftwo
 \fi
}%
\providecommand \@ifx [1]{%
 \ifx #1\expandafter \@firstoftwo
 \else \expandafter \@secondoftwo
 \fi
}%
\providecommand \natexlab [1]{#1}%
\providecommand \enquote  [1]{``#1''}%
\providecommand \bibnamefont  [1]{#1}%
\providecommand \bibfnamefont [1]{#1}%
\providecommand \citenamefont [1]{#1}%
\providecommand \href@noop [0]{\@secondoftwo}%
\providecommand \href [0]{\begingroup \@sanitize@url \@href}%
\providecommand \@href[1]{\@@startlink{#1}\@@href}%
\providecommand \@@href[1]{\endgroup#1\@@endlink}%
\providecommand \@sanitize@url [0]{\catcode `\\12\catcode `\$12\catcode
  `\&12\catcode `\#12\catcode `\^12\catcode `\_12\catcode `\%12\relax}%
\providecommand \@@startlink[1]{}%
\providecommand \@@endlink[0]{}%
\providecommand \url  [0]{\begingroup\@sanitize@url \@url }%
\providecommand \@url [1]{\endgroup\@href {#1}{\urlprefix }}%
\providecommand \urlprefix  [0]{URL }%
\providecommand \Eprint [0]{\href }%
\providecommand \doibase [0]{http://dx.doi.org/}%
\providecommand \selectlanguage [0]{\@gobble}%
\providecommand \bibinfo  [0]{\@secondoftwo}%
\providecommand \bibfield  [0]{\@secondoftwo}%
\providecommand \translation [1]{[#1]}%
\providecommand \BibitemOpen [0]{}%
\providecommand \bibitemStop [0]{}%
\providecommand \bibitemNoStop [0]{.\EOS\space}%
\providecommand \EOS [0]{\spacefactor3000\relax}%
\providecommand \BibitemShut  [1]{\csname bibitem#1\endcsname}%
\let\auto@bib@innerbib\@empty
%</preamble>
\bibitem [{\citenamefont {{Zeh}}(1970)}]{Decoherence1970}%
  \BibitemOpen
  \bibfield  {author} {\bibinfo {author} {\bibfnamefont {H.~D.}\ \bibnamefont
  {{Zeh}}},\ }\href {\doibase 10.1007/BF00708656} {\bibfield  {journal}
  {\bibinfo  {journal} {Foundations of Physics}\ }\textbf {\bibinfo {volume}
  {1}},\ \bibinfo {pages} {69} (\bibinfo {year} {1970})}\BibitemShut {NoStop}%
\bibitem [{\citenamefont {Schlosshauer}(2005)}]{RevModPhys.76.1267}%
  \BibitemOpen
  \bibfield  {author} {\bibinfo {author} {\bibfnamefont {M.}~\bibnamefont
  {Schlosshauer}},\ }\href {\doibase 10.1103/RevModPhys.76.1267} {\bibfield
  {journal} {\bibinfo  {journal} {Rev. Mod. Phys.}\ }\textbf {\bibinfo {volume}
  {76}},\ \bibinfo {pages} {1267} (\bibinfo {year} {2005})}\BibitemShut
  {NoStop}%
\bibitem [{\citenamefont {Joos}\ and\ \citenamefont {Zeh}(1985)}]{Joos85_2}%
  \BibitemOpen
  \bibfield  {author} {\bibinfo {author} {\bibfnamefont {E.}~\bibnamefont
  {Joos}}\ and\ \bibinfo {author} {\bibfnamefont {H.~D.}\ \bibnamefont {Zeh}},\
  }\href {\doibase 10.1007/BF01725541} {\bibfield  {journal} {\bibinfo
  {journal} {Zeitschrift f{\"u}r Physik B Condensed Matter}\ }\textbf {\bibinfo
  {volume} {59}},\ \bibinfo {pages} {223} (\bibinfo {year} {1985})}\BibitemShut
  {NoStop}%
\bibitem [{\citenamefont {Zurek}(1986)}]{Zurek1986}%
  \BibitemOpen
  \bibfield  {author} {\bibinfo {author} {\bibfnamefont {W.~H.}\ \bibnamefont
  {Zurek}},\ }\href {\doibase 10.1007/978-1-4613-2181-1_10} {\emph {\bibinfo
  {title} {Frontiers of Nonequilibrium Statistical Physics}}},\ edited by\
  \bibinfo {editor} {\bibfnamefont {G.~T.}\ \bibnamefont {Moore}}\ and\
  \bibinfo {editor} {\bibfnamefont {M.~O.}\ \bibnamefont {Scully}}\ (\bibinfo
  {publisher} {Springer US},\ \bibinfo {address} {Boston, MA},\ \bibinfo {year}
  {1986})\ Chap.\ \bibinfo {chapter} {Reduction of the Wavepacket: How Long
  Does it Take?}, pp.\ \bibinfo {pages} {145--149}\BibitemShut {NoStop}%
\bibitem [{\citenamefont {Zurek}(2003)}]{RevModPhys.75.715}%
  \BibitemOpen
  \bibfield  {author} {\bibinfo {author} {\bibfnamefont {W.~H.}\ \bibnamefont
  {Zurek}},\ }\href {\doibase 10.1103/RevModPhys.75.715} {\bibfield  {journal}
  {\bibinfo  {journal} {Rev. Mod. Phys.}\ }\textbf {\bibinfo {volume} {75}},\
  \bibinfo {pages} {715} (\bibinfo {year} {2003})}\BibitemShut {NoStop}%
\bibitem [{\citenamefont {{Zurek}}(2003)}]{2003quant.ph..6072Z}%
  \BibitemOpen
  \bibfield  {author} {\bibinfo {author} {\bibfnamefont {W.~H.}\ \bibnamefont
  {{Zurek}}},\ }\href@noop {} {\bibfield  {journal} {\bibinfo  {journal}
  {eprint arXiv:quant-ph/0306072}\ } (\bibinfo {year} {2003})},\ \Eprint
  {http://arxiv.org/abs/quant-ph/0306072} {quant-ph/0306072} \BibitemShut
  {NoStop}%
\bibitem [{\citenamefont {Cataliotti}\ \emph {et~al.}(2001)\citenamefont
  {Cataliotti}, \citenamefont {Burger}, \citenamefont {Fort}, \citenamefont
  {Maddaloni}, \citenamefont {Minardi}, \citenamefont {Trombettoni},
  \citenamefont {Smerzi},\ and\ \citenamefont {Inguscio}}]{Cataliotti03082001}%
  \BibitemOpen
  \bibfield  {author} {\bibinfo {author} {\bibfnamefont {F.~S.}\ \bibnamefont
  {Cataliotti}}, \bibinfo {author} {\bibfnamefont {S.}~\bibnamefont {Burger}},
  \bibinfo {author} {\bibfnamefont {C.}~\bibnamefont {Fort}}, \bibinfo {author}
  {\bibfnamefont {P.}~\bibnamefont {Maddaloni}}, \bibinfo {author}
  {\bibfnamefont {F.}~\bibnamefont {Minardi}}, \bibinfo {author} {\bibfnamefont
  {A.}~\bibnamefont {Trombettoni}}, \bibinfo {author} {\bibfnamefont
  {A.}~\bibnamefont {Smerzi}}, \ and\ \bibinfo {author} {\bibfnamefont
  {M.}~\bibnamefont {Inguscio}},\ }\href {\doibase 10.1126/science.1062612}
  {\bibfield  {journal} {\bibinfo  {journal} {Science}\ }\textbf {\bibinfo
  {volume} {293}},\ \bibinfo {pages} {843} (\bibinfo {year}
  {2001})}\BibitemShut {NoStop}%
\bibitem [{\citenamefont {{Salgueiro, A. N.}}\ \emph
  {et~al.}(2007)\citenamefont {{Salgueiro, A. N.}}, \citenamefont {{de Toledo
  Piza, A.F.R.}}, \citenamefont {{Lemos, G. B.}}, \citenamefont {{Drumond,
  R.}}, \citenamefont {{Nemes, M. C.}},\ and\ \citenamefont {{Weidem\"{u}ller,
  M.}}}]{refId0}%
  \BibitemOpen
  \bibfield  {author} {\bibinfo {author} {\bibnamefont {{Salgueiro, A. N.}}},
  \bibinfo {author} {\bibnamefont {{de Toledo Piza, A.F.R.}}}, \bibinfo
  {author} {\bibnamefont {{Lemos, G. B.}}}, \bibinfo {author} {\bibnamefont
  {{Drumond, R.}}}, \bibinfo {author} {\bibnamefont {{Nemes, M. C.}}}, \ and\
  \bibinfo {author} {\bibnamefont {{Weidem\"{u}ller, M.}}},\ }\href {\doibase
  10.1140/epjd/e2007-00224-4} {\bibfield  {journal} {\bibinfo  {journal} {Eur.
  Phys. J. D}\ }\textbf {\bibinfo {volume} {44}},\ \bibinfo {pages} {537}
  (\bibinfo {year} {2007})}\BibitemShut {NoStop}%
\bibitem [{\citenamefont {Hall}\ \emph
  {et~al.}(1998{\natexlab{a}})\citenamefont {Hall}, \citenamefont {Matthews},
  \citenamefont {Ensher}, \citenamefont {Wieman},\ and\ \citenamefont
  {Cornell}}]{PhysRevLett.81.1539}%
  \BibitemOpen
  \bibfield  {author} {\bibinfo {author} {\bibfnamefont {D.~S.}\ \bibnamefont
  {Hall}}, \bibinfo {author} {\bibfnamefont {M.~R.}\ \bibnamefont {Matthews}},
  \bibinfo {author} {\bibfnamefont {J.~R.}\ \bibnamefont {Ensher}}, \bibinfo
  {author} {\bibfnamefont {C.~E.}\ \bibnamefont {Wieman}}, \ and\ \bibinfo
  {author} {\bibfnamefont {E.~A.}\ \bibnamefont {Cornell}},\ }\href {\doibase
  10.1103/PhysRevLett.81.1539} {\bibfield  {journal} {\bibinfo  {journal}
  {Phys. Rev. Lett.}\ }\textbf {\bibinfo {volume} {81}},\ \bibinfo {pages}
  {1539} (\bibinfo {year} {1998}{\natexlab{a}})}\BibitemShut {NoStop}%
\bibitem [{\citenamefont {Hall}\ \emph
  {et~al.}(1998{\natexlab{b}})\citenamefont {Hall}, \citenamefont {Matthews},
  \citenamefont {Wieman},\ and\ \citenamefont {Cornell}}]{PhysRevLett.81.1543}%
  \BibitemOpen
  \bibfield  {author} {\bibinfo {author} {\bibfnamefont {D.~S.}\ \bibnamefont
  {Hall}}, \bibinfo {author} {\bibfnamefont {M.~R.}\ \bibnamefont {Matthews}},
  \bibinfo {author} {\bibfnamefont {C.~E.}\ \bibnamefont {Wieman}}, \ and\
  \bibinfo {author} {\bibfnamefont {E.~A.}\ \bibnamefont {Cornell}},\ }\href
  {\doibase 10.1103/PhysRevLett.81.1543} {\bibfield  {journal} {\bibinfo
  {journal} {Phys. Rev. Lett.}\ }\textbf {\bibinfo {volume} {81}},\ \bibinfo
  {pages} {1543} (\bibinfo {year} {1998}{\natexlab{b}})}\BibitemShut {NoStop}%
\bibitem [{\citenamefont {{Gross}}\ \emph {et~al.}(2010)\citenamefont
  {{Gross}}, \citenamefont {{Zibold}}, \citenamefont {{Nicklas}}, \citenamefont
  {{Est{\`e}ve}},\ and\ \citenamefont {{Oberthaler}}}]{2010Natur.464.1165G}%
  \BibitemOpen
  \bibfield  {author} {\bibinfo {author} {\bibfnamefont {C.}~\bibnamefont
  {{Gross}}}, \bibinfo {author} {\bibfnamefont {T.}~\bibnamefont {{Zibold}}},
  \bibinfo {author} {\bibfnamefont {E.}~\bibnamefont {{Nicklas}}}, \bibinfo
  {author} {\bibfnamefont {J.}~\bibnamefont {{Est{\`e}ve}}}, \ and\ \bibinfo
  {author} {\bibfnamefont {M.~K.}\ \bibnamefont {{Oberthaler}}},\ }\href
  {\doibase 10.1038/nature08919} {\bibfield  {journal} {\bibinfo  {journal}
  {Nature}\ }\textbf {\bibinfo {volume} {464}},\ \bibinfo {pages} {1165}
  (\bibinfo {year} {2010})}\BibitemShut {NoStop}%
\bibitem [{\citenamefont {Wang}\ \emph {et~al.}(2005)\citenamefont {Wang},
  \citenamefont {Anderson}, \citenamefont {Bright}, \citenamefont {Cornell},
  \citenamefont {Diot}, \citenamefont {Kishimoto}, \citenamefont {Prentiss},
  \citenamefont {Saravanan}, \citenamefont {Segal},\ and\ \citenamefont
  {Wu}}]{PhysRevLett.94.090405}%
  \BibitemOpen
  \bibfield  {author} {\bibinfo {author} {\bibfnamefont {Y.-J.}\ \bibnamefont
  {Wang}}, \bibinfo {author} {\bibfnamefont {D.~Z.}\ \bibnamefont {Anderson}},
  \bibinfo {author} {\bibfnamefont {V.~M.}\ \bibnamefont {Bright}}, \bibinfo
  {author} {\bibfnamefont {E.~A.}\ \bibnamefont {Cornell}}, \bibinfo {author}
  {\bibfnamefont {Q.}~\bibnamefont {Diot}}, \bibinfo {author} {\bibfnamefont
  {T.}~\bibnamefont {Kishimoto}}, \bibinfo {author} {\bibfnamefont
  {M.}~\bibnamefont {Prentiss}}, \bibinfo {author} {\bibfnamefont {R.~A.}\
  \bibnamefont {Saravanan}}, \bibinfo {author} {\bibfnamefont {S.~R.}\
  \bibnamefont {Segal}}, \ and\ \bibinfo {author} {\bibfnamefont
  {S.}~\bibnamefont {Wu}},\ }\href {\doibase 10.1103/PhysRevLett.94.090405}
  {\bibfield  {journal} {\bibinfo  {journal} {Phys. Rev. Lett.}\ }\textbf
  {\bibinfo {volume} {94}},\ \bibinfo {pages} {090405} (\bibinfo {year}
  {2005})}\BibitemShut {NoStop}%
\bibitem [{\citenamefont {{H{\"a}nsel}}\ \emph
  {et~al.}(2001{\natexlab{a}})\citenamefont {{H{\"a}nsel}}, \citenamefont
  {{Hommelhoff}}, \citenamefont {{H{\"a}nsch}},\ and\ \citenamefont
  {{Reichel}}}]{2001Natur.413..498H}%
  \BibitemOpen
  \bibfield  {author} {\bibinfo {author} {\bibfnamefont {W.}~\bibnamefont
  {{H{\"a}nsel}}}, \bibinfo {author} {\bibfnamefont {P.}~\bibnamefont
  {{Hommelhoff}}}, \bibinfo {author} {\bibfnamefont {T.~W.}\ \bibnamefont
  {{H{\"a}nsch}}}, \ and\ \bibinfo {author} {\bibfnamefont {J.}~\bibnamefont
  {{Reichel}}},\ }\href {\doibase 10.1038/35097032} {\bibfield  {journal}
  {\bibinfo  {journal} {Nature}\ }\textbf {\bibinfo {volume} {413}},\ \bibinfo
  {pages} {498} (\bibinfo {year} {2001}{\natexlab{a}})}\BibitemShut {NoStop}%
\bibitem [{\citenamefont {{Berrada}}\ \emph {et~al.}(2013)\citenamefont
  {{Berrada}}, \citenamefont {{van Frank}}, \citenamefont {{B{\"u}cker}},
  \citenamefont {{Schumm}}, \citenamefont {{Schaff}},\ and\ \citenamefont
  {{Schmiedmayer}}}]{2013NatCo...4E2077B}%
  \BibitemOpen
  \bibfield  {author} {\bibinfo {author} {\bibfnamefont {T.}~\bibnamefont
  {{Berrada}}}, \bibinfo {author} {\bibfnamefont {S.}~\bibnamefont {{van
  Frank}}}, \bibinfo {author} {\bibfnamefont {R.}~\bibnamefont {{B{\"u}cker}}},
  \bibinfo {author} {\bibfnamefont {T.}~\bibnamefont {{Schumm}}}, \bibinfo
  {author} {\bibfnamefont {J.-F.}\ \bibnamefont {{Schaff}}}, \ and\ \bibinfo
  {author} {\bibfnamefont {J.}~\bibnamefont {{Schmiedmayer}}},\ }\href
  {\doibase 10.1038/ncomms3077} {\bibfield  {journal} {\bibinfo  {journal}
  {Nature Communications}\ }\textbf {\bibinfo {volume} {4}},\ \bibinfo {eid}
  {2077} (\bibinfo {year} {2013})}\BibitemShut {NoStop}%
\bibitem [{\citenamefont {Fattori}\ \emph {et~al.}(2008)\citenamefont
  {Fattori}, \citenamefont {D'Errico}, \citenamefont {Roati}, \citenamefont
  {Zaccanti}, \citenamefont {Jona-Lasinio}, \citenamefont {Modugno},
  \citenamefont {Inguscio},\ and\ \citenamefont
  {Modugno}}]{PhysRevLett.100.080405}%
  \BibitemOpen
  \bibfield  {author} {\bibinfo {author} {\bibfnamefont {M.}~\bibnamefont
  {Fattori}}, \bibinfo {author} {\bibfnamefont {C.}~\bibnamefont {D'Errico}},
  \bibinfo {author} {\bibfnamefont {G.}~\bibnamefont {Roati}}, \bibinfo
  {author} {\bibfnamefont {M.}~\bibnamefont {Zaccanti}}, \bibinfo {author}
  {\bibfnamefont {M.}~\bibnamefont {Jona-Lasinio}}, \bibinfo {author}
  {\bibfnamefont {M.}~\bibnamefont {Modugno}}, \bibinfo {author} {\bibfnamefont
  {M.}~\bibnamefont {Inguscio}}, \ and\ \bibinfo {author} {\bibfnamefont
  {G.}~\bibnamefont {Modugno}},\ }\href {\doibase
  10.1103/PhysRevLett.100.080405} {\bibfield  {journal} {\bibinfo  {journal}
  {Phys. Rev. Lett.}\ }\textbf {\bibinfo {volume} {100}},\ \bibinfo {pages}
  {080405} (\bibinfo {year} {2008})}\BibitemShut {NoStop}%
\bibitem [{\citenamefont {Schumm}\ \emph {et~al.}(2006)\citenamefont {Schumm},
  \citenamefont {Krüger}, \citenamefont {Hofferberth}, \citenamefont
  {Lesanovsky}, \citenamefont {Wildermuth}, \citenamefont {Groth},
  \citenamefont {Bar-Joseph}, \citenamefont {Andersson},\ and\ \citenamefont
  {Schmiedmayer}}]{Schumm06}%
  \BibitemOpen
  \bibfield  {author} {\bibinfo {author} {\bibfnamefont {T.}~\bibnamefont
  {Schumm}}, \bibinfo {author} {\bibfnamefont {P.}~\bibnamefont {Krüger}},
  \bibinfo {author} {\bibfnamefont {S.}~\bibnamefont {Hofferberth}}, \bibinfo
  {author} {\bibfnamefont {I.}~\bibnamefont {Lesanovsky}}, \bibinfo {author}
  {\bibfnamefont {S.}~\bibnamefont {Wildermuth}}, \bibinfo {author}
  {\bibfnamefont {S.}~\bibnamefont {Groth}}, \bibinfo {author} {\bibfnamefont
  {I.}~\bibnamefont {Bar-Joseph}}, \bibinfo {author} {\bibfnamefont
  {L.}~\bibnamefont {Andersson}}, \ and\ \bibinfo {author} {\bibfnamefont
  {J.}~\bibnamefont {Schmiedmayer}},\ }\href {\doibase
  10.1007/s11128-006-0033-2} {\bibfield  {journal} {\bibinfo  {journal}
  {Quantum Information Processing}\ }\textbf {\bibinfo {volume} {5}},\ \bibinfo
  {pages} {537} (\bibinfo {year} {2006})}\BibitemShut {NoStop}%
\bibitem [{\citenamefont {{Folman}}\ \emph {et~al.}(2000)\citenamefont
  {{Folman}}, \citenamefont {{Kr{\"u}ger}}, \citenamefont {{Cassettari}},
  \citenamefont {{Hessmo}}, \citenamefont {{Maier}},\ and\ \citenamefont
  {{Schmiedmayer}}}]{2000PhRvL..84.4749F}%
  \BibitemOpen
  \bibfield  {author} {\bibinfo {author} {\bibfnamefont {R.}~\bibnamefont
  {{Folman}}}, \bibinfo {author} {\bibfnamefont {P.}~\bibnamefont
  {{Kr{\"u}ger}}}, \bibinfo {author} {\bibfnamefont {D.}~\bibnamefont
  {{Cassettari}}}, \bibinfo {author} {\bibfnamefont {B.}~\bibnamefont
  {{Hessmo}}}, \bibinfo {author} {\bibfnamefont {T.}~\bibnamefont {{Maier}}}, \
  and\ \bibinfo {author} {\bibfnamefont {J.}~\bibnamefont {{Schmiedmayer}}},\
  }\href {\doibase 10.1103/PhysRevLett.84.4749} {\bibfield  {journal} {\bibinfo
   {journal} {Physical Review Letters}\ }\textbf {\bibinfo {volume} {84}},\
  \bibinfo {pages} {4749} (\bibinfo {year} {2000})}\BibitemShut {NoStop}%
\bibitem [{\citenamefont {Pyrkov}\ and\ \citenamefont
  {Byrnes}(2013)}]{1367-2630-15-9-093019}%
  \BibitemOpen
  \bibfield  {author} {\bibinfo {author} {\bibfnamefont {A.~N.}\ \bibnamefont
  {Pyrkov}}\ and\ \bibinfo {author} {\bibfnamefont {T.}~\bibnamefont
  {Byrnes}},\ }\href {http://stacks.iop.org/1367-2630/15/i=9/a=093019}
  {\bibfield  {journal} {\bibinfo  {journal} {New Journal of Physics}\ }\textbf
  {\bibinfo {volume} {15}},\ \bibinfo {pages} {093019} (\bibinfo {year}
  {2013})}\BibitemShut {NoStop}%
\bibitem [{\citenamefont {Byrnes}\ \emph {et~al.}(2012)\citenamefont {Byrnes},
  \citenamefont {Wen},\ and\ \citenamefont {Yamamoto}}]{PhysRevA.85.040306}%
  \BibitemOpen
  \bibfield  {author} {\bibinfo {author} {\bibfnamefont {T.}~\bibnamefont
  {Byrnes}}, \bibinfo {author} {\bibfnamefont {K.}~\bibnamefont {Wen}}, \ and\
  \bibinfo {author} {\bibfnamefont {Y.}~\bibnamefont {Yamamoto}},\ }\href
  {\doibase 10.1103/PhysRevA.85.040306} {\bibfield  {journal} {\bibinfo
  {journal} {Phys. Rev. A}\ }\textbf {\bibinfo {volume} {85}},\ \bibinfo
  {pages} {040306} (\bibinfo {year} {2012})}\BibitemShut {NoStop}%
\bibitem [{\citenamefont {{H{\"a}nsel}}\ \emph
  {et~al.}(2001{\natexlab{b}})\citenamefont {{H{\"a}nsel}}, \citenamefont
  {{Hommelhoff}}, \citenamefont {{H{\"a}nsch}},\ and\ \citenamefont
  {{Reichel}}}]{2001Natur.413..498HComp}%
  \BibitemOpen
  \bibfield  {author} {\bibinfo {author} {\bibfnamefont {W.}~\bibnamefont
  {{H{\"a}nsel}}}, \bibinfo {author} {\bibfnamefont {P.}~\bibnamefont
  {{Hommelhoff}}}, \bibinfo {author} {\bibfnamefont {T.~W.}\ \bibnamefont
  {{H{\"a}nsch}}}, \ and\ \bibinfo {author} {\bibfnamefont {J.}~\bibnamefont
  {{Reichel}}},\ }\href {\doibase 10.1038/35097032} {\bibfield  {journal}
  {\bibinfo  {journal} {Nature}\ }\textbf {\bibinfo {volume} {413}},\ \bibinfo
  {pages} {498} (\bibinfo {year} {2001}{\natexlab{b}})}\BibitemShut {NoStop}%
\bibitem [{\citenamefont {{S{\o}rensen}}\ \emph {et~al.}(2001)\citenamefont
  {{S{\o}rensen}}, \citenamefont {{Duan}}, \citenamefont {{Cirac}},\ and\
  \citenamefont {{Zoller}}}]{2001Natur.409...63S}%
  \BibitemOpen
  \bibfield  {author} {\bibinfo {author} {\bibfnamefont {A.}~\bibnamefont
  {{S{\o}rensen}}}, \bibinfo {author} {\bibfnamefont {L.-M.}\ \bibnamefont
  {{Duan}}}, \bibinfo {author} {\bibfnamefont {J.~I.}\ \bibnamefont {{Cirac}}},
  \ and\ \bibinfo {author} {\bibfnamefont {P.}~\bibnamefont {{Zoller}}},\
  }\href {\doibase 10.1038/35051038} {\bibfield  {journal} {\bibinfo  {journal}
  {Nature}\ }\textbf {\bibinfo {volume} {409}},\ \bibinfo {pages} {63}
  (\bibinfo {year} {2001})}\BibitemShut {NoStop}%
\bibitem [{\citenamefont {Buluta}\ and\ \citenamefont
  {Nori}(2009)}]{Buluta02102009}%
  \BibitemOpen
  \bibfield  {author} {\bibinfo {author} {\bibfnamefont {I.}~\bibnamefont
  {Buluta}}\ and\ \bibinfo {author} {\bibfnamefont {F.}~\bibnamefont {Nori}},\
  }\href {\doibase 10.1126/science.1177838} {\bibfield  {journal} {\bibinfo
  {journal} {Science}\ }\textbf {\bibinfo {volume} {326}},\ \bibinfo {pages}
  {108} (\bibinfo {year} {2009})}\BibitemShut {NoStop}%
\bibitem [{\citenamefont {Fort\'agh}\ and\ \citenamefont
  {Zimmermann}(2007)}]{RevModPhys.79.235}%
  \BibitemOpen
  \bibfield  {author} {\bibinfo {author} {\bibfnamefont {J.}~\bibnamefont
  {Fort\'agh}}\ and\ \bibinfo {author} {\bibfnamefont {C.}~\bibnamefont
  {Zimmermann}},\ }\href {\doibase 10.1103/RevModPhys.79.235} {\bibfield
  {journal} {\bibinfo  {journal} {Rev. Mod. Phys.}\ }\textbf {\bibinfo {volume}
  {79}},\ \bibinfo {pages} {235} (\bibinfo {year} {2007})}\BibitemShut
  {NoStop}%
\bibitem [{\citenamefont {Hinds}\ \emph {et~al.}(1998)\citenamefont {Hinds},
  \citenamefont {Boshier},\ and\ \citenamefont {Hughes}}]{PhysRevLett.80.645}%
  \BibitemOpen
  \bibfield  {author} {\bibinfo {author} {\bibfnamefont {E.~A.}\ \bibnamefont
  {Hinds}}, \bibinfo {author} {\bibfnamefont {M.~G.}\ \bibnamefont {Boshier}},
  \ and\ \bibinfo {author} {\bibfnamefont {I.~G.}\ \bibnamefont {Hughes}},\
  }\href {\doibase 10.1103/PhysRevLett.80.645} {\bibfield  {journal} {\bibinfo
  {journal} {Phys. Rev. Lett.}\ }\textbf {\bibinfo {volume} {80}},\ \bibinfo
  {pages} {645} (\bibinfo {year} {1998})}\BibitemShut {NoStop}%
\bibitem [{\citenamefont {Dekker}\ \emph {et~al.}(2000)\citenamefont {Dekker},
  \citenamefont {Lee}, \citenamefont {Lorent}, \citenamefont {Thywissen},
  \citenamefont {Smith}, \citenamefont {Drndi\ifmmode~\acute{c}\else
  \'{c}\fi{}}, \citenamefont {Westervelt},\ and\ \citenamefont
  {Prentiss}}]{PhysRevLett.84.1124}%
  \BibitemOpen
  \bibfield  {author} {\bibinfo {author} {\bibfnamefont {N.~H.}\ \bibnamefont
  {Dekker}}, \bibinfo {author} {\bibfnamefont {C.~S.}\ \bibnamefont {Lee}},
  \bibinfo {author} {\bibfnamefont {V.}~\bibnamefont {Lorent}}, \bibinfo
  {author} {\bibfnamefont {J.~H.}\ \bibnamefont {Thywissen}}, \bibinfo {author}
  {\bibfnamefont {S.~P.}\ \bibnamefont {Smith}}, \bibinfo {author}
  {\bibfnamefont {M.}~\bibnamefont {Drndi\ifmmode~\acute{c}\else \'{c}\fi{}}},
  \bibinfo {author} {\bibfnamefont {R.~M.}\ \bibnamefont {Westervelt}}, \ and\
  \bibinfo {author} {\bibfnamefont {M.}~\bibnamefont {Prentiss}},\ }\href
  {\doibase 10.1103/PhysRevLett.84.1124} {\bibfield  {journal} {\bibinfo
  {journal} {Phys. Rev. Lett.}\ }\textbf {\bibinfo {volume} {84}},\ \bibinfo
  {pages} {1124} (\bibinfo {year} {2000})}\BibitemShut {NoStop}%
\bibitem [{\citenamefont {{Ahmadi}}\ \emph {et~al.}(2014)\citenamefont
  {{Ahmadi}}, \citenamefont {{Bruschi}}, \citenamefont {{Sab{\'{\i}}n}},
  \citenamefont {{Adesso}},\ and\ \citenamefont {{Fuentes}}}]{Accelerometer}%
  \BibitemOpen
  \bibfield  {author} {\bibinfo {author} {\bibfnamefont {M.}~\bibnamefont
  {{Ahmadi}}}, \bibinfo {author} {\bibfnamefont {D.~E.}\ \bibnamefont
  {{Bruschi}}}, \bibinfo {author} {\bibfnamefont {C.}~\bibnamefont
  {{Sab{\'{\i}}n}}}, \bibinfo {author} {\bibfnamefont {G.}~\bibnamefont
  {{Adesso}}}, \ and\ \bibinfo {author} {\bibfnamefont {I.}~\bibnamefont
  {{Fuentes}}},\ }\href {\doibase 10.1038/srep04996} {\bibfield  {journal}
  {\bibinfo  {journal} {Scientific Reports}\ }\textbf {\bibinfo {volume} {4}},\
  \bibinfo {eid} {4996} (\bibinfo {year} {2014})}\BibitemShut {NoStop}%
\bibitem [{\citenamefont {{Sab{\'{\i}}n}}\ \emph {et~al.}(2014)\citenamefont
  {{Sab{\'{\i}}n}}, \citenamefont {{Bruschi}}, \citenamefont {{Ahmadi}},\ and\
  \citenamefont {{Fuentes}}}]{GWDetectorFirst}%
  \BibitemOpen
  \bibfield  {author} {\bibinfo {author} {\bibfnamefont {C.}~\bibnamefont
  {{Sab{\'{\i}}n}}}, \bibinfo {author} {\bibfnamefont {D.~E.}\ \bibnamefont
  {{Bruschi}}}, \bibinfo {author} {\bibfnamefont {M.}~\bibnamefont {{Ahmadi}}},
  \ and\ \bibinfo {author} {\bibfnamefont {I.}~\bibnamefont {{Fuentes}}},\
  }\href {\doibase 10.1088/1367-2630/16/8/085003} {\bibfield  {journal}
  {\bibinfo  {journal} {New Journal of Physics}\ }\textbf {\bibinfo {volume}
  {16}},\ \bibinfo {eid} {085003} (\bibinfo {year} {2014})}\BibitemShut
  {NoStop}%
\bibitem [{\citenamefont {Sab{\'\i}n}\ \emph {et~al.}(2016)\citenamefont
  {Sab{\'\i}n}, \citenamefont {Kohlrus}, \citenamefont {Bruschi},\ and\
  \citenamefont {Fuentes}}]{GWDetectorThermal}%
  \BibitemOpen
  \bibfield  {author} {\bibinfo {author} {\bibfnamefont {C.}~\bibnamefont
  {Sab{\'\i}n}}, \bibinfo {author} {\bibfnamefont {J.}~\bibnamefont {Kohlrus}},
  \bibinfo {author} {\bibfnamefont {D.~E.}\ \bibnamefont {Bruschi}}, \ and\
  \bibinfo {author} {\bibfnamefont {I.}~\bibnamefont {Fuentes}},\ }\href
  {http://epjquantumtechnology.springeropen.com/articles/10.1140/epjqt/s40507-016-0046-4}
  {\bibfield  {journal} {\bibinfo  {journal} {EPJ Quantum Technology}\ }\textbf
  {\bibinfo {volume} {3}},\ \bibinfo {pages} {8} (\bibinfo {year}
  {2016})}\BibitemShut {NoStop}%
\bibitem [{\citenamefont {{Howl}}\ \emph {et~al.}(2016)\citenamefont {{Howl}},
  \citenamefont {{Hackermuller}}, \citenamefont {{Bruschi}},\ and\
  \citenamefont {{Fuentes}}}]{ReviewPaper}%
  \BibitemOpen
  \bibfield  {author} {\bibinfo {author} {\bibfnamefont {R.}~\bibnamefont
  {{Howl}}}, \bibinfo {author} {\bibfnamefont {L.}~\bibnamefont
  {{Hackermuller}}}, \bibinfo {author} {\bibfnamefont {D.~E.}\ \bibnamefont
  {{Bruschi}}}, \ and\ \bibinfo {author} {\bibfnamefont {I.}~\bibnamefont
  {{Fuentes}}},\ }\href {\doibase 10.1080/23746149.2017.1383184} {\bibfield
  {journal} {\bibinfo  {journal} {Advances in Physics: X. (In Press.)}\ }
  (\bibinfo {year} {2016}),\ 10.1080/23746149.2017.1383184},\ \Eprint
  {http://arxiv.org/abs/1607.06666} {arXiv:1607.06666 [quant-ph]} \BibitemShut
  {NoStop}%
\bibitem [{\citenamefont {Soykal}\ \emph {et~al.}(2011)\citenamefont {Soykal},
  \citenamefont {Ruskov},\ and\ \citenamefont
  {Tahan}}]{PhysRevLett.107.235502}%
  \BibitemOpen
  \bibfield  {author} {\bibinfo {author} {\bibfnamefont {O.~O.}\ \bibnamefont
  {Soykal}}, \bibinfo {author} {\bibfnamefont {R.}~\bibnamefont {Ruskov}}, \
  and\ \bibinfo {author} {\bibfnamefont {C.}~\bibnamefont {Tahan}},\ }\href
  {\doibase 10.1103/PhysRevLett.107.235502} {\bibfield  {journal} {\bibinfo
  {journal} {Phys. Rev. Lett.}\ }\textbf {\bibinfo {volume} {107}},\ \bibinfo
  {pages} {235502} (\bibinfo {year} {2011})}\BibitemShut {NoStop}%
\bibitem [{\citenamefont {Ruskov}\ and\ \citenamefont
  {Tahan}(2013)}]{PhysRevB.88.064308}%
  \BibitemOpen
  \bibfield  {author} {\bibinfo {author} {\bibfnamefont {R.}~\bibnamefont
  {Ruskov}}\ and\ \bibinfo {author} {\bibfnamefont {C.}~\bibnamefont {Tahan}},\
  }\href {\doibase 10.1103/PhysRevB.88.064308} {\bibfield  {journal} {\bibinfo
  {journal} {Phys. Rev. B}\ }\textbf {\bibinfo {volume} {88}},\ \bibinfo
  {pages} {064308} (\bibinfo {year} {2013})}\BibitemShut {NoStop}%
\bibitem [{\citenamefont {Habraken}\ \emph {et~al.}(2012)\citenamefont
  {Habraken}, \citenamefont {Stannigel}, \citenamefont {Lukin}, \citenamefont
  {Zoller},\ and\ \citenamefont {Rabl}}]{PhononQuantumNetworks}%
  \BibitemOpen
  \bibfield  {author} {\bibinfo {author} {\bibfnamefont {S.~J.~M.}\
  \bibnamefont {Habraken}}, \bibinfo {author} {\bibfnamefont {K.}~\bibnamefont
  {Stannigel}}, \bibinfo {author} {\bibfnamefont {M.~D.}\ \bibnamefont
  {Lukin}}, \bibinfo {author} {\bibfnamefont {P.}~\bibnamefont {Zoller}}, \
  and\ \bibinfo {author} {\bibfnamefont {P.}~\bibnamefont {Rabl}},\ }\href
  {http://stacks.iop.org/1367-2630/14/i=11/a=115004} {\bibfield  {journal}
  {\bibinfo  {journal} {New Journal of Physics}\ }\textbf {\bibinfo {volume}
  {14}},\ \bibinfo {pages} {115004} (\bibinfo {year} {2012})}\BibitemShut
  {NoStop}%
\bibitem [{\citenamefont {Schuetz}\ \emph {et~al.}(2015)\citenamefont
  {Schuetz}, \citenamefont {Kessler}, \citenamefont {Giedke}, \citenamefont
  {Vandersypen}, \citenamefont {Lukin},\ and\ \citenamefont
  {Cirac}}]{PhysRevX.5.031031}%
  \BibitemOpen
  \bibfield  {author} {\bibinfo {author} {\bibfnamefont {M.~J.~A.}\
  \bibnamefont {Schuetz}}, \bibinfo {author} {\bibfnamefont {E.~M.}\
  \bibnamefont {Kessler}}, \bibinfo {author} {\bibfnamefont {G.}~\bibnamefont
  {Giedke}}, \bibinfo {author} {\bibfnamefont {L.~M.~K.}\ \bibnamefont
  {Vandersypen}}, \bibinfo {author} {\bibfnamefont {M.~D.}\ \bibnamefont
  {Lukin}}, \ and\ \bibinfo {author} {\bibfnamefont {J.~I.}\ \bibnamefont
  {Cirac}},\ }\href {\doibase 10.1103/PhysRevX.5.031031} {\bibfield  {journal}
  {\bibinfo  {journal} {Phys. Rev. X}\ }\textbf {\bibinfo {volume} {5}},\
  \bibinfo {pages} {031031} (\bibinfo {year} {2015})}\BibitemShut {NoStop}%
\bibitem [{\citenamefont {Gustafsson}\ \emph {et~al.}(2014)\citenamefont
  {Gustafsson}, \citenamefont {Aref}, \citenamefont {Kockum}, \citenamefont
  {Ekstr{\"o}m}, \citenamefont {Johansson},\ and\ \citenamefont
  {Delsing}}]{Gustafsson207}%
  \BibitemOpen
  \bibfield  {author} {\bibinfo {author} {\bibfnamefont {M.~V.}\ \bibnamefont
  {Gustafsson}}, \bibinfo {author} {\bibfnamefont {T.}~\bibnamefont {Aref}},
  \bibinfo {author} {\bibfnamefont {A.~F.}\ \bibnamefont {Kockum}}, \bibinfo
  {author} {\bibfnamefont {M.~K.}\ \bibnamefont {Ekstr{\"o}m}}, \bibinfo
  {author} {\bibfnamefont {G.}~\bibnamefont {Johansson}}, \ and\ \bibinfo
  {author} {\bibfnamefont {P.}~\bibnamefont {Delsing}},\ }\href {\doibase
  10.1126/science.1257219} {\bibfield  {journal} {\bibinfo  {journal}
  {Science}\ }\textbf {\bibinfo {volume} {346}},\ \bibinfo {pages} {207}
  (\bibinfo {year} {2014})}\BibitemShut {NoStop}%
\bibitem [{\citenamefont {Stannigel}\ \emph {et~al.}(2012)\citenamefont
  {Stannigel}, \citenamefont {Komar}, \citenamefont {Habraken}, \citenamefont
  {Bennett}, \citenamefont {Lukin}, \citenamefont {Zoller},\ and\ \citenamefont
  {Rabl}}]{PhysRevLett.109.013603}%
  \BibitemOpen
  \bibfield  {author} {\bibinfo {author} {\bibfnamefont {K.}~\bibnamefont
  {Stannigel}}, \bibinfo {author} {\bibfnamefont {P.}~\bibnamefont {Komar}},
  \bibinfo {author} {\bibfnamefont {S.~J.~M.}\ \bibnamefont {Habraken}},
  \bibinfo {author} {\bibfnamefont {S.~D.}\ \bibnamefont {Bennett}}, \bibinfo
  {author} {\bibfnamefont {M.~D.}\ \bibnamefont {Lukin}}, \bibinfo {author}
  {\bibfnamefont {P.}~\bibnamefont {Zoller}}, \ and\ \bibinfo {author}
  {\bibfnamefont {P.}~\bibnamefont {Rabl}},\ }\href {\doibase
  10.1103/PhysRevLett.109.013603} {\bibfield  {journal} {\bibinfo  {journal}
  {Phys. Rev. Lett.}\ }\textbf {\bibinfo {volume} {109}},\ \bibinfo {pages}
  {013603} (\bibinfo {year} {2012})}\BibitemShut {NoStop}%
\bibitem [{\citenamefont {Sklan}(2015)}]{PhononicComputingReview}%
  \BibitemOpen
  \bibfield  {author} {\bibinfo {author} {\bibfnamefont {S.~R.}\ \bibnamefont
  {Sklan}},\ }\href
  {http://scitation.aip.org/content/aip/journal/adva/5/5/10.1063/1.4919584}
  {\bibfield  {journal} {\bibinfo  {journal} {AIP Advances}\ }\textbf {\bibinfo
  {volume} {5}},\ \bibinfo {eid} {053302} (\bibinfo {year} {2015})}\BibitemShut
  {NoStop}%
\bibitem [{\citenamefont {{Lahav}}\ \emph {et~al.}(2010)\citenamefont
  {{Lahav}}, \citenamefont {{Itah}}, \citenamefont {{Blumkin}}, \citenamefont
  {{Gordon}}, \citenamefont {{Rinott}}, \citenamefont {{Zayats}},\ and\
  \citenamefont {{Steinhauer}}}]{2010PhRvL.105x0401L}%
  \BibitemOpen
  \bibfield  {author} {\bibinfo {author} {\bibfnamefont {O.}~\bibnamefont
  {{Lahav}}}, \bibinfo {author} {\bibfnamefont {A.}~\bibnamefont {{Itah}}},
  \bibinfo {author} {\bibfnamefont {A.}~\bibnamefont {{Blumkin}}}, \bibinfo
  {author} {\bibfnamefont {C.}~\bibnamefont {{Gordon}}}, \bibinfo {author}
  {\bibfnamefont {S.}~\bibnamefont {{Rinott}}}, \bibinfo {author}
  {\bibfnamefont {A.}~\bibnamefont {{Zayats}}}, \ and\ \bibinfo {author}
  {\bibfnamefont {J.}~\bibnamefont {{Steinhauer}}},\ }\href {\doibase
  10.1103/PhysRevLett.105.240401} {\bibfield  {journal} {\bibinfo  {journal}
  {Physical Review Letters}\ }\textbf {\bibinfo {volume} {105}},\ \bibinfo
  {eid} {240401} (\bibinfo {year} {2010})}\BibitemShut {NoStop}%
\bibitem [{\citenamefont {Steinhauer}(2014)}]{Steinhauer2014}%
  \BibitemOpen
  \bibfield  {author} {\bibinfo {author} {\bibfnamefont {J.}~\bibnamefont
  {Steinhauer}},\ }\href {http://dx.doi.org/10.1038/nphys3104} {\bibfield
  {journal} {\bibinfo  {journal} {Nat Phys}\ }\textbf {\bibinfo {volume}
  {10}},\ \bibinfo {pages} {864} (\bibinfo {year} {2014})}\BibitemShut
  {NoStop}%
\bibitem [{\citenamefont {Steinhauer}(2016)}]{Steinhauer2016}%
  \BibitemOpen
  \bibfield  {author} {\bibinfo {author} {\bibfnamefont {J.}~\bibnamefont
  {Steinhauer}},\ }\href {http://dx.doi.org/10.1038/nphys3863} {\bibfield
  {journal} {\bibinfo  {journal} {Nat Phys}\ }\textbf {\bibinfo {volume}
  {12}},\ \bibinfo {pages} {959} (\bibinfo {year} {2016})}\BibitemShut
  {NoStop}%
\bibitem [{\citenamefont {Landau}(1946)}]{Landau:1946jc}%
  \BibitemOpen
  \bibfield  {author} {\bibinfo {author} {\bibfnamefont {L.~D.}\ \bibnamefont
  {Landau}},\ }\href@noop {} {\bibfield  {journal} {\bibinfo  {journal} {J.
  Phys.(USSR)}\ }\textbf {\bibinfo {volume} {10}},\ \bibinfo {pages} {25}
  (\bibinfo {year} {1946})},\ \bibinfo {note} {[Zh. Eksp. Teor.
  Fiz.16,574(1946)]}\BibitemShut {NoStop}%
%%CITATION = JOPYA,10,25;%%
\bibitem [{\citenamefont {Beliaev}(1958{\natexlab{a}})}]{Beliaev1958aSet2}%
  \BibitemOpen
  \bibfield  {author} {\bibinfo {author} {\bibfnamefont {S.~T.}\ \bibnamefont
  {Beliaev}},\ }\href@noop {} {\bibfield  {journal} {\bibinfo  {journal} {Sov.
  Phys. JETP}\ }\textbf {\bibinfo {volume} {34}},\ \bibinfo {pages} {323}
  (\bibinfo {year} {1958}{\natexlab{a}})}\BibitemShut {NoStop}%
\bibitem [{\citenamefont {Beliaev}(1958{\natexlab{b}})}]{Beliaev1958bSet2}%
  \BibitemOpen
  \bibfield  {author} {\bibinfo {author} {\bibfnamefont {S.}~\bibnamefont
  {Beliaev}},\ }\href@noop {} {\bibfield  {journal} {\bibinfo  {journal} {Sov.
  Phys. JETP}\ }\textbf {\bibinfo {volume} {7}},\ \bibinfo {pages} {299}
  (\bibinfo {year} {1958}{\natexlab{b}})}\BibitemShut {NoStop}%
\bibitem [{\citenamefont {Hohenberg}\ and\ \citenamefont
  {Martin}(1965)}]{Hohenberg1965291}%
  \BibitemOpen
  \bibfield  {author} {\bibinfo {author} {\bibfnamefont {P.}~\bibnamefont
  {Hohenberg}}\ and\ \bibinfo {author} {\bibfnamefont {P.}~\bibnamefont
  {Martin}},\ }\href {\doibase 10.1016/0003-4916(65)90280-0} {\bibfield
  {journal} {\bibinfo  {journal} {Annals of Physics}\ }\textbf {\bibinfo
  {volume} {34}},\ \bibinfo {pages} {291 } (\bibinfo {year}
  {1965})}\BibitemShut {NoStop}%
\bibitem [{\citenamefont {Lifshitz}\ and\ \citenamefont
  {Pitaevskii}(1981)}]{PhysicalKinetics}%
  \BibitemOpen
  \bibfield  {author} {\bibinfo {author} {\bibfnamefont {E.}~\bibnamefont
  {Lifshitz}}\ and\ \bibinfo {author} {\bibfnamefont {L.}~\bibnamefont
  {Pitaevskii}},\ }\href {\doibase
  https://doi.org/10.1016/B978-0-08-057049-5.50002-3} {\emph {\bibinfo {title}
  {Physical Kinetics}}}\ (\bibinfo  {publisher} {Pergamon Press, Oxford},\
  \bibinfo {year} {1981})\BibitemShut {NoStop}%
\bibitem [{\citenamefont {Kondor}\ and\ \citenamefont
  {Szépfalusy}(1974)}]{KONDOR1974393}%
  \BibitemOpen
  \bibfield  {author} {\bibinfo {author} {\bibfnamefont {I.}~\bibnamefont
  {Kondor}}\ and\ \bibinfo {author} {\bibfnamefont {P.}~\bibnamefont
  {Szépfalusy}},\ }\href {\doibase 10.1016/0375-9601(74)90143-1} {\bibfield
  {journal} {\bibinfo  {journal} {Physics Letters A}\ }\textbf {\bibinfo
  {volume} {47}},\ \bibinfo {pages} {393 } (\bibinfo {year}
  {1974})}\BibitemShut {NoStop}%
\bibitem [{\citenamefont {Liu}\ and\ \citenamefont
  {Shieve}(1997)}]{LiuAndShieve}%
  \BibitemOpen
  \bibfield  {author} {\bibinfo {author} {\bibfnamefont {W.}~\bibnamefont
  {Liu}}\ and\ \bibinfo {author} {\bibfnamefont {W.}~\bibnamefont {Shieve}},\
  }\href {http://arxiv.org/abs/cond-mat/9702122v1} {\bibfield  {journal}
  {\bibinfo  {journal} {cond-mat/9702122}\ } (\bibinfo {year}
  {1997})}\BibitemShut {NoStop}%
\bibitem [{\citenamefont {Pitaevskii}\ and\ \citenamefont
  {Stringari}(1997)}]{PitaevskiiDamping}%
  \BibitemOpen
  \bibfield  {author} {\bibinfo {author} {\bibfnamefont {L.}~\bibnamefont
  {Pitaevskii}}\ and\ \bibinfo {author} {\bibfnamefont {S.}~\bibnamefont
  {Stringari}},\ }\href {\doibase 10.1016/S0375-9601(97)00666-X} {\bibfield
  {journal} {\bibinfo  {journal} {Physics Letters A}\ }\textbf {\bibinfo
  {volume} {235}},\ \bibinfo {pages} {398 } (\bibinfo {year}
  {1997})}\BibitemShut {NoStop}%
\bibitem [{\citenamefont {Vincent~Liu}(1997)}]{Liu}%
  \BibitemOpen
  \bibfield  {author} {\bibinfo {author} {\bibfnamefont {W.}~\bibnamefont
  {Vincent~Liu}},\ }\href {\doibase 10.1103/PhysRevLett.79.4056} {\bibfield
  {journal} {\bibinfo  {journal} {Phys. Rev. Lett.}\ }\textbf {\bibinfo
  {volume} {79}},\ \bibinfo {pages} {4056} (\bibinfo {year}
  {1997})}\BibitemShut {NoStop}%
\bibitem [{\citenamefont {Giorgini}(1998)}]{Giorgini1998}%
  \BibitemOpen
  \bibfield  {author} {\bibinfo {author} {\bibfnamefont {S.}~\bibnamefont
  {Giorgini}},\ }\href {\doibase 10.1103/PhysRevA.57.2949} {\bibfield
  {journal} {\bibinfo  {journal} {Phys. Rev. A}\ }\textbf {\bibinfo {volume}
  {57}},\ \bibinfo {pages} {2949} (\bibinfo {year} {1998})}\BibitemShut
  {NoStop}%
\bibitem [{\citenamefont {Fedichev}\ \emph {et~al.}(1998)\citenamefont
  {Fedichev}, \citenamefont {Shlyapnikov},\ and\ \citenamefont
  {Walraven}}]{DampingTrapped}%
  \BibitemOpen
  \bibfield  {author} {\bibinfo {author} {\bibfnamefont {P.~O.}\ \bibnamefont
  {Fedichev}}, \bibinfo {author} {\bibfnamefont {G.~V.}\ \bibnamefont
  {Shlyapnikov}}, \ and\ \bibinfo {author} {\bibfnamefont {J.~T.~M.}\
  \bibnamefont {Walraven}},\ }\href {\doibase 10.1103/PhysRevLett.80.2269}
  {\bibfield  {journal} {\bibinfo  {journal} {Phys. Rev. Lett.}\ }\textbf
  {\bibinfo {volume} {80}},\ \bibinfo {pages} {2269} (\bibinfo {year}
  {1998})}\BibitemShut {NoStop}%
\bibitem [{\citenamefont {Fedichev}\ and\ \citenamefont
  {Shlyapnikov}(1998)}]{PhysRevA.58.3146}%
  \BibitemOpen
  \bibfield  {author} {\bibinfo {author} {\bibfnamefont {P.~O.}\ \bibnamefont
  {Fedichev}}\ and\ \bibinfo {author} {\bibfnamefont {G.~V.}\ \bibnamefont
  {Shlyapnikov}},\ }\href {\doibase 10.1103/PhysRevA.58.3146} {\bibfield
  {journal} {\bibinfo  {journal} {Phys. Rev. A}\ }\textbf {\bibinfo {volume}
  {58}},\ \bibinfo {pages} {3146} (\bibinfo {year} {1998})}\BibitemShut
  {NoStop}%
\bibitem [{\citenamefont {Bene}\ and\ \citenamefont
  {Sz\'epfalusy}(1998)}]{PhysRevA.58.R3391}%
  \BibitemOpen
  \bibfield  {author} {\bibinfo {author} {\bibfnamefont {G.}~\bibnamefont
  {Bene}}\ and\ \bibinfo {author} {\bibfnamefont {P.}~\bibnamefont
  {Sz\'epfalusy}},\ }\href {\doibase 10.1103/PhysRevA.58.R3391} {\bibfield
  {journal} {\bibinfo  {journal} {Phys. Rev. A}\ }\textbf {\bibinfo {volume}
  {58}},\ \bibinfo {pages} {R3391} (\bibinfo {year} {1998})}\BibitemShut
  {NoStop}%
\bibitem [{\citenamefont {Graham}(2000{\natexlab{a}})}]{Langevin}%
  \BibitemOpen
  \bibfield  {author} {\bibinfo {author} {\bibfnamefont {R.}~\bibnamefont
  {Graham}},\ }\href {\doibase 10.1023/A:1026453803436} {\bibfield  {journal}
  {\bibinfo  {journal} {Journal of Statistical Physics}\ }\textbf {\bibinfo
  {volume} {101}},\ \bibinfo {pages} {243} (\bibinfo {year}
  {2000}{\natexlab{a}})}\BibitemShut {NoStop}%
\bibitem [{\citenamefont {Rusch}\ and\ \citenamefont
  {Burnett}(1999)}]{PhysRevA.59.3851}%
  \BibitemOpen
  \bibfield  {author} {\bibinfo {author} {\bibfnamefont {M.}~\bibnamefont
  {Rusch}}\ and\ \bibinfo {author} {\bibfnamefont {K.}~\bibnamefont
  {Burnett}},\ }\href {\doibase 10.1103/PhysRevA.59.3851} {\bibfield  {journal}
  {\bibinfo  {journal} {Phys. Rev. A}\ }\textbf {\bibinfo {volume} {59}},\
  \bibinfo {pages} {3851} (\bibinfo {year} {1999})}\BibitemShut {NoStop}%
\bibitem [{\citenamefont {Jackson}\ and\ \citenamefont
  {Zaremba}(2003)}]{LBReview}%
  \BibitemOpen
  \bibfield  {author} {\bibinfo {author} {\bibfnamefont {B.}~\bibnamefont
  {Jackson}}\ and\ \bibinfo {author} {\bibfnamefont {E.}~\bibnamefont
  {Zaremba}},\ }\href {http://stacks.iop.org/1367-2630/5/i=1/a=388} {\bibfield
  {journal} {\bibinfo  {journal} {New Journal of Physics}\ }\textbf {\bibinfo
  {volume} {5}},\ \bibinfo {pages} {88} (\bibinfo {year} {2003})}\BibitemShut
  {NoStop}%
\bibitem [{\citenamefont {Sinatra}\ \emph {et~al.}(2007)\citenamefont
  {Sinatra}, \citenamefont {Castin},\ and\ \citenamefont
  {Witkowska}}]{CastinAppendix}%
  \BibitemOpen
  \bibfield  {author} {\bibinfo {author} {\bibfnamefont {A.}~\bibnamefont
  {Sinatra}}, \bibinfo {author} {\bibfnamefont {Y.}~\bibnamefont {Castin}}, \
  and\ \bibinfo {author} {\bibfnamefont {E.}~\bibnamefont {Witkowska}},\ }\href
  {\doibase 10.1103/PhysRevA.75.033616} {\bibfield  {journal} {\bibinfo
  {journal} {Phys. Rev. A}\ }\textbf {\bibinfo {volume} {75}},\ \bibinfo
  {pages} {033616} (\bibinfo {year} {2007})}\BibitemShut {NoStop}%
\bibitem [{\citenamefont {Graham}(1998)}]{PhysRevLett.81.5262}%
  \BibitemOpen
  \bibfield  {author} {\bibinfo {author} {\bibfnamefont {R.}~\bibnamefont
  {Graham}},\ }\href {\doibase 10.1103/PhysRevLett.81.5262} {\bibfield
  {journal} {\bibinfo  {journal} {Phys. Rev. Lett.}\ }\textbf {\bibinfo
  {volume} {81}},\ \bibinfo {pages} {5262} (\bibinfo {year}
  {1998})}\BibitemShut {NoStop}%
\bibitem [{\citenamefont {Graham}(2000{\natexlab{b}})}]{PhysRevA.62.023609}%
  \BibitemOpen
  \bibfield  {author} {\bibinfo {author} {\bibfnamefont {R.}~\bibnamefont
  {Graham}},\ }\href {\doibase 10.1103/PhysRevA.62.023609} {\bibfield
  {journal} {\bibinfo  {journal} {Phys. Rev. A}\ }\textbf {\bibinfo {volume}
  {62}},\ \bibinfo {pages} {023609} (\bibinfo {year}
  {2000}{\natexlab{b}})}\BibitemShut {NoStop}%
\bibitem [{\citenamefont {Castin}\ and\ \citenamefont
  {Sinatra}(2013)}]{2012arXiv1203.4458C}%
  \BibitemOpen
  \bibfield  {author} {\bibinfo {author} {\bibfnamefont {Y.}~\bibnamefont
  {Castin}}\ and\ \bibinfo {author} {\bibfnamefont {A.}~\bibnamefont
  {Sinatra}},\ }in\ \href {\doibase
  https://doi.org/10.1007/978-3-642-37569-9_15} {\emph {\bibinfo {booktitle}
  {Physics of Quantum Fluids}}}\ (\bibinfo  {publisher} {Springer},\ \bibinfo
  {year} {2013})\ pp.\ \bibinfo {pages} {315--339}\BibitemShut {NoStop}%
\bibitem [{\citenamefont {Katz}\ \emph {et~al.}(2004)\citenamefont {Katz},
  \citenamefont {Ozeri}, \citenamefont {Rowen}, \citenamefont {Gershnabel},\
  and\ \citenamefont {Davidson}}]{PhysRevA.70.033615}%
  \BibitemOpen
  \bibfield  {author} {\bibinfo {author} {\bibfnamefont {N.}~\bibnamefont
  {Katz}}, \bibinfo {author} {\bibfnamefont {R.}~\bibnamefont {Ozeri}},
  \bibinfo {author} {\bibfnamefont {E.}~\bibnamefont {Rowen}}, \bibinfo
  {author} {\bibfnamefont {E.}~\bibnamefont {Gershnabel}}, \ and\ \bibinfo
  {author} {\bibfnamefont {N.}~\bibnamefont {Davidson}},\ }\href {\doibase
  10.1103/PhysRevA.70.033615} {\bibfield  {journal} {\bibinfo  {journal} {Phys.
  Rev. A}\ }\textbf {\bibinfo {volume} {70}},\ \bibinfo {pages} {033615}
  (\bibinfo {year} {2004})}\BibitemShut {NoStop}%
\bibitem [{\citenamefont {{Rowen}}(2008)}]{2008PhDT........70E}%
  \BibitemOpen
  \bibfield  {author} {\bibinfo {author} {\bibfnamefont {E.~E.}\ \bibnamefont
  {{Rowen}}},\ }\emph {\bibinfo {title} {Coherence and decoherence of
  excitations in a trapped Bose-Einstein condensate}},\ \href@noop {} {Ph.D.
  thesis},\ \bibinfo  {school} {The Weizmann Institute of Science (Israel)}
  (\bibinfo {year} {2008})\BibitemShut {NoStop}%
\bibitem [{\citenamefont {Bar-Gill}\ \emph {et~al.}(2009)\citenamefont
  {Bar-Gill}, \citenamefont {Rowen}, \citenamefont {Kurizki},\ and\
  \citenamefont {Davidson}}]{PhysRevLett.102.110401}%
  \BibitemOpen
  \bibfield  {author} {\bibinfo {author} {\bibfnamefont {N.}~\bibnamefont
  {Bar-Gill}}, \bibinfo {author} {\bibfnamefont {E.~E.}\ \bibnamefont {Rowen}},
  \bibinfo {author} {\bibfnamefont {G.}~\bibnamefont {Kurizki}}, \ and\
  \bibinfo {author} {\bibfnamefont {N.}~\bibnamefont {Davidson}},\ }\href
  {\doibase 10.1103/PhysRevLett.102.110401} {\bibfield  {journal} {\bibinfo
  {journal} {Phys. Rev. Lett.}\ }\textbf {\bibinfo {volume} {102}},\ \bibinfo
  {pages} {110401} (\bibinfo {year} {2009})}\BibitemShut {NoStop}%
\bibitem [{\citenamefont {Gri\ifmmode~\check{s}\else \v{s}\fi{}ins}\ \emph
  {et~al.}(2016)\citenamefont {Gri\ifmmode~\check{s}\else \v{s}\fi{}ins},
  \citenamefont {Rauer}, \citenamefont {Langen}, \citenamefont {Schmiedmayer},\
  and\ \citenamefont {Mazets}}]{PhysRevA.93.033634}%
  \BibitemOpen
  \bibfield  {author} {\bibinfo {author} {\bibfnamefont {P.}~\bibnamefont
  {Gri\ifmmode~\check{s}\else \v{s}\fi{}ins}}, \bibinfo {author} {\bibfnamefont
  {B.}~\bibnamefont {Rauer}}, \bibinfo {author} {\bibfnamefont
  {T.}~\bibnamefont {Langen}}, \bibinfo {author} {\bibfnamefont
  {J.}~\bibnamefont {Schmiedmayer}}, \ and\ \bibinfo {author} {\bibfnamefont
  {I.~E.}\ \bibnamefont {Mazets}},\ }\href {\doibase
  10.1103/PhysRevA.93.033634} {\bibfield  {journal} {\bibinfo  {journal} {Phys.
  Rev. A}\ }\textbf {\bibinfo {volume} {93}},\ \bibinfo {pages} {033634}
  (\bibinfo {year} {2016})}\BibitemShut {NoStop}%
\bibitem [{\citenamefont {Pitaevskii}\ and\ \citenamefont
  {Stringari}(2003)}]{PitaevskiiBook}%
  \BibitemOpen
  \bibfield  {author} {\bibinfo {author} {\bibfnamefont {L.}~\bibnamefont
  {Pitaevskii}}\ and\ \bibinfo {author} {\bibfnamefont {S.}~\bibnamefont
  {Stringari}},\ }\href@noop {} {\emph {\bibinfo {title} {Bose-Einstein
  Condensation}}}\ (\bibinfo  {publisher} {Oxford University Press},\ \bibinfo
  {year} {2003})\BibitemShut {NoStop}%
\bibitem [{\citenamefont {Gaunt}\ \emph {et~al.}(2013)\citenamefont {Gaunt},
  \citenamefont {Schmidutz}, \citenamefont {Gotlibovych}, \citenamefont
  {Smith},\ and\ \citenamefont {Hadzibabic}}]{PhysRevLett.110.200406}%
  \BibitemOpen
  \bibfield  {author} {\bibinfo {author} {\bibfnamefont {A.~L.}\ \bibnamefont
  {Gaunt}}, \bibinfo {author} {\bibfnamefont {T.~F.}\ \bibnamefont
  {Schmidutz}}, \bibinfo {author} {\bibfnamefont {I.}~\bibnamefont
  {Gotlibovych}}, \bibinfo {author} {\bibfnamefont {R.~P.}\ \bibnamefont
  {Smith}}, \ and\ \bibinfo {author} {\bibfnamefont {Z.}~\bibnamefont
  {Hadzibabic}},\ }\href {\doibase 10.1103/PhysRevLett.110.200406} {\bibfield
  {journal} {\bibinfo  {journal} {Phys. Rev. Lett.}\ }\textbf {\bibinfo
  {volume} {110}},\ \bibinfo {pages} {200406} (\bibinfo {year}
  {2013})}\BibitemShut {NoStop}%
\bibitem [{\citenamefont {Dalfovo}\ \emph {et~al.}(1999)\citenamefont
  {Dalfovo}, \citenamefont {Giorgini}, \citenamefont {Pitaevskii},\ and\
  \citenamefont {Stringari}}]{RevModPhys.71.463}%
  \BibitemOpen
  \bibfield  {author} {\bibinfo {author} {\bibfnamefont {F.}~\bibnamefont
  {Dalfovo}}, \bibinfo {author} {\bibfnamefont {S.}~\bibnamefont {Giorgini}},
  \bibinfo {author} {\bibfnamefont {L.~P.}\ \bibnamefont {Pitaevskii}}, \ and\
  \bibinfo {author} {\bibfnamefont {S.}~\bibnamefont {Stringari}},\ }\href
  {\doibase 10.1103/RevModPhys.71.463} {\bibfield  {journal} {\bibinfo
  {journal} {Rev. Mod. Phys.}\ }\textbf {\bibinfo {volume} {71}},\ \bibinfo
  {pages} {463} (\bibinfo {year} {1999})}\BibitemShut {NoStop}%
\bibitem [{\citenamefont {Smith}\ \emph {et~al.}(2011)\citenamefont {Smith},
  \citenamefont {Tammuz}, \citenamefont {Campbell}, \citenamefont {Holzmann},\
  and\ \citenamefont {Hadzibabic}}]{PhysRevLett.107.190403}%
  \BibitemOpen
  \bibfield  {author} {\bibinfo {author} {\bibfnamefont {R.~P.}\ \bibnamefont
  {Smith}}, \bibinfo {author} {\bibfnamefont {N.}~\bibnamefont {Tammuz}},
  \bibinfo {author} {\bibfnamefont {R.~L.~D.}\ \bibnamefont {Campbell}},
  \bibinfo {author} {\bibfnamefont {M.}~\bibnamefont {Holzmann}}, \ and\
  \bibinfo {author} {\bibfnamefont {Z.}~\bibnamefont {Hadzibabic}},\ }\href
  {\doibase 10.1103/PhysRevLett.107.190403} {\bibfield  {journal} {\bibinfo
  {journal} {Phys. Rev. Lett.}\ }\textbf {\bibinfo {volume} {107}},\ \bibinfo
  {pages} {190403} (\bibinfo {year} {2011})}\BibitemShut {NoStop}%
\bibitem [{\citenamefont {Drake}\ \emph {et~al.}(2012)\citenamefont {Drake},
  \citenamefont {Sagi}, \citenamefont {Paudel}, \citenamefont {Stewart},
  \citenamefont {Gaebler},\ and\ \citenamefont {Jin}}]{PhysRevA.86.031601}%
  \BibitemOpen
  \bibfield  {author} {\bibinfo {author} {\bibfnamefont {T.~E.}\ \bibnamefont
  {Drake}}, \bibinfo {author} {\bibfnamefont {Y.}~\bibnamefont {Sagi}},
  \bibinfo {author} {\bibfnamefont {R.}~\bibnamefont {Paudel}}, \bibinfo
  {author} {\bibfnamefont {J.~T.}\ \bibnamefont {Stewart}}, \bibinfo {author}
  {\bibfnamefont {J.~P.}\ \bibnamefont {Gaebler}}, \ and\ \bibinfo {author}
  {\bibfnamefont {D.~S.}\ \bibnamefont {Jin}},\ }\href {\doibase
  10.1103/PhysRevA.86.031601} {\bibfield  {journal} {\bibinfo  {journal} {Phys.
  Rev. A}\ }\textbf {\bibinfo {volume} {86}},\ \bibinfo {pages} {031601}
  (\bibinfo {year} {2012})}\BibitemShut {NoStop}%
\bibitem [{\citenamefont {Sagi}\ \emph {et~al.}(2012)\citenamefont {Sagi},
  \citenamefont {Drake}, \citenamefont {Paudel},\ and\ \citenamefont
  {Jin}}]{PhysRevLett.109.220402}%
  \BibitemOpen
  \bibfield  {author} {\bibinfo {author} {\bibfnamefont {Y.}~\bibnamefont
  {Sagi}}, \bibinfo {author} {\bibfnamefont {T.~E.}\ \bibnamefont {Drake}},
  \bibinfo {author} {\bibfnamefont {R.}~\bibnamefont {Paudel}}, \ and\ \bibinfo
  {author} {\bibfnamefont {D.~S.}\ \bibnamefont {Jin}},\ }\href {\doibase
  10.1103/PhysRevLett.109.220402} {\bibfield  {journal} {\bibinfo  {journal}
  {Phys. Rev. Lett.}\ }\textbf {\bibinfo {volume} {109}},\ \bibinfo {pages}
  {220402} (\bibinfo {year} {2012})}\BibitemShut {NoStop}%
\bibitem [{\citenamefont {Bogoliubov}(1947)}]{Bogo1947}%
  \BibitemOpen
  \bibfield  {author} {\bibinfo {author} {\bibfnamefont {N.~N.}\ \bibnamefont
  {Bogoliubov}},\ }\href@noop {} {\bibfield  {journal} {\bibinfo  {journal} {J.
  Phys. (Moscow)}\ }\textbf {\bibinfo {volume} {11}},\ \bibinfo {pages} {23}
  (\bibinfo {year} {1947})}\BibitemShut {NoStop}%
\bibitem [{\citenamefont {Landau}\ and\ \citenamefont
  {Lifshitz}(1987)}]{LandauLifshitzQM}%
  \BibitemOpen
  \bibfield  {author} {\bibinfo {author} {\bibfnamefont {L.~D.}\ \bibnamefont
  {Landau}}\ and\ \bibinfo {author} {\bibfnamefont {E.~M.}\ \bibnamefont
  {Lifshitz}},\ }\href@noop {} {\emph {\bibinfo {title} {Quantum Mechanics}}},\
  \bibinfo {edition} {third edition}\ ed.\ (\bibinfo  {publisher} {Pregamon,
  Oxford},\ \bibinfo {year} {1987})\BibitemShut {NoStop}%
\bibitem [{\citenamefont {Burnham}\ and\ \citenamefont
  {Weinberg}(1970)}]{SPDC}%
  \BibitemOpen
  \bibfield  {author} {\bibinfo {author} {\bibfnamefont {D.~C.}\ \bibnamefont
  {Burnham}}\ and\ \bibinfo {author} {\bibfnamefont {D.~L.}\ \bibnamefont
  {Weinberg}},\ }\href {\doibase 10.1103/PhysRevLett.25.84} {\bibfield
  {journal} {\bibinfo  {journal} {Phys. Rev. Lett.}\ }\textbf {\bibinfo
  {volume} {25}},\ \bibinfo {pages} {84} (\bibinfo {year} {1970})}\BibitemShut
  {NoStop}%
\bibitem [{\citenamefont {Stamper-Kurn}\ \emph {et~al.}(1999)\citenamefont
  {Stamper-Kurn}, \citenamefont {Chikkatur}, \citenamefont {G\"orlitz},
  \citenamefont {Inouye}, \citenamefont {Gupta}, \citenamefont {Pritchard},\
  and\ \citenamefont {Ketterle}}]{PhysRevLett.83.2876}%
  \BibitemOpen
  \bibfield  {author} {\bibinfo {author} {\bibfnamefont {D.~M.}\ \bibnamefont
  {Stamper-Kurn}}, \bibinfo {author} {\bibfnamefont {A.~P.}\ \bibnamefont
  {Chikkatur}}, \bibinfo {author} {\bibfnamefont {A.}~\bibnamefont
  {G\"orlitz}}, \bibinfo {author} {\bibfnamefont {S.}~\bibnamefont {Inouye}},
  \bibinfo {author} {\bibfnamefont {S.}~\bibnamefont {Gupta}}, \bibinfo
  {author} {\bibfnamefont {D.~E.}\ \bibnamefont {Pritchard}}, \ and\ \bibinfo
  {author} {\bibfnamefont {W.}~\bibnamefont {Ketterle}},\ }\href {\doibase
  10.1103/PhysRevLett.83.2876} {\bibfield  {journal} {\bibinfo  {journal}
  {Phys. Rev. Lett.}\ }\textbf {\bibinfo {volume} {83}},\ \bibinfo {pages}
  {2876} (\bibinfo {year} {1999})}\BibitemShut {NoStop}%
\bibitem [{\citenamefont {Vogels}\ \emph {et~al.}(2002)\citenamefont {Vogels},
  \citenamefont {Xu}, \citenamefont {Raman}, \citenamefont {Abo-Shaeer},\ and\
  \citenamefont {Ketterle}}]{PhysRevLett.88.060402}%
  \BibitemOpen
  \bibfield  {author} {\bibinfo {author} {\bibfnamefont {J.~M.}\ \bibnamefont
  {Vogels}}, \bibinfo {author} {\bibfnamefont {K.}~\bibnamefont {Xu}}, \bibinfo
  {author} {\bibfnamefont {C.}~\bibnamefont {Raman}}, \bibinfo {author}
  {\bibfnamefont {J.~R.}\ \bibnamefont {Abo-Shaeer}}, \ and\ \bibinfo {author}
  {\bibfnamefont {W.}~\bibnamefont {Ketterle}},\ }\href {\doibase
  10.1103/PhysRevLett.88.060402} {\bibfield  {journal} {\bibinfo  {journal}
  {Phys. Rev. Lett.}\ }\textbf {\bibinfo {volume} {88}},\ \bibinfo {pages}
  {060402} (\bibinfo {year} {2002})}\BibitemShut {NoStop}%
\bibitem [{\citenamefont {Steinhauer}\ \emph {et~al.}(2002)\citenamefont
  {Steinhauer}, \citenamefont {Ozeri}, \citenamefont {Katz},\ and\
  \citenamefont {Davidson}}]{PhysRevLett.88.120407}%
  \BibitemOpen
  \bibfield  {author} {\bibinfo {author} {\bibfnamefont {J.}~\bibnamefont
  {Steinhauer}}, \bibinfo {author} {\bibfnamefont {R.}~\bibnamefont {Ozeri}},
  \bibinfo {author} {\bibfnamefont {N.}~\bibnamefont {Katz}}, \ and\ \bibinfo
  {author} {\bibfnamefont {N.}~\bibnamefont {Davidson}},\ }\href {\doibase
  10.1103/PhysRevLett.88.120407} {\bibfield  {journal} {\bibinfo  {journal}
  {Phys. Rev. Lett.}\ }\textbf {\bibinfo {volume} {88}},\ \bibinfo {pages}
  {120407} (\bibinfo {year} {2002})}\BibitemShut {NoStop}%
\bibitem [{\citenamefont {Andrews}\ \emph {et~al.}(1997)\citenamefont
  {Andrews}, \citenamefont {Kurn}, \citenamefont {Miesner}, \citenamefont
  {Durfee}, \citenamefont {Townsend}, \citenamefont {Inouye},\ and\
  \citenamefont {Ketterle}}]{FirstSoundWaves1}%
  \BibitemOpen
  \bibfield  {author} {\bibinfo {author} {\bibfnamefont {M.~R.}\ \bibnamefont
  {Andrews}}, \bibinfo {author} {\bibfnamefont {D.~M.}\ \bibnamefont {Kurn}},
  \bibinfo {author} {\bibfnamefont {H.-J.}\ \bibnamefont {Miesner}}, \bibinfo
  {author} {\bibfnamefont {D.~S.}\ \bibnamefont {Durfee}}, \bibinfo {author}
  {\bibfnamefont {C.~G.}\ \bibnamefont {Townsend}}, \bibinfo {author}
  {\bibfnamefont {S.}~\bibnamefont {Inouye}}, \ and\ \bibinfo {author}
  {\bibfnamefont {W.}~\bibnamefont {Ketterle}},\ }\href {\doibase
  10.1103/PhysRevLett.79.553} {\bibfield  {journal} {\bibinfo  {journal} {Phys.
  Rev. Lett.}\ }\textbf {\bibinfo {volume} {79}},\ \bibinfo {pages} {553}
  (\bibinfo {year} {1997})}\BibitemShut {NoStop}%
\bibitem [{\citenamefont {Andrews}\ \emph {et~al.}(1998)\citenamefont
  {Andrews}, \citenamefont {Stamper-Kurn}, \citenamefont {Miesner},
  \citenamefont {Durfee}, \citenamefont {Townsend}, \citenamefont {Inouye},\
  and\ \citenamefont {Ketterle}}]{FirstSoundWaves2}%
  \BibitemOpen
  \bibfield  {author} {\bibinfo {author} {\bibfnamefont {M.~R.}\ \bibnamefont
  {Andrews}}, \bibinfo {author} {\bibfnamefont {D.~M.}\ \bibnamefont
  {Stamper-Kurn}}, \bibinfo {author} {\bibfnamefont {H.-J.}\ \bibnamefont
  {Miesner}}, \bibinfo {author} {\bibfnamefont {D.~S.}\ \bibnamefont {Durfee}},
  \bibinfo {author} {\bibfnamefont {C.~G.}\ \bibnamefont {Townsend}}, \bibinfo
  {author} {\bibfnamefont {S.}~\bibnamefont {Inouye}}, \ and\ \bibinfo {author}
  {\bibfnamefont {W.}~\bibnamefont {Ketterle}},\ }\href {\doibase
  10.1103/PhysRevLett.80.2967} {\bibfield  {journal} {\bibinfo  {journal}
  {Phys. Rev. Lett.}\ }\textbf {\bibinfo {volume} {80}},\ \bibinfo {pages}
  {2967} (\bibinfo {year} {1998})}\BibitemShut {NoStop}%
\bibitem [{\citenamefont {Carusotto}\ \emph {et~al.}(2010)\citenamefont
  {Carusotto}, \citenamefont {Balbinot}, \citenamefont {Fabbri},\ and\
  \citenamefont {Recati}}]{DCETheory}%
  \BibitemOpen
  \bibfield  {author} {\bibinfo {author} {\bibfnamefont {I.}~\bibnamefont
  {Carusotto}}, \bibinfo {author} {\bibfnamefont {R.}~\bibnamefont {Balbinot}},
  \bibinfo {author} {\bibfnamefont {A.}~\bibnamefont {Fabbri}}, \ and\ \bibinfo
  {author} {\bibfnamefont {A.}~\bibnamefont {Recati}},\ }\href {\doibase
  10.1140/epjd/e2009-00314-3} {\bibfield  {journal} {\bibinfo  {journal} {The
  European Physical Journal D}\ }\textbf {\bibinfo {volume} {56}},\ \bibinfo
  {pages} {391} (\bibinfo {year} {2010})}\BibitemShut {NoStop}%
\bibitem [{\citenamefont {Jaskula}\ \emph {et~al.}(2012)\citenamefont
  {Jaskula}, \citenamefont {Partridge}, \citenamefont {Bonneau}, \citenamefont
  {Lopes}, \citenamefont {Ruaudel}, \citenamefont {Boiron},\ and\ \citenamefont
  {Westbrook}}]{DCEWestbrook}%
  \BibitemOpen
  \bibfield  {author} {\bibinfo {author} {\bibfnamefont {J.-C.}\ \bibnamefont
  {Jaskula}}, \bibinfo {author} {\bibfnamefont {G.~B.}\ \bibnamefont
  {Partridge}}, \bibinfo {author} {\bibfnamefont {M.}~\bibnamefont {Bonneau}},
  \bibinfo {author} {\bibfnamefont {R.}~\bibnamefont {Lopes}}, \bibinfo
  {author} {\bibfnamefont {J.}~\bibnamefont {Ruaudel}}, \bibinfo {author}
  {\bibfnamefont {D.}~\bibnamefont {Boiron}}, \ and\ \bibinfo {author}
  {\bibfnamefont {C.~I.}\ \bibnamefont {Westbrook}},\ }\href {\doibase
  10.1103/PhysRevLett.109.220401} {\bibfield  {journal} {\bibinfo  {journal}
  {Phys. Rev. Lett.}\ }\textbf {\bibinfo {volume} {109}},\ \bibinfo {pages}
  {220401} (\bibinfo {year} {2012})}\BibitemShut {NoStop}%
\bibitem [{\citenamefont {Isar}\ \emph {et~al.}(1994)\citenamefont {Isar},
  \citenamefont {Sandulescu}, \citenamefont {Scutaru}, \citenamefont
  {Stefanescu},\ and\ \citenamefont {Sceid}}]{Isar1994}%
  \BibitemOpen
  \bibfield  {author} {\bibinfo {author} {\bibfnamefont {A.}~\bibnamefont
  {Isar}}, \bibinfo {author} {\bibfnamefont {A.}~\bibnamefont {Sandulescu}},
  \bibinfo {author} {\bibfnamefont {H.}~\bibnamefont {Scutaru}}, \bibinfo
  {author} {\bibfnamefont {E.}~\bibnamefont {Stefanescu}}, \ and\ \bibinfo
  {author} {\bibfnamefont {W.}~\bibnamefont {Sceid}},\ }\href {\doibase
  10.1142/S0218301394000164} {\bibfield  {journal} {\bibinfo  {journal}
  {International Journal of Modern Physics E}\ }\textbf {\bibinfo {volume}
  {03}},\ \bibinfo {pages} {635} (\bibinfo {year} {1994})}\BibitemShut
  {NoStop}%
\bibitem [{\citenamefont {Sandulescu}\ and\ \citenamefont
  {Scutaru}(1987)}]{Sandulescu1987}%
  \BibitemOpen
  \bibfield  {author} {\bibinfo {author} {\bibfnamefont {A.}~\bibnamefont
  {Sandulescu}}\ and\ \bibinfo {author} {\bibfnamefont {H.}~\bibnamefont
  {Scutaru}},\ }\href {\doibase 10.1016/0003-4916(87)90162-X} {\bibfield
  {journal} {\bibinfo  {journal} {Annals of Physics}\ }\textbf {\bibinfo
  {volume} {173}},\ \bibinfo {pages} {277 } (\bibinfo {year}
  {1987})}\BibitemShut {NoStop}%
\bibitem [{\citenamefont {Genoni}\ \emph {et~al.}(2013)\citenamefont {Genoni},
  \citenamefont {Mancini},\ and\ \citenamefont
  {Serafini}}]{OptimalFeedbackControl}%
  \BibitemOpen
  \bibfield  {author} {\bibinfo {author} {\bibfnamefont {M.~G.}\ \bibnamefont
  {Genoni}}, \bibinfo {author} {\bibfnamefont {S.}~\bibnamefont {Mancini}}, \
  and\ \bibinfo {author} {\bibfnamefont {A.}~\bibnamefont {Serafini}},\ }\href
  {\doibase 10.1103/PhysRevA.87.042333} {\bibfield  {journal} {\bibinfo
  {journal} {Phys. Rev. A}\ }\textbf {\bibinfo {volume} {87}},\ \bibinfo
  {pages} {042333} (\bibinfo {year} {2013})}\BibitemShut {NoStop}%
\bibitem [{\citenamefont {Schumaker}(1986)}]{Schumaker1986317}%
  \BibitemOpen
  \bibfield  {author} {\bibinfo {author} {\bibfnamefont {B.~L.}\ \bibnamefont
  {Schumaker}},\ }\href {\doibase 10.1016/0370-1573(86)90179-1} {\bibfield
  {journal} {\bibinfo  {journal} {Physics Reports}\ }\textbf {\bibinfo {volume}
  {135}},\ \bibinfo {pages} {317 } (\bibinfo {year} {1986})}\BibitemShut
  {NoStop}%
\bibitem [{\citenamefont {Carlini}\ \emph {et~al.}(2014)\citenamefont
  {Carlini}, \citenamefont {Mari},\ and\ \citenamefont
  {Giovannetti}}]{OptimalThermalization}%
  \BibitemOpen
  \bibfield  {author} {\bibinfo {author} {\bibfnamefont {A.}~\bibnamefont
  {Carlini}}, \bibinfo {author} {\bibfnamefont {A.}~\bibnamefont {Mari}}, \
  and\ \bibinfo {author} {\bibfnamefont {V.}~\bibnamefont {Giovannetti}},\
  }\href {\doibase 10.1103/PhysRevA.90.052324} {\bibfield  {journal} {\bibinfo
  {journal} {Phys. Rev. A}\ }\textbf {\bibinfo {volume} {90}},\ \bibinfo
  {pages} {052324} (\bibinfo {year} {2014})}\BibitemShut {NoStop}%
\bibitem [{\citenamefont {Gajic}\ and\ \citenamefont
  {Qureshi}(1995)}]{LyapunovBook}%
  \BibitemOpen
  \bibfield  {author} {\bibinfo {author} {\bibfnamefont {Z.}~\bibnamefont
  {Gajic}}\ and\ \bibinfo {author} {\bibfnamefont {M.~T.~J.}\ \bibnamefont
  {Qureshi}},\ }\href {\doibase
  https://doi.org/10.1016/B978-0-08-057049-5.50002-3} {\emph {\bibinfo {title}
  {{Lyapunov matrix equation in system stability and control}}}},\ Mathematics
  in science and engineering\ (\bibinfo  {publisher} {Elsevier},\ \bibinfo
  {address} {Burlington, MA},\ \bibinfo {year} {1995})\BibitemShut {NoStop}%
\bibitem [{\citenamefont {Axelsson}\ and\ \citenamefont
  {Gustafsson}(2015)}]{CtsTimeLyapunov}%
  \BibitemOpen
  \bibfield  {author} {\bibinfo {author} {\bibfnamefont {P.}~\bibnamefont
  {Axelsson}}\ and\ \bibinfo {author} {\bibfnamefont {F.}~\bibnamefont
  {Gustafsson}},\ }\href {\doibase 10.1109/TAC.2014.2353112} {\bibfield
  {journal} {\bibinfo  {journal} {Automatic Control, IEEE Transactions on}\
  }\textbf {\bibinfo {volume} {60}},\ \bibinfo {pages} {632} (\bibinfo {year}
  {2015})}\BibitemShut {NoStop}%
\bibitem [{\citenamefont {Rome}(1969)}]{DiagonalSolution}%
  \BibitemOpen
  \bibfield  {author} {\bibinfo {author} {\bibfnamefont {H.}~\bibnamefont
  {Rome}},\ }\href {\doibase 10.1109/TAC.1969.1099271} {\bibfield  {journal}
  {\bibinfo  {journal} {Automatic Control, IEEE Transactions on}\ }\textbf
  {\bibinfo {volume} {14}},\ \bibinfo {pages} {592} (\bibinfo {year}
  {1969})}\BibitemShut {NoStop}%
\bibitem [{\citenamefont {Breuer}\ and\ \citenamefont
  {Petruccione}(2002)}]{OpenBook}%
  \BibitemOpen
  \bibfield  {author} {\bibinfo {author} {\bibfnamefont {H.-P.}\ \bibnamefont
  {Breuer}}\ and\ \bibinfo {author} {\bibfnamefont {F.}~\bibnamefont
  {Petruccione}},\ }\href@noop {} {\emph {\bibinfo {title} {The Theory of Open
  Qunatum Systems}}},\ \bibinfo {edition} {second edition}\ ed.\ (\bibinfo
  {publisher} {Oxford University Press},\ \bibinfo {year} {2002})\BibitemShut
  {NoStop}%
\bibitem [{\citenamefont {Olivares}(2012)}]{QOInPhaseSpace}%
  \BibitemOpen
  \bibfield  {author} {\bibinfo {author} {\bibfnamefont {S.}~\bibnamefont
  {Olivares}},\ }\href {\doibase 10.1140/epjst/e2012-01532-4} {\bibfield
  {journal} {\bibinfo  {journal} {The European Physical Journal Special
  Topics}\ }\textbf {\bibinfo {volume} {203}},\ \bibinfo {pages} {3} (\bibinfo
  {year} {2012})}\BibitemShut {NoStop}%
\bibitem [{\citenamefont {Ferraro}\ \emph {et~al.}(2005)\citenamefont
  {Ferraro}, \citenamefont {Olivares},\ and\ \citenamefont
  {Paris}}]{GaussianStatesInContinuousQI}%
  \BibitemOpen
  \bibfield  {author} {\bibinfo {author} {\bibfnamefont {A.}~\bibnamefont
  {Ferraro}}, \bibinfo {author} {\bibfnamefont {S.}~\bibnamefont {Olivares}}, \
  and\ \bibinfo {author} {\bibfnamefont {M.}~\bibnamefont {Paris}},\
  }\href@noop {} {\emph {\bibinfo {title} {Gaussian states in continuous
  variable quantum information}}}\ (\bibinfo  {publisher} {Napoli Series on
  Physics and Astrophysics ed. Bibliopolis, Napoli},\ \bibinfo {year}
  {2005})\BibitemShut {NoStop}%
\bibitem [{\citenamefont {Serafini}\ \emph
  {et~al.}(2005{\natexlab{a}})\citenamefont {Serafini}, \citenamefont {Paris},
  \citenamefont {Illuminati},\ and\ \citenamefont
  {Siena}}]{QuantifyingDecoherenceRef}%
  \BibitemOpen
  \bibfield  {author} {\bibinfo {author} {\bibfnamefont {A.}~\bibnamefont
  {Serafini}}, \bibinfo {author} {\bibfnamefont {M.~G.~A.}\ \bibnamefont
  {Paris}}, \bibinfo {author} {\bibfnamefont {F.}~\bibnamefont {Illuminati}}, \
  and\ \bibinfo {author} {\bibfnamefont {S.~D.}\ \bibnamefont {Siena}},\ }\href
  {http://stacks.iop.org/1464-4266/7/i=4/a=R01} {\bibfield  {journal} {\bibinfo
   {journal} {Journal of Optics B: Quantum and Semiclassical Optics}\ }\textbf
  {\bibinfo {volume} {7}},\ \bibinfo {pages} {R19} (\bibinfo {year}
  {2005}{\natexlab{a}})}\BibitemShut {NoStop}%
\bibitem [{\citenamefont {Demoen}\ \emph {et~al.}(1977)\citenamefont {Demoen},
  \citenamefont {Vanheuverzwijn},\ and\ \citenamefont {Verbeure}}]{Demoen97}%
  \BibitemOpen
  \bibfield  {author} {\bibinfo {author} {\bibfnamefont {B.}~\bibnamefont
  {Demoen}}, \bibinfo {author} {\bibfnamefont {P.}~\bibnamefont
  {Vanheuverzwijn}}, \ and\ \bibinfo {author} {\bibfnamefont {A.}~\bibnamefont
  {Verbeure}},\ }\href {\doibase 10.1007/BF00398582} {\bibfield  {journal}
  {\bibinfo  {journal} {Letters in Mathematical Physics}\ }\textbf {\bibinfo
  {volume} {2}},\ \bibinfo {pages} {161} (\bibinfo {year} {1977})}\BibitemShut
  {NoStop}%
\bibitem [{\citenamefont {Caves}\ and\ \citenamefont
  {Drummond}(1994)}]{Caves94}%
  \BibitemOpen
  \bibfield  {author} {\bibinfo {author} {\bibfnamefont {C.~M.}\ \bibnamefont
  {Caves}}\ and\ \bibinfo {author} {\bibfnamefont {P.~D.}\ \bibnamefont
  {Drummond}},\ }\href {\doibase 10.1103/RevModPhys.66.481} {\bibfield
  {journal} {\bibinfo  {journal} {Rev. Mod. Phys.}\ }\textbf {\bibinfo {volume}
  {66}},\ \bibinfo {pages} {481} (\bibinfo {year} {1994})}\BibitemShut
  {NoStop}%
\bibitem [{\citenamefont {Lindblad}(2000)}]{Lindblad2000}%
  \BibitemOpen
  \bibfield  {author} {\bibinfo {author} {\bibfnamefont {G.}~\bibnamefont
  {Lindblad}},\ }\href {http://stacks.iop.org/0305-4470/33/i=28/a=310}
  {\bibfield  {journal} {\bibinfo  {journal} {Journal of Physics A:
  Mathematical and General}\ }\textbf {\bibinfo {volume} {33}},\ \bibinfo
  {pages} {5059} (\bibinfo {year} {2000})}\BibitemShut {NoStop}%
\bibitem [{\citenamefont {Harrington}\ and\ \citenamefont
  {Preskill}(2001)}]{Preskill2001}%
  \BibitemOpen
  \bibfield  {author} {\bibinfo {author} {\bibfnamefont {J.}~\bibnamefont
  {Harrington}}\ and\ \bibinfo {author} {\bibfnamefont {J.}~\bibnamefont
  {Preskill}},\ }\href {\doibase 10.1103/PhysRevA.64.062301} {\bibfield
  {journal} {\bibinfo  {journal} {Phys. Rev. A}\ }\textbf {\bibinfo {volume}
  {64}},\ \bibinfo {pages} {062301} (\bibinfo {year} {2001})}\BibitemShut
  {NoStop}%
\bibitem [{\citenamefont {Eisert}\ and\ \citenamefont
  {Plenio}(2002)}]{Eisert02}%
  \BibitemOpen
  \bibfield  {author} {\bibinfo {author} {\bibfnamefont {J.}~\bibnamefont
  {Eisert}}\ and\ \bibinfo {author} {\bibfnamefont {M.~B.}\ \bibnamefont
  {Plenio}},\ }\href {\doibase 10.1103/PhysRevLett.89.097901} {\bibfield
  {journal} {\bibinfo  {journal} {Phys. Rev. Lett.}\ }\textbf {\bibinfo
  {volume} {89}},\ \bibinfo {pages} {097901} (\bibinfo {year}
  {2002})}\BibitemShut {NoStop}%
\bibitem [{\citenamefont {Serafini}\ \emph {et~al.}(2004)\citenamefont
  {Serafini}, \citenamefont {Illuminati}, \citenamefont {Paris},\ and\
  \citenamefont {De~Siena}}]{Serafini04}%
  \BibitemOpen
  \bibfield  {author} {\bibinfo {author} {\bibfnamefont {A.}~\bibnamefont
  {Serafini}}, \bibinfo {author} {\bibfnamefont {F.}~\bibnamefont
  {Illuminati}}, \bibinfo {author} {\bibfnamefont {M.~G.~A.}\ \bibnamefont
  {Paris}}, \ and\ \bibinfo {author} {\bibfnamefont {S.}~\bibnamefont
  {De~Siena}},\ }\href {\doibase 10.1103/PhysRevA.69.022318} {\bibfield
  {journal} {\bibinfo  {journal} {Phys. Rev. A}\ }\textbf {\bibinfo {volume}
  {69}},\ \bibinfo {pages} {022318} (\bibinfo {year} {2004})}\BibitemShut
  {NoStop}%
\bibitem [{\citenamefont {Holevo}\ \emph {et~al.}(1999)\citenamefont {Holevo},
  \citenamefont {Sohma},\ and\ \citenamefont {Hirota}}]{Holevo99}%
  \BibitemOpen
  \bibfield  {author} {\bibinfo {author} {\bibfnamefont {A.~S.}\ \bibnamefont
  {Holevo}}, \bibinfo {author} {\bibfnamefont {M.}~\bibnamefont {Sohma}}, \
  and\ \bibinfo {author} {\bibfnamefont {O.}~\bibnamefont {Hirota}},\ }\href
  {\doibase 10.1103/PhysRevA.59.1820} {\bibfield  {journal} {\bibinfo
  {journal} {Phys. Rev. A}\ }\textbf {\bibinfo {volume} {59}},\ \bibinfo
  {pages} {1820} (\bibinfo {year} {1999})}\BibitemShut {NoStop}%
\bibitem [{\citenamefont {Holevo}\ and\ \citenamefont
  {Werner}(2001)}]{Holevo01}%
  \BibitemOpen
  \bibfield  {author} {\bibinfo {author} {\bibfnamefont {A.~S.}\ \bibnamefont
  {Holevo}}\ and\ \bibinfo {author} {\bibfnamefont {R.~F.}\ \bibnamefont
  {Werner}},\ }\href {\doibase 10.1103/PhysRevA.63.032312} {\bibfield
  {journal} {\bibinfo  {journal} {Phys. Rev. A}\ }\textbf {\bibinfo {volume}
  {63}},\ \bibinfo {pages} {032312} (\bibinfo {year} {2001})}\BibitemShut
  {NoStop}%
\bibitem [{\citenamefont {Giovannetti}\ \emph
  {et~al.}(2003{\natexlab{a}})\citenamefont {Giovannetti}, \citenamefont
  {Lloyd}, \citenamefont {Maccone},\ and\ \citenamefont
  {Shor}}]{Giovannetti03}%
  \BibitemOpen
  \bibfield  {author} {\bibinfo {author} {\bibfnamefont {V.}~\bibnamefont
  {Giovannetti}}, \bibinfo {author} {\bibfnamefont {S.}~\bibnamefont {Lloyd}},
  \bibinfo {author} {\bibfnamefont {L.}~\bibnamefont {Maccone}}, \ and\
  \bibinfo {author} {\bibfnamefont {P.~W.}\ \bibnamefont {Shor}},\ }\href
  {\doibase 10.1103/PhysRevLett.91.047901} {\bibfield  {journal} {\bibinfo
  {journal} {Phys. Rev. Lett.}\ }\textbf {\bibinfo {volume} {91}},\ \bibinfo
  {pages} {047901} (\bibinfo {year} {2003}{\natexlab{a}})}\BibitemShut
  {NoStop}%
\bibitem [{\citenamefont {Giovannetti}\ \emph
  {et~al.}(2003{\natexlab{b}})\citenamefont {Giovannetti}, \citenamefont
  {Lloyd}, \citenamefont {Maccone},\ and\ \citenamefont
  {Shor}}]{Giovannetti03b}%
  \BibitemOpen
  \bibfield  {author} {\bibinfo {author} {\bibfnamefont {V.}~\bibnamefont
  {Giovannetti}}, \bibinfo {author} {\bibfnamefont {S.}~\bibnamefont {Lloyd}},
  \bibinfo {author} {\bibfnamefont {L.}~\bibnamefont {Maccone}}, \ and\
  \bibinfo {author} {\bibfnamefont {P.~W.}\ \bibnamefont {Shor}},\ }\href
  {\doibase 10.1103/PhysRevA.68.062323} {\bibfield  {journal} {\bibinfo
  {journal} {Phys. Rev. A}\ }\textbf {\bibinfo {volume} {68}},\ \bibinfo
  {pages} {062323} (\bibinfo {year} {2003}{\natexlab{b}})}\BibitemShut
  {NoStop}%
\bibitem [{\citenamefont {Serafini}\ \emph
  {et~al.}(2005{\natexlab{b}})\citenamefont {Serafini}, \citenamefont
  {Eisert},\ and\ \citenamefont {Wolf}}]{Multiplicativity}%
  \BibitemOpen
  \bibfield  {author} {\bibinfo {author} {\bibfnamefont {A.}~\bibnamefont
  {Serafini}}, \bibinfo {author} {\bibfnamefont {J.}~\bibnamefont {Eisert}}, \
  and\ \bibinfo {author} {\bibfnamefont {M.~M.}\ \bibnamefont {Wolf}},\ }\href
  {\doibase 10.1103/PhysRevA.71.012320} {\bibfield  {journal} {\bibinfo
  {journal} {Phys. Rev. A}\ }\textbf {\bibinfo {volume} {71}},\ \bibinfo
  {pages} {012320} (\bibinfo {year} {2005}{\natexlab{b}})}\BibitemShut
  {NoStop}%
\bibitem [{\citenamefont {Weedbrook}\ \emph {et~al.}(2012)\citenamefont
  {Weedbrook}, \citenamefont {Pirandola}, \citenamefont {Garc\'{\i}a-Patr\'on},
  \citenamefont {Cerf}, \citenamefont {Ralph}, \citenamefont {Shapiro},\ and\
  \citenamefont {Lloyd}}]{GQI}%
  \BibitemOpen
  \bibfield  {author} {\bibinfo {author} {\bibfnamefont {C.}~\bibnamefont
  {Weedbrook}}, \bibinfo {author} {\bibfnamefont {S.}~\bibnamefont
  {Pirandola}}, \bibinfo {author} {\bibfnamefont {R.}~\bibnamefont
  {Garc\'{\i}a-Patr\'on}}, \bibinfo {author} {\bibfnamefont {N.~J.}\
  \bibnamefont {Cerf}}, \bibinfo {author} {\bibfnamefont {T.~C.}\ \bibnamefont
  {Ralph}}, \bibinfo {author} {\bibfnamefont {J.~H.}\ \bibnamefont {Shapiro}},
  \ and\ \bibinfo {author} {\bibfnamefont {S.}~\bibnamefont {Lloyd}},\ }\href
  {\doibase 10.1103/RevModPhys.84.621} {\bibfield  {journal} {\bibinfo
  {journal} {Rev. Mod. Phys.}\ }\textbf {\bibinfo {volume} {84}},\ \bibinfo
  {pages} {621} (\bibinfo {year} {2012})}\BibitemShut {NoStop}%
\bibitem [{\citenamefont {Marian}\ and\ \citenamefont
  {Marian}(1993{\natexlab{a}})}]{PhysRevA.47.4474}%
  \BibitemOpen
  \bibfield  {author} {\bibinfo {author} {\bibfnamefont {P.}~\bibnamefont
  {Marian}}\ and\ \bibinfo {author} {\bibfnamefont {T.~A.}\ \bibnamefont
  {Marian}},\ }\href {\doibase 10.1103/PhysRevA.47.4474} {\bibfield  {journal}
  {\bibinfo  {journal} {Phys. Rev. A}\ }\textbf {\bibinfo {volume} {47}},\
  \bibinfo {pages} {4474} (\bibinfo {year} {1993}{\natexlab{a}})}\BibitemShut
  {NoStop}%
\bibitem [{\citenamefont {Marian}\ and\ \citenamefont
  {Marian}(1993{\natexlab{b}})}]{PhysRevA.47.4487}%
  \BibitemOpen
  \bibfield  {author} {\bibinfo {author} {\bibfnamefont {P.}~\bibnamefont
  {Marian}}\ and\ \bibinfo {author} {\bibfnamefont {T.~A.}\ \bibnamefont
  {Marian}},\ }\href {\doibase 10.1103/PhysRevA.47.4487} {\bibfield  {journal}
  {\bibinfo  {journal} {Phys. Rev. A}\ }\textbf {\bibinfo {volume} {47}},\
  \bibinfo {pages} {4487} (\bibinfo {year} {1993}{\natexlab{b}})}\BibitemShut
  {NoStop}%
\bibitem [{\citenamefont {Paris}\ \emph {et~al.}(2003)\citenamefont {Paris},
  \citenamefont {Illuminati}, \citenamefont {Serafini},\ and\ \citenamefont
  {De~Siena}}]{Purity}%
  \BibitemOpen
  \bibfield  {author} {\bibinfo {author} {\bibfnamefont {M.~G.~A.}\
  \bibnamefont {Paris}}, \bibinfo {author} {\bibfnamefont {F.}~\bibnamefont
  {Illuminati}}, \bibinfo {author} {\bibfnamefont {A.}~\bibnamefont
  {Serafini}}, \ and\ \bibinfo {author} {\bibfnamefont {S.}~\bibnamefont
  {De~Siena}},\ }\href {\doibase 10.1103/PhysRevA.68.012314} {\bibfield
  {journal} {\bibinfo  {journal} {Phys. Rev. A}\ }\textbf {\bibinfo {volume}
  {68}},\ \bibinfo {pages} {012314} (\bibinfo {year} {2003})}\BibitemShut
  {NoStop}%
\bibitem [{\citenamefont {Williamson}(1936)}]{Williamson}%
  \BibitemOpen
  \bibfield  {author} {\bibinfo {author} {\bibfnamefont {J.}~\bibnamefont
  {Williamson}},\ }\href@noop {} {\bibfield  {journal} {\bibinfo  {journal}
  {Am. J. of Math.}\ }\textbf {\bibinfo {volume} {58}},\ \bibinfo {pages} {141}
  (\bibinfo {year} {1936})}\BibitemShut {NoStop}%
\bibitem [{\citenamefont {Adam}(1995)}]{Adam95}%
  \BibitemOpen
  \bibfield  {author} {\bibinfo {author} {\bibfnamefont {G.}~\bibnamefont
  {Adam}},\ }\href {\doibase 10.1080/09500349514551141} {\bibfield  {journal}
  {\bibinfo  {journal} {J. Mod. Opt.}\ }\textbf {\bibinfo {volume} {1311}},\
  \bibinfo {pages} {052115} (\bibinfo {year} {1995})}\BibitemShut {NoStop}%
\bibitem [{\citenamefont {Lee}(1991)}]{NonclassicalDepth}%
  \BibitemOpen
  \bibfield  {author} {\bibinfo {author} {\bibfnamefont {C.~T.}\ \bibnamefont
  {Lee}},\ }\href {\doibase 10.1103/PhysRevA.44.R2775} {\bibfield  {journal}
  {\bibinfo  {journal} {Phys. Rev. A}\ }\textbf {\bibinfo {volume} {44}},\
  \bibinfo {pages} {R2775} (\bibinfo {year} {1991})}\BibitemShut {NoStop}%
\bibitem [{\citenamefont {{Lee}}(1992)}]{Lee1992}%
  \BibitemOpen
  \bibfield  {author} {\bibinfo {author} {\bibfnamefont {C.~T.}\ \bibnamefont
  {{Lee}}},\ }in\ \href@noop {} {\emph {\bibinfo {booktitle} {Squeezed States
  and Uncertainty Relations}}},\ \bibinfo {editor} {edited by\ \bibinfo
  {editor} {\bibfnamefont {D.}~\bibnamefont {{Han}}}, \bibinfo {editor}
  {\bibfnamefont {Y.~S.}\ \bibnamefont {{Kim}}}, \ and\ \bibinfo {editor}
  {\bibfnamefont {W.~W.}\ \bibnamefont {{Zachary}}}}\ (\bibinfo  {publisher}
  {NASA. Goddard Space Flight Center, Workshop on Squeezed States and
  Uncertainty Relations},\ \bibinfo {year} {1992})\ pp.\ \bibinfo {pages}
  {365--367}\BibitemShut {NoStop}%
\bibitem [{\citenamefont {Wade}\ \emph {et~al.}(2015)\citenamefont {Wade},
  \citenamefont {Sherson},\ and\ \citenamefont
  {M\o{}lmer}}]{PhysRevLett.115.060401}%
  \BibitemOpen
  \bibfield  {author} {\bibinfo {author} {\bibfnamefont {A.~C.~J.}\
  \bibnamefont {Wade}}, \bibinfo {author} {\bibfnamefont {J.~F.}\ \bibnamefont
  {Sherson}}, \ and\ \bibinfo {author} {\bibfnamefont {K.}~\bibnamefont
  {M\o{}lmer}},\ }\href {\doibase 10.1103/PhysRevLett.115.060401} {\bibfield
  {journal} {\bibinfo  {journal} {Phys. Rev. Lett.}\ }\textbf {\bibinfo
  {volume} {115}},\ \bibinfo {pages} {060401} (\bibinfo {year}
  {2015})}\BibitemShut {NoStop}%
\bibitem [{\citenamefont {Wade}\ \emph {et~al.}(2016)\citenamefont {Wade},
  \citenamefont {Sherson},\ and\ \citenamefont
  {M\o{}lmer}}]{PhysRevA.93.023610}%
  \BibitemOpen
  \bibfield  {author} {\bibinfo {author} {\bibfnamefont {A.~C.~J.}\
  \bibnamefont {Wade}}, \bibinfo {author} {\bibfnamefont {J.~F.}\ \bibnamefont
  {Sherson}}, \ and\ \bibinfo {author} {\bibfnamefont {K.}~\bibnamefont
  {M\o{}lmer}},\ }\href {\doibase 10.1103/PhysRevA.93.023610} {\bibfield
  {journal} {\bibinfo  {journal} {Phys. Rev. A}\ }\textbf {\bibinfo {volume}
  {93}},\ \bibinfo {pages} {023610} (\bibinfo {year} {2016})}\BibitemShut
  {NoStop}%
\bibitem [{\citenamefont {Rogel-Salazar}\ \emph {et~al.}(2002)\citenamefont
  {Rogel-Salazar}, \citenamefont {New}, \citenamefont {Choi},\ and\
  \citenamefont {Burnett}}]{BeliaevEntanglement}%
  \BibitemOpen
  \bibfield  {author} {\bibinfo {author} {\bibfnamefont {J.}~\bibnamefont
  {Rogel-Salazar}}, \bibinfo {author} {\bibfnamefont {G.~H.~C.}\ \bibnamefont
  {New}}, \bibinfo {author} {\bibfnamefont {S.}~\bibnamefont {Choi}}, \ and\
  \bibinfo {author} {\bibfnamefont {K.}~\bibnamefont {Burnett}},\ }\href
  {\doibase 10.1103/PhysRevA.65.023601} {\bibfield  {journal} {\bibinfo
  {journal} {Phys. Rev. A}\ }\textbf {\bibinfo {volume} {65}},\ \bibinfo
  {pages} {023601} (\bibinfo {year} {2002})}\BibitemShut {NoStop}%
\bibitem [{\citenamefont {Braunstein}\ and\ \citenamefont
  {Caves}(1994)}]{PhysRevLett.72.3439}%
  \BibitemOpen
  \bibfield  {author} {\bibinfo {author} {\bibfnamefont {S.~L.}\ \bibnamefont
  {Braunstein}}\ and\ \bibinfo {author} {\bibfnamefont {C.~M.}\ \bibnamefont
  {Caves}},\ }\href {\doibase 10.1103/PhysRevLett.72.3439} {\bibfield
  {journal} {\bibinfo  {journal} {Phys. Rev. Lett.}\ }\textbf {\bibinfo
  {volume} {72}},\ \bibinfo {pages} {3439} (\bibinfo {year}
  {1994})}\BibitemShut {NoStop}%
\bibitem [{\citenamefont {Barcel{\'o}}\ \emph {et~al.}(2005)\citenamefont
  {Barcel{\'o}}, \citenamefont {Liberati},\ and\ \citenamefont
  {Visser}}]{AnalogueGravityReview}%
  \BibitemOpen
  \bibfield  {author} {\bibinfo {author} {\bibfnamefont {C.}~\bibnamefont
  {Barcel{\'o}}}, \bibinfo {author} {\bibfnamefont {S.}~\bibnamefont
  {Liberati}}, \ and\ \bibinfo {author} {\bibfnamefont {M.}~\bibnamefont
  {Visser}},\ }\href {\doibase 10.12942/lrr-2005-12} {\bibfield  {journal}
  {\bibinfo  {journal} {Living Reviews in Relativity}\ }\textbf {\bibinfo
  {volume} {8}},\ \bibinfo {pages} {12} (\bibinfo {year} {2005})}\BibitemShut
  {NoStop}%
\bibitem [{\citenamefont {Fagnocchi}\ \emph {et~al.}(2010)\citenamefont
  {Fagnocchi}, \citenamefont {Finazzi}, \citenamefont {Liberati}, \citenamefont
  {Kormos},\ and\ \citenamefont {Trombettoni}}]{RelBECs}%
  \BibitemOpen
  \bibfield  {author} {\bibinfo {author} {\bibfnamefont {S.}~\bibnamefont
  {Fagnocchi}}, \bibinfo {author} {\bibfnamefont {S.}~\bibnamefont {Finazzi}},
  \bibinfo {author} {\bibfnamefont {S.}~\bibnamefont {Liberati}}, \bibinfo
  {author} {\bibfnamefont {M.}~\bibnamefont {Kormos}}, \ and\ \bibinfo {author}
  {\bibfnamefont {A.}~\bibnamefont {Trombettoni}},\ }\href
  {http://stacks.iop.org/1367-2630/12/i=9/a=095012} {\bibfield  {journal}
  {\bibinfo  {journal} {New Journal of Physics}\ }\textbf {\bibinfo {volume}
  {12}},\ \bibinfo {pages} {095012} (\bibinfo {year} {2010})}\BibitemShut
  {NoStop}%
\bibitem [{\citenamefont {Martynov}\ \emph {et~al.}(2016)\citenamefont
  {Martynov}, \citenamefont {Hall}, \citenamefont {Abbott}, \citenamefont
  {Abbott}, \citenamefont {Abbott}, \citenamefont {Adams}, \citenamefont
  {Adhikari}, \citenamefont {Anderson}, \citenamefont {Anderson}, \citenamefont
  {Arai}, \citenamefont {Arain}, \citenamefont {Aston}, \citenamefont {Austin},
  \citenamefont {Ballmer}, \citenamefont {Barbet} \emph
  {et~al.}}]{PhysRevD.93.112004Truncated}%
  \BibitemOpen
  \bibfield  {author} {\bibinfo {author} {\bibfnamefont {D.~V.}\ \bibnamefont
  {Martynov}}, \bibinfo {author} {\bibfnamefont {E.~D.}\ \bibnamefont {Hall}},
  \bibinfo {author} {\bibfnamefont {B.~P.}\ \bibnamefont {Abbott}}, \bibinfo
  {author} {\bibfnamefont {R.}~\bibnamefont {Abbott}}, \bibinfo {author}
  {\bibfnamefont {T.~D.}\ \bibnamefont {Abbott}}, \bibinfo {author}
  {\bibfnamefont {C.}~\bibnamefont {Adams}}, \bibinfo {author} {\bibfnamefont
  {R.~X.}\ \bibnamefont {Adhikari}}, \bibinfo {author} {\bibfnamefont {R.~A.}\
  \bibnamefont {Anderson}}, \bibinfo {author} {\bibfnamefont {S.~B.}\
  \bibnamefont {Anderson}}, \bibinfo {author} {\bibfnamefont {K.}~\bibnamefont
  {Arai}}, \bibinfo {author} {\bibfnamefont {M.~A.}\ \bibnamefont {Arain}},
  \bibinfo {author} {\bibfnamefont {S.~M.}\ \bibnamefont {Aston}}, \bibinfo
  {author} {\bibfnamefont {L.}~\bibnamefont {Austin}}, \bibinfo {author}
  {\bibfnamefont {S.~W.}\ \bibnamefont {Ballmer}}, \bibinfo {author}
  {\bibnamefont {Barbet}},  \emph {et~al.},\ }\href {\doibase
  10.1103/PhysRevD.93.112004} {\bibfield  {journal} {\bibinfo  {journal} {Phys.
  Rev. D}\ }\textbf {\bibinfo {volume} {93}},\ \bibinfo {pages} {112004}
  (\bibinfo {year} {2016})}\BibitemShut {NoStop}%
\bibitem [{\citenamefont {Leanhardt}\ \emph {et~al.}(2003)\citenamefont
  {Leanhardt}, \citenamefont {Pasquini}, \citenamefont {Saba}, \citenamefont
  {Schirotzek}, \citenamefont {Shin}, \citenamefont {Kielpinski}, \citenamefont
  {Pritchard},\ and\ \citenamefont {Ketterle}}]{Leanhardt1513}%
  \BibitemOpen
  \bibfield  {author} {\bibinfo {author} {\bibfnamefont {A.~E.}\ \bibnamefont
  {Leanhardt}}, \bibinfo {author} {\bibfnamefont {T.~A.}\ \bibnamefont
  {Pasquini}}, \bibinfo {author} {\bibfnamefont {M.}~\bibnamefont {Saba}},
  \bibinfo {author} {\bibfnamefont {A.}~\bibnamefont {Schirotzek}}, \bibinfo
  {author} {\bibfnamefont {Y.}~\bibnamefont {Shin}}, \bibinfo {author}
  {\bibfnamefont {D.}~\bibnamefont {Kielpinski}}, \bibinfo {author}
  {\bibfnamefont {D.~E.}\ \bibnamefont {Pritchard}}, \ and\ \bibinfo {author}
  {\bibfnamefont {W.}~\bibnamefont {Ketterle}},\ }\href {\doibase
  10.1126/science.1088827} {\bibfield  {journal} {\bibinfo  {journal}
  {Science}\ }\textbf {\bibinfo {volume} {301}},\ \bibinfo {pages} {1513}
  (\bibinfo {year} {2003})}\BibitemShut {NoStop}%
\bibitem [{\citenamefont {Shammass}\ \emph {et~al.}(2012)\citenamefont
  {Shammass}, \citenamefont {Rinott}, \citenamefont {Berkovitz}, \citenamefont
  {Schley},\ and\ \citenamefont {Steinhauer}}]{PhysRevLett.109.195301}%
  \BibitemOpen
  \bibfield  {author} {\bibinfo {author} {\bibfnamefont {I.}~\bibnamefont
  {Shammass}}, \bibinfo {author} {\bibfnamefont {S.}~\bibnamefont {Rinott}},
  \bibinfo {author} {\bibfnamefont {A.}~\bibnamefont {Berkovitz}}, \bibinfo
  {author} {\bibfnamefont {R.}~\bibnamefont {Schley}}, \ and\ \bibinfo {author}
  {\bibfnamefont {J.}~\bibnamefont {Steinhauer}},\ }\href {\doibase
  10.1103/PhysRevLett.109.195301} {\bibfield  {journal} {\bibinfo  {journal}
  {Phys. Rev. Lett.}\ }\textbf {\bibinfo {volume} {109}},\ \bibinfo {pages}
  {195301} (\bibinfo {year} {2012})}\BibitemShut {NoStop}%
\bibitem [{\citenamefont {Yuen}\ \emph {et~al.}(2015)\citenamefont {Yuen},
  \citenamefont {Barr}, \citenamefont {Cotter}, \citenamefont {Butler},\ and\
  \citenamefont {Hinds}}]{1367-2630-17-9-093041}%
  \BibitemOpen
  \bibfield  {author} {\bibinfo {author} {\bibfnamefont {B.}~\bibnamefont
  {Yuen}}, \bibinfo {author} {\bibfnamefont {I.~J.~M.}\ \bibnamefont {Barr}},
  \bibinfo {author} {\bibfnamefont {J.~P.}\ \bibnamefont {Cotter}}, \bibinfo
  {author} {\bibfnamefont {E.}~\bibnamefont {Butler}}, \ and\ \bibinfo {author}
  {\bibfnamefont {E.~A.}\ \bibnamefont {Hinds}},\ }\href
  {http://stacks.iop.org/1367-2630/17/i=9/a=093041} {\bibfield  {journal}
  {\bibinfo  {journal} {New Journal of Physics}\ }\textbf {\bibinfo {volume}
  {17}},\ \bibinfo {pages} {093041} (\bibinfo {year} {2015})}\BibitemShut
  {NoStop}%
\bibitem [{\citenamefont {{Ma}}\ \emph {et~al.}(2011)\citenamefont {{Ma}},
  \citenamefont {{Yang}}, \citenamefont {{Lu}},\ and\ \citenamefont
  {{Wei}}}]{2011ChPhB..20g0307M}%
  \BibitemOpen
  \bibfield  {author} {\bibinfo {author} {\bibfnamefont {X.-D.}\ \bibnamefont
  {{Ma}}}, \bibinfo {author} {\bibfnamefont {Z.-J.}\ \bibnamefont {{Yang}}},
  \bibinfo {author} {\bibfnamefont {J.-Z.}\ \bibnamefont {{Lu}}}, \ and\
  \bibinfo {author} {\bibfnamefont {W.}~\bibnamefont {{Wei}}},\ }\href
  {\doibase 10.1088/1674-1056/20/7/070307} {\bibfield  {journal} {\bibinfo
  {journal} {Chinese Physics B}\ }\textbf {\bibinfo {volume} {20}},\ \bibinfo
  {eid} {070307} (\bibinfo {year} {2011})}\BibitemShut {NoStop}%
\bibitem [{\citenamefont {Trallero-Giner}\ \emph {et~al.}(2015)\citenamefont
  {Trallero-Giner}, \citenamefont {Santiago-P\'erez}, \citenamefont {Chung},
  \citenamefont {Marques},\ and\ \citenamefont
  {Cipolatti}}]{2015arXiv150308884T}%
  \BibitemOpen
  \bibfield  {author} {\bibinfo {author} {\bibfnamefont {C.}~\bibnamefont
  {Trallero-Giner}}, \bibinfo {author} {\bibfnamefont {D.~G.}\ \bibnamefont
  {Santiago-P\'erez}}, \bibinfo {author} {\bibfnamefont {M.-C.}\ \bibnamefont
  {Chung}}, \bibinfo {author} {\bibfnamefont {G.~E.}\ \bibnamefont {Marques}},
  \ and\ \bibinfo {author} {\bibfnamefont {R.}~\bibnamefont {Cipolatti}},\
  }\href {\doibase 10.1103/PhysRevA.92.042502} {\bibfield  {journal} {\bibinfo
  {journal} {Phys. Rev. A}\ }\textbf {\bibinfo {volume} {92}},\ \bibinfo
  {pages} {042502} (\bibinfo {year} {2015})}\BibitemShut {NoStop}%
\bibitem [{\citenamefont {Arahata}\ and\ \citenamefont
  {Nikuni}(2008)}]{PhysRevA.77.033610}%
  \BibitemOpen
  \bibfield  {author} {\bibinfo {author} {\bibfnamefont {E.}~\bibnamefont
  {Arahata}}\ and\ \bibinfo {author} {\bibfnamefont {T.}~\bibnamefont
  {Nikuni}},\ }\href {\doibase 10.1103/PhysRevA.77.033610} {\bibfield
  {journal} {\bibinfo  {journal} {Phys. Rev. A}\ }\textbf {\bibinfo {volume}
  {77}},\ \bibinfo {pages} {033610} (\bibinfo {year} {2008})}\BibitemShut
  {NoStop}%
\bibitem [{\citenamefont {Mazets}(2011)}]{PhysRevA.83.043625}%
  \BibitemOpen
  \bibfield  {author} {\bibinfo {author} {\bibfnamefont {I.~E.}\ \bibnamefont
  {Mazets}},\ }\href {\doibase 10.1103/PhysRevA.83.043625} {\bibfield
  {journal} {\bibinfo  {journal} {Phys. Rev. A}\ }\textbf {\bibinfo {volume}
  {83}},\ \bibinfo {pages} {043625} (\bibinfo {year} {2011})}\BibitemShut
  {NoStop}%
\bibitem [{\citenamefont {Yuen}(2013)}]{YuenThesis}%
  \BibitemOpen
  \bibfield  {author} {\bibinfo {author} {\bibfnamefont {B.}~\bibnamefont
  {Yuen}},\ }\emph {\bibinfo {title} {{Production and Oscillations of a Bose
  Einstein Condensate on an Atom Chip}}},\ \href
  {https://spiral.imperial.ac.uk:8443/handle/10044/1/18833} {Ph.D. thesis},\
  \bibinfo  {school} {Imperial College London} (\bibinfo {year}
  {2013})\BibitemShut {NoStop}%
\bibitem [{\citenamefont {Ziń}\ and\ \citenamefont
  {Pylak}(2017)}]{2016arXiv160609400Z}%
  \BibitemOpen
  \bibfield  {author} {\bibinfo {author} {\bibfnamefont {P.}~\bibnamefont
  {Ziń}}\ and\ \bibinfo {author} {\bibfnamefont {M.}~\bibnamefont {Pylak}},\
  }\href {http://stacks.iop.org/0953-4075/50/i=8/a=085301} {\bibfield
  {journal} {\bibinfo  {journal} {Journal of Physics B: Atomic, Molecular and
  Optical Physics}\ }\textbf {\bibinfo {volume} {50}},\ \bibinfo {pages}
  {085301} (\bibinfo {year} {2017})}\BibitemShut {NoStop}%
\bibitem [{\citenamefont {Moerdijk}\ \emph {et~al.}(1996)\citenamefont
  {Moerdijk}, \citenamefont {Boesten},\ and\ \citenamefont
  {Verhaar}}]{PhysRevA.53.916}%
  \BibitemOpen
  \bibfield  {author} {\bibinfo {author} {\bibfnamefont {A.~J.}\ \bibnamefont
  {Moerdijk}}, \bibinfo {author} {\bibfnamefont {H.~M. J.~M.}\ \bibnamefont
  {Boesten}}, \ and\ \bibinfo {author} {\bibfnamefont {B.~J.}\ \bibnamefont
  {Verhaar}},\ }\href {\doibase 10.1103/PhysRevA.53.916} {\bibfield  {journal}
  {\bibinfo  {journal} {Phys. Rev. A}\ }\textbf {\bibinfo {volume} {53}},\
  \bibinfo {pages} {916} (\bibinfo {year} {1996})}\BibitemShut {NoStop}%
\bibitem [{\citenamefont {Burt}\ \emph {et~al.}(1997)\citenamefont {Burt},
  \citenamefont {Ghrist}, \citenamefont {Myatt}, \citenamefont {Holland},
  \citenamefont {Cornell},\ and\ \citenamefont {Wieman}}]{Rb3bodyRate}%
  \BibitemOpen
  \bibfield  {author} {\bibinfo {author} {\bibfnamefont {E.~A.}\ \bibnamefont
  {Burt}}, \bibinfo {author} {\bibfnamefont {R.~W.}\ \bibnamefont {Ghrist}},
  \bibinfo {author} {\bibfnamefont {C.~J.}\ \bibnamefont {Myatt}}, \bibinfo
  {author} {\bibfnamefont {M.~J.}\ \bibnamefont {Holland}}, \bibinfo {author}
  {\bibfnamefont {E.~A.}\ \bibnamefont {Cornell}}, \ and\ \bibinfo {author}
  {\bibfnamefont {C.~E.}\ \bibnamefont {Wieman}},\ }\href {\doibase
  10.1103/PhysRevLett.79.337} {\bibfield  {journal} {\bibinfo  {journal} {Phys.
  Rev. Lett.}\ }\textbf {\bibinfo {volume} {79}},\ \bibinfo {pages} {337}
  (\bibinfo {year} {1997})}\BibitemShut {NoStop}%
\bibitem [{\citenamefont {Esry}\ \emph {et~al.}(2001)\citenamefont {Esry},
  \citenamefont {Greene},\ and\ \citenamefont {Suno}}]{PhysRevA.65.010705}%
  \BibitemOpen
  \bibfield  {author} {\bibinfo {author} {\bibfnamefont {B.~D.}\ \bibnamefont
  {Esry}}, \bibinfo {author} {\bibfnamefont {C.~H.}\ \bibnamefont {Greene}}, \
  and\ \bibinfo {author} {\bibfnamefont {H.}~\bibnamefont {Suno}},\ }\href
  {\doibase 10.1103/PhysRevA.65.010705} {\bibfield  {journal} {\bibinfo
  {journal} {Phys. Rev. A}\ }\textbf {\bibinfo {volume} {65}},\ \bibinfo
  {pages} {010705} (\bibinfo {year} {2001})}\BibitemShut {NoStop}%
\bibitem [{\citenamefont {Mehta}\ \emph {et~al.}(2007)\citenamefont {Mehta},
  \citenamefont {Esry},\ and\ \citenamefont {Greene}}]{PhysRevA.76.022711}%
  \BibitemOpen
  \bibfield  {author} {\bibinfo {author} {\bibfnamefont {N.~P.}\ \bibnamefont
  {Mehta}}, \bibinfo {author} {\bibfnamefont {B.~D.}\ \bibnamefont {Esry}}, \
  and\ \bibinfo {author} {\bibfnamefont {C.~H.}\ \bibnamefont {Greene}},\
  }\href {\doibase 10.1103/PhysRevA.76.022711} {\bibfield  {journal} {\bibinfo
  {journal} {Phys. Rev. A}\ }\textbf {\bibinfo {volume} {76}},\ \bibinfo
  {pages} {022711} (\bibinfo {year} {2007})}\BibitemShut {NoStop}%
\bibitem [{\citenamefont {Helfrich}\ and\ \citenamefont
  {Hammer}(2011)}]{PhysRevA.83.052703}%
  \BibitemOpen
  \bibfield  {author} {\bibinfo {author} {\bibfnamefont {K.}~\bibnamefont
  {Helfrich}}\ and\ \bibinfo {author} {\bibfnamefont {H.-W.}\ \bibnamefont
  {Hammer}},\ }\href {\doibase 10.1103/PhysRevA.83.052703} {\bibfield
  {journal} {\bibinfo  {journal} {Phys. Rev. A}\ }\textbf {\bibinfo {volume}
  {83}},\ \bibinfo {pages} {052703} (\bibinfo {year} {2011})}\BibitemShut
  {NoStop}%
\bibitem [{\citenamefont {D'Incao}\ \emph {et~al.}(2015)\citenamefont
  {D'Incao}, \citenamefont {Anis},\ and\ \citenamefont
  {Esry}}]{PhysRevA.91.062710}%
  \BibitemOpen
  \bibfield  {author} {\bibinfo {author} {\bibfnamefont {J.~P.}\ \bibnamefont
  {D'Incao}}, \bibinfo {author} {\bibfnamefont {F.}~\bibnamefont {Anis}}, \
  and\ \bibinfo {author} {\bibfnamefont {B.~D.}\ \bibnamefont {Esry}},\ }\href
  {\doibase 10.1103/PhysRevA.91.062710} {\bibfield  {journal} {\bibinfo
  {journal} {Phys. Rev. A}\ }\textbf {\bibinfo {volume} {91}},\ \bibinfo
  {pages} {062710} (\bibinfo {year} {2015})}\BibitemShut {NoStop}%
\bibitem [{\citenamefont {Dziarmaga}\ and\ \citenamefont
  {Sacha}(2003)}]{PhysRevA.68.043607}%
  \BibitemOpen
  \bibfield  {author} {\bibinfo {author} {\bibfnamefont {J.}~\bibnamefont
  {Dziarmaga}}\ and\ \bibinfo {author} {\bibfnamefont {K.}~\bibnamefont
  {Sacha}},\ }\href {\doibase 10.1103/PhysRevA.68.043607} {\bibfield  {journal}
  {\bibinfo  {journal} {Phys. Rev. A}\ }\textbf {\bibinfo {volume} {68}},\
  \bibinfo {pages} {043607} (\bibinfo {year} {2003})}\BibitemShut {NoStop}%
\bibitem [{\citenamefont {Finazzi}\ and\ \citenamefont
  {Carusotto}(2014)}]{EntangledPhonons}%
  \BibitemOpen
  \bibfield  {author} {\bibinfo {author} {\bibfnamefont {S.}~\bibnamefont
  {Finazzi}}\ and\ \bibinfo {author} {\bibfnamefont {I.}~\bibnamefont
  {Carusotto}},\ }\href {\doibase 10.1103/PhysRevA.90.033607} {\bibfield
  {journal} {\bibinfo  {journal} {Phys. Rev. A}\ }\textbf {\bibinfo {volume}
  {90}},\ \bibinfo {pages} {033607} (\bibinfo {year} {2014})}\BibitemShut
  {NoStop}%
\bibitem [{\citenamefont {Steinhauer}(2015)}]{PhysRevD.92.024043}%
  \BibitemOpen
  \bibfield  {author} {\bibinfo {author} {\bibfnamefont {J.}~\bibnamefont
  {Steinhauer}},\ }\href {\doibase 10.1103/PhysRevD.92.024043} {\bibfield
  {journal} {\bibinfo  {journal} {Phys. Rev. D}\ }\textbf {\bibinfo {volume}
  {92}},\ \bibinfo {pages} {024043} (\bibinfo {year} {2015})}\BibitemShut
  {NoStop}%
\bibitem [{\citenamefont {Busch}\ \emph {et~al.}(2014)\citenamefont {Busch},
  \citenamefont {Parentani},\ and\ \citenamefont
  {Robertson}}]{PhysRevA.89.063606}%
  \BibitemOpen
  \bibfield  {author} {\bibinfo {author} {\bibfnamefont {X.}~\bibnamefont
  {Busch}}, \bibinfo {author} {\bibfnamefont {R.}~\bibnamefont {Parentani}}, \
  and\ \bibinfo {author} {\bibfnamefont {S.}~\bibnamefont {Robertson}},\ }\href
  {\doibase 10.1103/PhysRevA.89.063606} {\bibfield  {journal} {\bibinfo
  {journal} {Phys. Rev. A}\ }\textbf {\bibinfo {volume} {89}},\ \bibinfo
  {pages} {063606} (\bibinfo {year} {2014})}\BibitemShut {NoStop}%
\bibitem [{\citenamefont {Busch}\ and\ \citenamefont
  {Parentani}(2014)}]{PhysRevD.89.105024}%
  \BibitemOpen
  \bibfield  {author} {\bibinfo {author} {\bibfnamefont {X.}~\bibnamefont
  {Busch}}\ and\ \bibinfo {author} {\bibfnamefont {R.}~\bibnamefont
  {Parentani}},\ }\href {\doibase 10.1103/PhysRevD.89.105024} {\bibfield
  {journal} {\bibinfo  {journal} {Phys. Rev. D}\ }\textbf {\bibinfo {volume}
  {89}},\ \bibinfo {pages} {105024} (\bibinfo {year} {2014})}\BibitemShut
  {NoStop}%
\bibitem [{\citenamefont {Anderson}\ \emph {et~al.}(2013)\citenamefont
  {Anderson}, \citenamefont {Balbinot}, \citenamefont {Fabbri},\ and\
  \citenamefont {Parentani}}]{PhysRevD.87.124018}%
  \BibitemOpen
  \bibfield  {author} {\bibinfo {author} {\bibfnamefont {P.~R.}\ \bibnamefont
  {Anderson}}, \bibinfo {author} {\bibfnamefont {R.}~\bibnamefont {Balbinot}},
  \bibinfo {author} {\bibfnamefont {A.}~\bibnamefont {Fabbri}}, \ and\ \bibinfo
  {author} {\bibfnamefont {R.}~\bibnamefont {Parentani}},\ }\href {\doibase
  10.1103/PhysRevD.87.124018} {\bibfield  {journal} {\bibinfo  {journal} {Phys.
  Rev. D}\ }\textbf {\bibinfo {volume} {87}},\ \bibinfo {pages} {124018}
  (\bibinfo {year} {2013})}\BibitemShut {NoStop}%
\bibitem [{\citenamefont {Robertson}\ \emph {et~al.}(2017)\citenamefont
  {Robertson}, \citenamefont {Michel},\ and\ \citenamefont
  {Parentani}}]{2016arXiv161103904R}%
  \BibitemOpen
  \bibfield  {author} {\bibinfo {author} {\bibfnamefont {S.}~\bibnamefont
  {Robertson}}, \bibinfo {author} {\bibfnamefont {F.}~\bibnamefont {Michel}}, \
  and\ \bibinfo {author} {\bibfnamefont {R.}~\bibnamefont {Parentani}},\ }\href
  {\doibase 10.1103/PhysRevD.95.065020} {\bibfield  {journal} {\bibinfo
  {journal} {Phys. Rev. D}\ }\textbf {\bibinfo {volume} {95}},\ \bibinfo
  {pages} {065020} (\bibinfo {year} {2017})}\BibitemShut {NoStop}%
\bibitem [{\citenamefont {{Aberg}}(2006)}]{2006quant.ph.12146A}%
  \BibitemOpen
  \bibfield  {author} {\bibinfo {author} {\bibfnamefont {J.}~\bibnamefont
  {{Aberg}}},\ }\href@noop {} {\bibfield  {journal} {\bibinfo  {journal}
  {eprint arXiv:quant-ph/0612146}\ } (\bibinfo {year} {2006})},\ \Eprint
  {http://arxiv.org/abs/quant-ph/0612146} {quant-ph/0612146} \BibitemShut
  {NoStop}%
\bibitem [{\citenamefont {Baumgratz}\ \emph {et~al.}(2014)\citenamefont
  {Baumgratz}, \citenamefont {Cramer},\ and\ \citenamefont
  {Plenio}}]{QuantifyingCoherence}%
  \BibitemOpen
  \bibfield  {author} {\bibinfo {author} {\bibfnamefont {T.}~\bibnamefont
  {Baumgratz}}, \bibinfo {author} {\bibfnamefont {M.}~\bibnamefont {Cramer}}, \
  and\ \bibinfo {author} {\bibfnamefont {M.~B.}\ \bibnamefont {Plenio}},\
  }\href {\doibase 10.1103/PhysRevLett.113.140401} {\bibfield  {journal}
  {\bibinfo  {journal} {Phys. Rev. Lett.}\ }\textbf {\bibinfo {volume} {113}},\
  \bibinfo {pages} {140401} (\bibinfo {year} {2014})}\BibitemShut {NoStop}%
\bibitem [{\citenamefont {Yuan}\ \emph {et~al.}(2015)\citenamefont {Yuan},
  \citenamefont {Zhou}, \citenamefont {Cao},\ and\ \citenamefont
  {Ma}}]{PhysRevA.92.022124}%
  \BibitemOpen
  \bibfield  {author} {\bibinfo {author} {\bibfnamefont {X.}~\bibnamefont
  {Yuan}}, \bibinfo {author} {\bibfnamefont {H.}~\bibnamefont {Zhou}}, \bibinfo
  {author} {\bibfnamefont {Z.}~\bibnamefont {Cao}}, \ and\ \bibinfo {author}
  {\bibfnamefont {X.}~\bibnamefont {Ma}},\ }\href {\doibase
  10.1103/PhysRevA.92.022124} {\bibfield  {journal} {\bibinfo  {journal} {Phys.
  Rev. A}\ }\textbf {\bibinfo {volume} {92}},\ \bibinfo {pages} {022124}
  (\bibinfo {year} {2015})}\BibitemShut {NoStop}%
\bibitem [{\citenamefont {Streltsov}\ \emph {et~al.}(2015)\citenamefont
  {Streltsov}, \citenamefont {Singh}, \citenamefont {Dhar}, \citenamefont
  {Bera},\ and\ \citenamefont {Adesso}}]{PhysRevLett.115.020403}%
  \BibitemOpen
  \bibfield  {author} {\bibinfo {author} {\bibfnamefont {A.}~\bibnamefont
  {Streltsov}}, \bibinfo {author} {\bibfnamefont {U.}~\bibnamefont {Singh}},
  \bibinfo {author} {\bibfnamefont {H.~S.}\ \bibnamefont {Dhar}}, \bibinfo
  {author} {\bibfnamefont {M.~N.}\ \bibnamefont {Bera}}, \ and\ \bibinfo
  {author} {\bibfnamefont {G.}~\bibnamefont {Adesso}},\ }\href {\doibase
  10.1103/PhysRevLett.115.020403} {\bibfield  {journal} {\bibinfo  {journal}
  {Phys. Rev. Lett.}\ }\textbf {\bibinfo {volume} {115}},\ \bibinfo {pages}
  {020403} (\bibinfo {year} {2015})}\BibitemShut {NoStop}%
\bibitem [{\citenamefont {Yu}\ \emph {et~al.}(2016)\citenamefont {Yu},
  \citenamefont {Zhang}, \citenamefont {Xu},\ and\ \citenamefont
  {Tong}}]{2016arXiv160603181Y}%
  \BibitemOpen
  \bibfield  {author} {\bibinfo {author} {\bibfnamefont {X.-D.}\ \bibnamefont
  {Yu}}, \bibinfo {author} {\bibfnamefont {D.-J.}\ \bibnamefont {Zhang}},
  \bibinfo {author} {\bibfnamefont {G.~F.}\ \bibnamefont {Xu}}, \ and\ \bibinfo
  {author} {\bibfnamefont {D.~M.}\ \bibnamefont {Tong}},\ }\href {\doibase
  10.1103/PhysRevA.94.060302} {\bibfield  {journal} {\bibinfo  {journal} {Phys.
  Rev. A}\ }\textbf {\bibinfo {volume} {94}},\ \bibinfo {pages} {060302}
  (\bibinfo {year} {2016})}\BibitemShut {NoStop}%
\bibitem [{\citenamefont {Zhang}\ \emph {et~al.}(2016)\citenamefont {Zhang},
  \citenamefont {Shao}, \citenamefont {Li},\ and\ \citenamefont
  {Fan}}]{PhysRevA.93.012334}%
  \BibitemOpen
  \bibfield  {author} {\bibinfo {author} {\bibfnamefont {Y.-R.}\ \bibnamefont
  {Zhang}}, \bibinfo {author} {\bibfnamefont {L.-H.}\ \bibnamefont {Shao}},
  \bibinfo {author} {\bibfnamefont {Y.}~\bibnamefont {Li}}, \ and\ \bibinfo
  {author} {\bibfnamefont {H.}~\bibnamefont {Fan}},\ }\href {\doibase
  10.1103/PhysRevA.93.012334} {\bibfield  {journal} {\bibinfo  {journal} {Phys.
  Rev. A}\ }\textbf {\bibinfo {volume} {93}},\ \bibinfo {pages} {012334}
  (\bibinfo {year} {2016})}\BibitemShut {NoStop}%
\bibitem [{\citenamefont {Xu}(2016)}]{PhysRevA.93.032111}%
  \BibitemOpen
  \bibfield  {author} {\bibinfo {author} {\bibfnamefont {J.}~\bibnamefont
  {Xu}},\ }\href {\doibase 10.1103/PhysRevA.93.032111} {\bibfield  {journal}
  {\bibinfo  {journal} {Phys. Rev. A}\ }\textbf {\bibinfo {volume} {93}},\
  \bibinfo {pages} {032111} (\bibinfo {year} {2016})}\BibitemShut {NoStop}%
\bibitem [{\citenamefont {Kiefer}(2001)}]{Kiefer2001}%
  \BibitemOpen
  \bibfield  {author} {\bibinfo {author} {\bibfnamefont {C.}~\bibnamefont
  {Kiefer}},\ }\href {\doibase 10.1007/b13745} {\bibfield  {journal} {\bibinfo
  {journal} {Classical and Quantum Gravity}\ }\textbf {\bibinfo {volume}
  {18}},\ \bibinfo {pages} {L151} (\bibinfo {year} {2001})}\BibitemShut
  {NoStop}%
\bibitem [{\citenamefont {Kiefer}(2004)}]{Kiefer2004}%
  \BibitemOpen
  \bibfield  {author} {\bibinfo {author} {\bibfnamefont {C.}~\bibnamefont
  {Kiefer}},\ }in\ \href {\doibase 10.1007/978-3-540-40968-7_6} {\emph
  {\bibinfo {booktitle} {Decoherence and Entropy in Complex Systems: Selected
  Lectures from DICE 2002}}},\ \bibinfo {editor} {edited by\ \bibinfo {editor}
  {\bibfnamefont {H.-T.}\ \bibnamefont {Elze}}}\ (\bibinfo  {publisher}
  {Springer Berlin Heidelberg},\ \bibinfo {year} {2004})\ pp.\ \bibinfo {pages}
  {84--95}\BibitemShut {NoStop}%
\end{thebibliography}
\end{document}